\documentclass{aa}  

\usepackage{float}
\usepackage{graphicx}
\usepackage{txfonts}
\usepackage{natbib}
\bibpunct{(}{)}{,}{a}{}{,}  
\usepackage{hyperref}
\hypersetup{breaklinks=true,
            colorlinks,
            filecolor=blue,
            linkcolor=blue,
            citecolor=blue,
            urlcolor=blue,
            anchorcolor=blue} 
\begin{document}

   \title{A Walk on the Retrograde Side (WRS) project}

   \subtitle{I. Tidying-up the retrograde halo with high-resolution spectroscopy}
   
   \titlerunning{A Walk on the Retrograde Side. I}
   
   \author{E. Ceccarelli
          \inst{1,2}
          \and
          D. Massari
          \inst{1}
          \and
          A. Mucciarelli
          \inst{2,1}
          \and
          M. Bellazzini
          \inst{1}
          \and
          A. Nunnari
          \inst{3,4}
          \and
          F. Cusano
          \inst{1}
          \and          
          C. Lardo
          \inst{2,1}  
          \and
          D. Romano
          \inst{1}           
          \and
          I. Ilyin
          \inst{5}            
          \and
          A. Stokholm
          \inst{1,2,6,7}            
          }

   \institute{INAF - Astrophysics and Space Science Observatory of 
              Bologna, Via Gobetti 93/3, 40129 Bologna, Italy\\ \email{edoardo.ceccarelli3@unibo.it}
         \and
             Department of Physics and Astronomy, University of 
             Bologna, Via Gobetti 93/2, 40129 Bologna, Italy 
         \and
             INAF - Astronomic Observatory of Rome, Via Frascati 33, 00078 Monte Porzio Catone, Italy
         \and
             Department of Physics, University of Roma Tor Vergata, Via della Ricerca Scientifica 1, 00133 Roma, Italy
         \and
             Leibniz-Institut f\"ur Astrophysik Potsdam (AIP), An der Sternwarte 16, 14482 Potsdam, Germany
         \and
             School of Physics and Astronomy, University of Birmingham, Birmingham B15 2TT, UK
         \and
             Stellar Astrophysics Centre, Department of Physics and Astronomy, Aarhus University, Ny Munkegade 120, DK-8000 Aarhus C, Denmark         
             %\thanks{Just to show the usage
             %of the elements in the author field}
             }

 %  \date{Received XX; accepted YY}

% \abstract{}{}{}{}{} 
% 5 {} token are mandatory
 
  \abstract
  {Relics of ancient accretion events experienced by the Milky Way are predominantly located within the stellar halo of our Galaxy. However, debris from different objects display overlapping distributions in dynamical spaces, making it extremely challenging to properly disentangle their contribution to the build-up of the Galaxy. To shed light on this chaotic context, we started a program aimed at the homogeneous chemical tagging of the local halo of the Milky Way, focusing on the component in retrograde motion, since this is expected to host a large fraction of stars accreted from past mergers. The {\em A Walk on the Retrograde Side} (WRS) project targets retrograde halo stars in the Solar Neighborhood having accurate $6$-D phase space information available, measuring the precise chemical abundance of several chemical elements from high-resolution spectroscopy.
  %Stars with all the $6$-D phase space information available are only a limited fraction of the Solar Neighborhood retrograde halo population and most of them are still lacking a detailed chemical characterization, since this region of the Galaxy remains widely underexplored by large spectroscopic surveys. Thus, we delve into the retrograde halo, adding the information on the detailed chemistry of the dynamical substructures. We use this additional ingredient to tell independent mergers apart, or to recognise substructures associated to the same former progenitor. 
  In this first paper, we present the project and the analysis of high-resolution spectra obtained with UVES at VLT and PEPSI at LBT for $186$ stars. Accurate radial velocity and chemical abundance of several elements have been obtained for all the target stars.
  %These stars were selected by using an innovative approach that takes advantage of tangential velocities only to select retrograde stars with high-fidelity. First, we derived line-of-sight velocities from the spectroscopic analysis and we combined them with Gaia kinematic information to infer the orbital properties of the stars in our sample. Then, we obtained chemical abundances for 16 different elements. 
  In particular we focus on the chemical composition of a specific subset of substructures identified dynamically in the literature. Our study reveals that two among the more recently discovered structures in the retrograde halo, namely Antaeus / L-RL$64$ and ED-$3$, have identical chemical patterns and similar integrals of motion, suggesting a common origin. In turn, the abundance patterns of this unified system differ from that of Gaia-Enceladus, confirming that it is an independent structure. Finally, Sequoia exhibits a different chemistry with respect to that of Gaia-Enceladus at $\mathrm{[Fe/H]} < -1.5$ dex, showcasing an excess of stars with lower Mg and Ca in the common metallicity range.}
  % context heading (optional)
  % {}
  % {} leave it empty if necessary  
  % {}
  % aims heading (mandatory)
  % {}
  % methods heading (mandatory)
  % {}
  % results heading (mandatory)
  % {}
  % conclusions heading (optional), leave it empty if necessary 
  % {} 


   \maketitle
%
%-------------------------------------------------------------------

\section{Introduction}

  The $\Lambda$ cold dark matter ($\Lambda$CDM) cosmological paradigm predicts that galaxies grow their masses hierarchically through merging events with smaller systems \citep{white&frenk1991}. The discovery of the ongoing accretion event of the Sagittarius dwarf galaxy \citep{ibata94} demonstrated for the first time that the Milky Way (MW) is not an exception. As extensively shown by N-body simulations of galaxy formation \citep{newton18}, the stellar halo of the surviving galaxy comprises mainly stars that have been accreted from smaller, disrupted satellite galaxies. Nevertheless, it has been established that the stellar halo is not exclusively composed of accreted objects \citep{pillepich2015,monachesi2019,font2020,belokurov22,khoperskov2023}. Detecting stellar substructures in the Galactic halo yields valuable insights into the formation and evolution of the MW, since the overall configuration of the halo has been deeply affected by the merger history of our Galaxy. Hence, comprehending the characteristics and mechanisms of formation of these substructures plays a pivotal role in reconstructing the evolutionary history of the MW. The search for substructures in phase space is currently the most common and successful method used to unravel ongoing or recent accretion event \citep{ibata94, helmi99, bonaca12, helmi2020, belokurov2020}. However, phase-mixing has proven some limitations to this approach, making it challenging to identify merging events that took place in the early stages of the MW's formation. Instead, N-body simulations have shown that debris from the same progenitor maintain a dynamical coherency in the space defined by the Integrals of Motion (IoM, e.g., energy and angular momentum) in a slowly evolving potential \citep{johnston1996, helmidezeeuw00, gomez13}. In this context, the advent of the ESA/Gaia mission \citep{GC16} has provided a deeper understanding of the MW overall structure, shedding light on its formation and evolution, thanks to the extremely accurate measurements of $3$-D position and velocity for $\approx$ 33.8 million of stars brighter than G = 14.0. According to the widely accepted scenario (see \citealt{helmi2020} for a review), a substantial fraction of the MW stellar halo is believed to have formed as a consequence of a merging event with a dwarf galaxy, known as Gaia-Enceladus / Sausage \citep[GES hereafter,][]{belokurov2018,helmi2018}.
  %\textbf{This merger had a significant impact on the pre-existing MW disc, resulting in an increase of its kinematic dispersion and the formation of the thick disc \citep{helmi2020}. However, the role of GES in influencing the thick disc star formation during early epochs remains uncertain. There are differing perspectives regarding whether this event either enhanced or inhibited star formation \citep{conroy2022,ciuca2023}. Additionally, the chronological sequence of thick disc and thin disc formation is still under debate \citep{prantzos2023}. While the prevailing assumption is that the thick disc formed prior to the thin disc, the possibility of their coeval formation has also been suggested \citep{beraldo2021}. Do you think we should also talk about the bimodality of the [$\alpha$/Fe] trend. Or maybe should we move this entire paragraph about the disc in the section where we compare results with APOGEE and NS10? We can also remove this part on the MW disc.} 
    
  Recent studies revealed that the stellar halo contains an increasing number of other smaller substructures, including streams, clumps and overdensities that are believed to be the remnants of past accretion events \citep{koppelman19,massari19,myeong19,kruijssen2020,naidu20,yuan2020,horta2021,ibata2021,malhna2022,myeong22,oria2022,ruiz-lara22,tenachi22,belokurov23,dodd23,mikkola2023}. However, the interpretation of these numerous recent discoveries can be extremely challenging. In fact, numerical simulations of a major merger between a MW-like galaxy and a GES-like progenitor have revealed that remnants from the same merging event can result in overdensities at various locations in the space defined by the IoM, that can be mistaken for independent substructures \citep{koppelman2020,amarante2022,belokurov23,davies2023}. On the other hand, relics with different origin may significantly overlap in the IoM space \citep{helmidezeeuw00,naidu20,lovdal22,dodd23} and unrelated sources can drastically contaminate samples of candidate members of a given substructure selected kinematically \citep{buder22,rey23}.
  
  A possible solution to this dynamical degeneracy can be achieved by adding an entirely independent dimension to the parameter space, that is the chemical composition of stars \citep[chemical tagging,][]{freeman02}. Indeed, the chemical abundance patterns of individual stars exhibit characteristics that reflect the star formation and chemical enrichment histories of the site where they were born. Therefore, they can greatly help in understanding the origin of a given substructure, e.g., distinguishing between accreted and in-situ formation, and / or to infer the properties of the (now dissolved) progenitor satellite \citep{gallart05,tolstoy09,nissen&schuster2010,mackereth19,kruijssen2020,minelli21,montalban2021,mucciarelli2021NatAs}. 
  
  Large spectroscopic surveys, like, e.g., APOGEE \citep{apogee22}, GALAH \citep{galah2021}, H3 \citep{conroy19}, have made it possible to conduct in-depth studies in this field, providing detailed elemental abundances for an enormous number of stars. Many recent works took advantage of these large chemical datasets, focusing primarily on the chemical patterns of GES \citep{aguado21,bonifacio21,feuillet21,hasselquist21,limberg22}. Still, there are interesting regions of the parameter space where the sampling of chemical abundances is sparse and occasional, as, for example, the highly retrograde halo (RH) that is expected to host a very large fraction of accreted stars \citep{naidu20,myeong22,horta23}. Also, merging of and comparison between chemical abundances from different sources has proven somehow problematic due to the non-homogeneity in the atmospheric parameters and abundances scales of the various surveys \citep[see, e.g.,][]{helmi2020}. 
  
  %Several studies based on large spectroscopic surveys suggest that stars associated with different halo substructures follow separate patterns in the chemical planes. However, as discussed by \citet{helmi2020}, these findings stem from dissimilar and non-homogeneous sources. In other words, it has been difficult to draw a precise picture of all the early merging events experienced by the MW due to the small differences in chemical paths between various progenitors, which are comparable to the typical uncertainties of elemental abundances, as well as the existence of numerous approaches to dinamically unravel substructures. Therefore, high-precision chemical abundance obtained from high-resolution and high signal-to-noise (S/N) spectra are necessary to reliably identify any abundance differences between stars born in independent progenitors. In particular, given that these differences are subtle and blurred by intrinsic spread, homogeneity and internal consistency in the analysis are crucial for identifying distinctions among substructures in the chemical space. \\

  Within this framework, here we present a small spectroscopic survey aimed at delivering detailed chemical tagging of RH stars in the Solar Neighborhood with accurate kinematics, based on the precise measurements of several elements (e.g., Fe, Na, Mg, Al, Ca, Sc, Ti, Mn, Ni, Zn, Y), the {\em A Walk on the Retrograde Side} (WRS) project. Our goal is to get chemical abundances from high-resolution spectroscopy for at least a few hundred stars, using accessible observational facilities. Spectroscopic observations of relatively bright stars (G $\la 15.0$) can be an excellent use of nights with sub-optimal seeing and lunar illumination at large telescopes. This is the niche that is being used for WRS observations.
  
  The multi-dimensional dataset that we plan to obtain will provide deep insight on the origin of each target star, by looking to similarities in the orbits and abundance patterns among them and with known relics of merging events. The approach we propose is complementary to that of large-scale general surveys such as H3 \citep{naidu20, naidu2022} and APOGEE \citep{horta23}. 
  
  The most relevant factors characterizing the WRS project are:
  \begin{itemize}
    \item  WRS is highly focused, targeting only RH stars in the Solar Neighborhood, within $D_{\sun}\le 1.0$ kpc. Retrograde stars are particularly interesting in this context, as it is more likely that they come from the accretion of ancient satellites, with respect to their prograde counterparts.

    \item WRS targets are selected among those having very accurate astrometry from the Gaia mission \citep{GC16}. In addition to the constraint in distance and to the accurate radial velocity measure obtained within WRS, this ensure the most reliable determination of orbital parameters and IoM for the considered stars.
    
      \item The stellar spectra obtained for WRS have higher spectral resolution (R $\approx$ 40,000) than large spectroscopic surveys. This enhanced spectral resolution facilitates the precise determination of chemical abundances and offers finer insights into stellar characteristics.
  \end{itemize}
  
  %The combination of these peculiarities sets this approach as a competitive alternative to the large scale surveys, particularly for achieving the specific research objectives outlined in this study.
  
  In this first paper of the series, we describe our sample selection process and present the results of the analysis of a first batch of 186 stars. In particular, we conduct a high-precision and homogeneous abundance analysis on RH stars that are associated with different dynamical substructures, according to \citet{dodd23}. We compare the results obtained for these substructures with the aim of uncovering their true nature and to understand whether they are all independent accreted components, or if some have originated from the same progenitor.
  %For example, we find that two of the newly discovered substructures, namely L-RL64 and ED-3, might be relics of the same accretion event as they show the same abundance pattern for every chemical element that we analyzed and that these patterns are  different from those of GES.
  
  The paper is structured as follows. In Section \ref{selection} we describe the selection criteria used to identify RH stars and we present the high-resolution spectroscopic dataset. In Section \ref{orbital_parameters} we present the dynamical characteristics of our sample of stars. In Section \ref{identification} we provide a detailed description of the criteria that were employed to define each individual substructure in the RH. In Section \ref{abu} we provide a comprehensive account of the detailed procedure employed to conduct the chemical analysis. In Section \ref{comparison_literature} we show the comparison with literature results. In Section \ref{chemical_discussion} we showcase an extensive examination of the halo substructures in different chemical planes, focusing in particular on odd-Z elements (Section \ref{oddz}), $\alpha$-elements (Section \ref{alpha}), iron-peak elements (Section \ref{ironpeak}) and neutron capture elements (Section \ref{neutronecapture}). In Section \ref{conclusion} we summarize and discuss the main results of our work.
%--------------------------------------------------------------------
\section{Selection of the sample and observations}
%\label{data}
%\subsection{Selection of RH stars}
\label{selection}
\subsection{Data selection}

  At the epoch when the WRS project was conceived, we had to face the problem of selecting retrograde stars from catalogues, the Gaia DR2 \citep{GaiaDR2} and EDR3 \citep{GC21} databases\footnote{That contained $V_{\mathrm{los}}$ for $\simeq 7.2\times 10^6$ stars with G $\la 13.0$, while the most recent release, DR3 \citep{GC23}, includes $V_{\mathrm{los}}$ for $\simeq 33.8\times 10^6$ stars down to G $\la 14.0$}, in which most of the stars were lacking a line of sight velocity ($V_{\mathrm{los}}$) measure in the range of apparent magnitudes of interest (G $\le 14.0$, for the time being). We tackled this issue by realising that a selection in tangential velocity $V_{\mathrm{T}} = \sqrt{V_{\mathrm{ra}}^{2} + V_{\mathrm{dec}}^{2}}$, that is the linear velocity in the plane of the sky, can be used to effectively select very likely retrograde stars.

\begin{figure}
   \centering
   \includegraphics[width=\hsize]{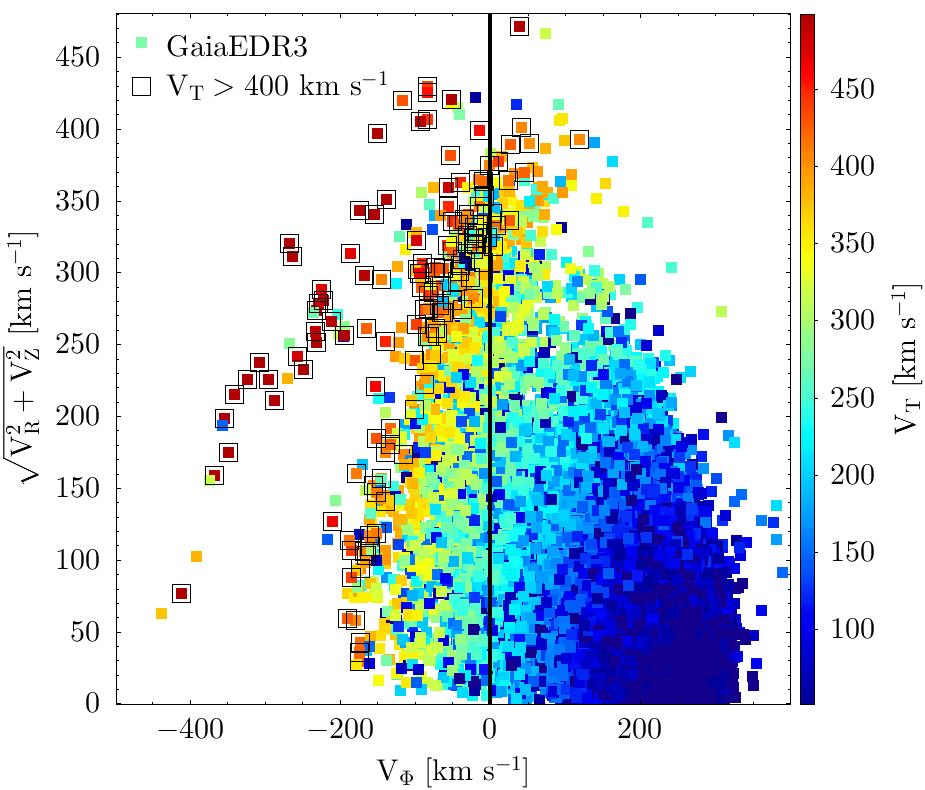}
      \caption{Toomre Diagram of the $\simeq 2.4$ million stars within 
      $1.0$~kpc from the Sun, having $V_{\mathrm{los}}$ measured in Gaia EDR3 and passing all the quality criteria we adopted for the WRS sample.  Stars are color-coded according to their tangential velocity and stars with  $V_{\mathrm{T}}>$ 400 km $\mathrm{s^{-1}}$ are highlighted with a black empty square.}
         \label{Fig_toomre}
   \end{figure}

  The technique is illustrated in Fig.~\ref{Fig_toomre}, where the well-measured Gaia EDR3 stars within 1.0 kpc from the Sun with the complete 6-D phase space information available are plotted in a Toomre diagram color-coded according to $V_{\mathrm{T}}$ \footnote{The adopted solar motion is from \citet{schonrich2010}, while the distance between the Sun and the Galactic Center and the rotational velocity of the Sun are the same used in \citet{mcmillan17}.}. It is quite clear that, in the vicinity of the Sun, high $V_{\mathrm{T}}$ values are displayed only by stars on retrograde orbits or on slightly prograde orbits but high $\sqrt{V_{\mathrm{R}}^2+V_{\mathrm{Z}}^2}$, the latter being equally interesting in our view, as they are typically associated with GES. Hence, we adopted $V_{\mathrm{T}} >$ 400 km $\mathrm{s^{-1}}$ as our criterion to select likely retrograde stars. The criterion proved to be very effective as only 16 out of the 186 stars considered here and selected in this way are not retrograde, all of them having $L_{\mathrm{z}}< 800$ kpc km $\mathrm{s^{-1}}$, while 123 of them have $L_{\mathrm{z}}\le -500$ kpc km $\mathrm{s^{-1}}$. In future observational campaigns we plan to select retrograde targets directly in $V_{\Phi}$, among the stars having $V_{\mathrm{los}}$ measures in the Gaia DR3 catalogue, or in catalogues from future Gaia data releases. This will allow us to sample also the mildly retrograde stars at relatively low total velocity ($-100~\mathrm{km~s^{-1}} \la V_{\Phi}<0.0 ~\mathrm{km~s^{-1}}$ and $\sqrt{V_{\mathrm{R}}^2+V_{\mathrm{Z}}^2}\la 200~\mathrm{km~s^{-1}}$) that are now excluded by our selection in $V_{\mathrm{T}}$, as well as to target specific substructures.

  The additional selection criteria we adopted were intended to select well-measured stars in the Solar Neighborhood. In particular we selected our targets among $V_{\mathrm{T}} >$~400~km~$\mathrm{s^{-1}}$ stars having:

\begin{enumerate}
    \item Magnitude G $\la 14.0$. This suitable magnitude range was selected in order to get high signal to noise (S/N) high resolution spectra with 8~m class telescopes with relatively short exposures (typically $\le 3000$ s).
    \item Error on the parallax lower than 10\%. This implies small errors in the distance, here always derived as $1/{\tt parallax}$, and in the proper motion.
    \item Distance from the Sun $D_{\sun}\le 1.0$ kpc.
    \item Renormalized Unit Weight Error, ${\tt RUWE}\le 1.3$ \citep{GC21} and Corrected BP and RP flux excess factor, $C^{\star} < 3\sigma$ \citep{riello2021}. These criteria ensure that the Gaia astrometry and photometry, respectively, of the target stars are of the highest quality.
    The first criterion is also effective in excluding astrometric binaries \citep[see][and references therein]{shion22}.
    \item Finally, inspecting the color-magnitude diagram, we excluded by hand a few dwarf stars that may be unresolved binaries, being located slightly above the main sequence.
    
\end{enumerate}

   The most relevant outcome of the selections above, complemented by uncertainties in the $V_{\mathrm{los}}$ measured from high-resolution spectra lower than 0.9~km~$\mathrm{s^{-1}}$, is that the typical uncertainty on the IoM is just a few per cent (see Section ~\ref{orbital_parameters}), hence the noise in the phase space is minimal in our sample. Basic information on our target stars is listed in Table \ref{TabMain}.

%-------------------------------------- Figure
%--------------------------------------- 
%-------------------------------------- Figure
   \begin{figure*}
   \centering
   \includegraphics[width=\hsize]{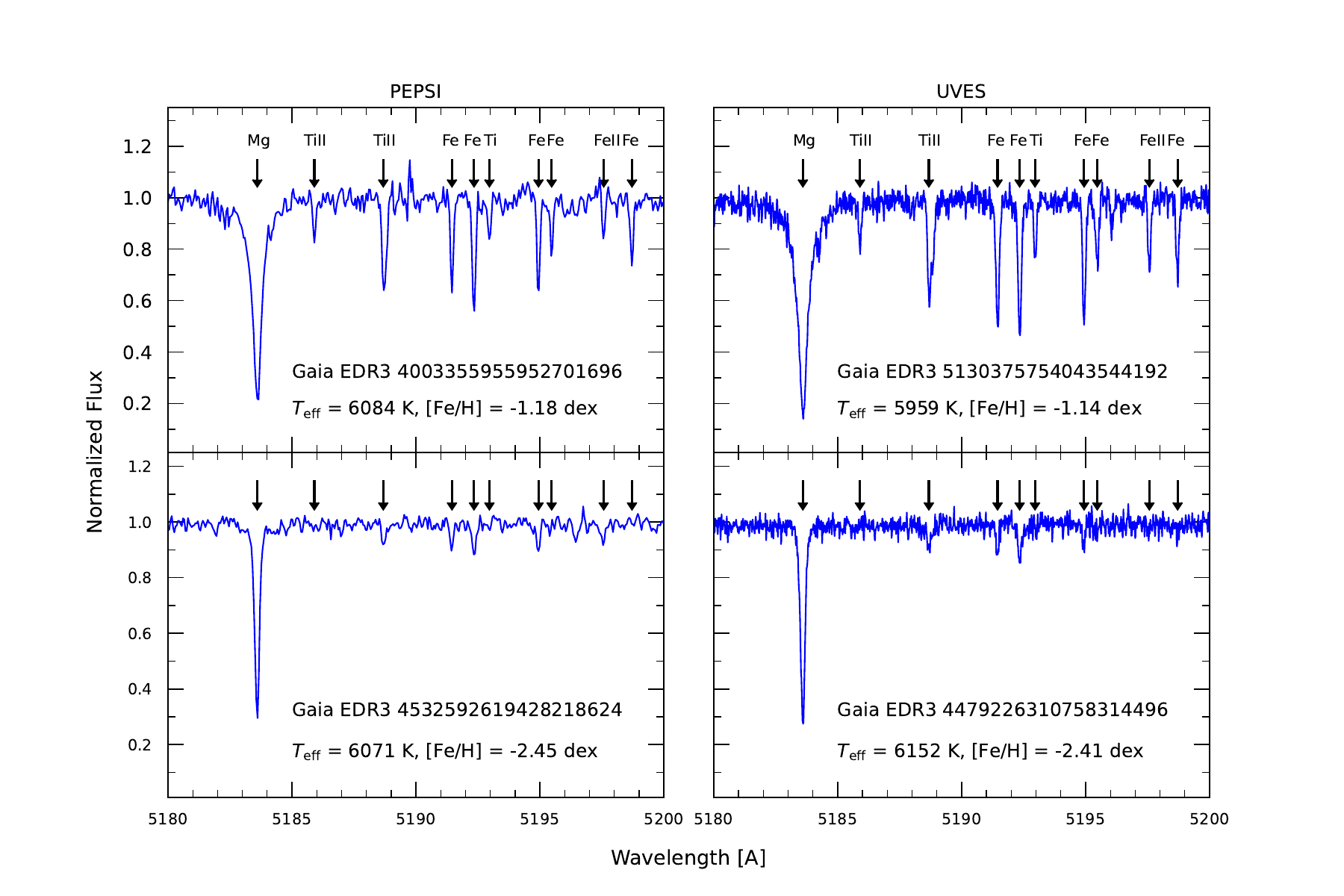}
      \caption{Left column: comparison between the spectra of two stars observed with PEPSI at LBT. Right column: same but for two stars observed with UVES at VLT. These stars have very similar $T_{\mathrm{eff}}$ and log $g$ but they differ in the iron content. Black arrows indicate the position of some atomic lines of interest.}
         \label{FigSpectra}
   \end{figure*}
%--------------------------------------- 
%------------------------ Figure ---------------------------
   \begin{figure*}
   \centering
   \begin{minipage}{0.33\textwidth}
        \centering
        \includegraphics[width=1.0\textwidth]{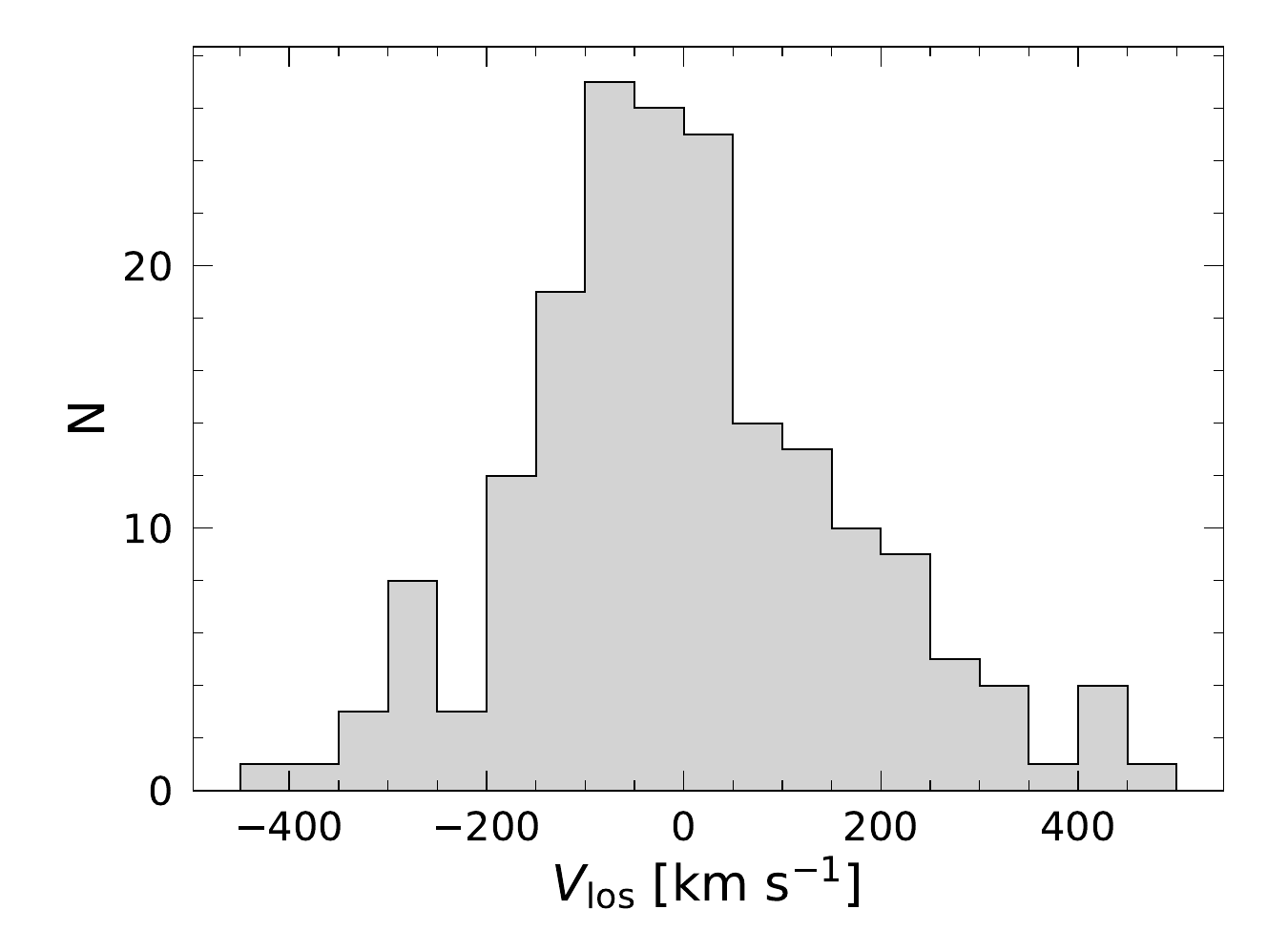} 
   \end{minipage}
   \begin{minipage}{0.33\textwidth}
        \centering
        \includegraphics[width=1.0\textwidth]{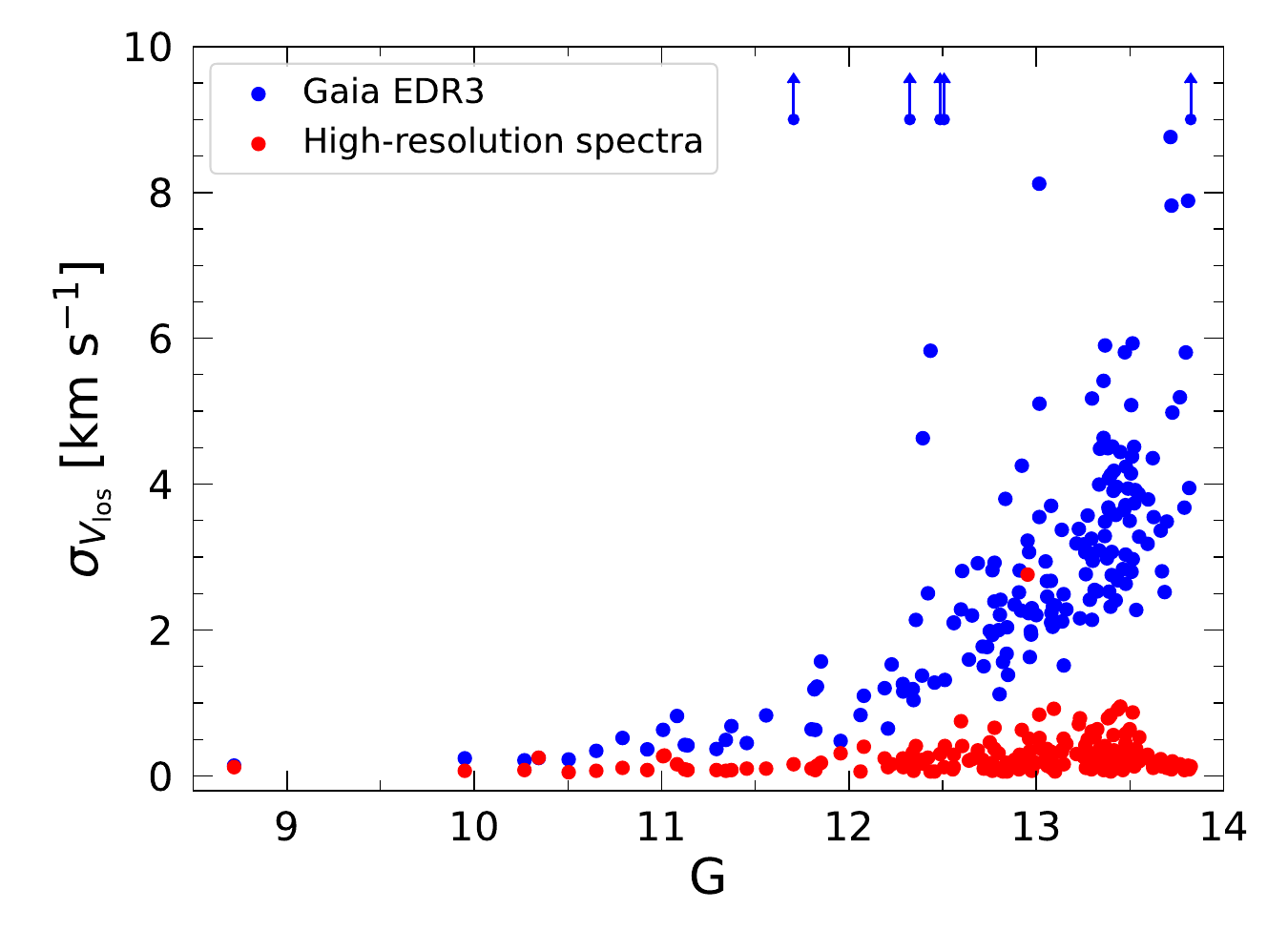}
   \end{minipage}
   \begin{minipage}{0.33\textwidth}
        \centering
        \includegraphics[width=1.0\textwidth]{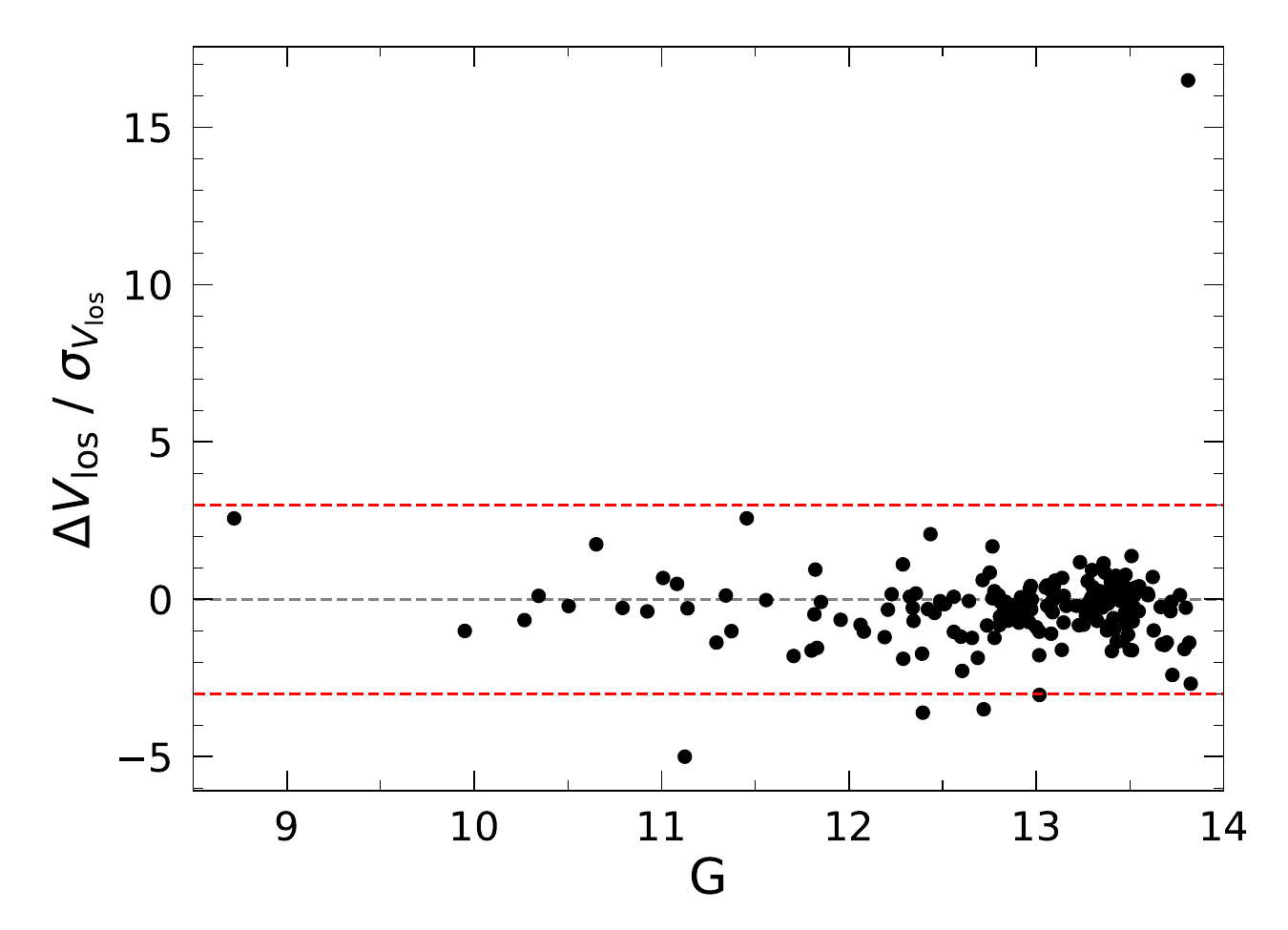}
   \end{minipage}
   \caption{Left panel: line-of-sight velocity distribution for the entire stellar sample. Central panel: comparison between the uncertainty on the line-of-sight velocity provided by Gaia (blue filled circles) and the one measured with high-resolution spectroscopy (red filled circles) as a function of observed G magnitude. The blue arrows point out the outliers in the distribution, specifically representing target stars with errors in the Gaia velocity measurement larger than 10 km $\mathrm{s^{-1}}$. Right panel: difference between the two measurements of the line-of-sight velocity. The red dashed lines mark the $\pm 3 \sigma$ threshold.}
              \label{vlos}%
    \end{figure*} 
%%-------------------------------------- Figure
%   \begin{figure}
%   \centering
%   \includegraphics[width=\hsize]{figures/vlos/GR.pdf}
%      \caption{Corrections on the line-of-sight velocity measurements due to gravitational redshift effect for the entire target sample. Stars are color coded by their effective temperature.}
%         \label{vlos_GR}
%   \end{figure}
%-------------------------- Table -------------------------
%
\begin{table*}
\caption{Main target information: ID, coordinates, apparent magnitudes and parallax from Gaia EDR3, color excess from the \texttt{EXPLORE} tool and the instrument used to collect the spectrum. The entire table is available in electronic form.}          
\label{TabMain}      
\centering          
\begin{tabular}{c c c c c c c c c}  
\hline      
ID Gaia EDR3 & RA & DEC & G & BP & RP & E(B-V) & $\varpi$ & Instrument \\ 
 & (deg) & (deg) & (mag) & (mag) & (mag) & (mag) & (mas) &   \\ 
\hline 
1156158780372565120 	 & 	229.213078	 & 	 4.816404	 & 	13.0165	 & 	13.2591	 & 	12.6141	 & 	0.04	 & 	1.5420	 & 	    UVES/PEPSI	  \\
6155896330944952576 	 & 	194.454739	 & 	-35.140162	 & 	13.0569	 & 	13.3303	 & 	12.6182	 & 	0.04	 & 	2.3972	 & 	    UVES	  \\
3028486001397877120 	 & 	111.278546	 & 	-15.569461	 & 	12.9617	 & 	13.2308	 & 	12.5292	 & 	0.08	 & 	2.0761	 & 	    UVES	  \\
2391446689585357568 	 & 	351.905432	 & 	-21.672762	 & 	13.0159	 & 	13.2788	 & 	12.5879	 & 	0.02	 & 	1.4260	 & 	    UVES	  \\
6661345365288720000 	 & 	285.309488	 & 	-48.777078	 & 	12.9113	 & 	13.1568	 & 	12.5012	 & 	0.04	 & 	1.6681	 & 	    UVES	  \\
6378884813840372864 	 & 	347.851094	 & 	-74.444135	 & 	13.0780	 & 	13.3274	 & 	12.6651	 & 	0.03	 & 	1.5581	 & 	    UVES	  \\
4479226310758314496 	 & 	277.960058	 & 	 8.598236	 & 	12.9231	 & 	13.2411	 & 	12.4282	 & 	0.11	 & 	2.8563	 & 	    UVES/PEPSI	  \\
4453220730438373504 	 & 	242.562536	 & 	 9.139019	 & 	12.9657	 & 	13.2278	 & 	12.5389	 & 	0.04	 & 	2.0442	 & 	    UVES/PEPSI	  \\
4752251952905666048 	 & 	42.700412	 & 	-47.853414	 & 	12.9728	 & 	13.1988	 & 	12.5982	 & 	0.01	 & 	1.3873	 & 	    UVES	  \\
4855735169813450624 	 & 	57.095446	 & 	-39.144457	 & 	13.0002	 & 	13.2695	 & 	12.5622	 & 	0.01	 & 	2.8394	 & 	    UVES	  \\
... 	 & ...	 & ...	 & ...	 & ...	 & ...	 & ...	 & ...	 & ...  \\

\hline                  
\end{tabular}
\end{table*}
%
%-------------------------------------------------------------

\subsection{High-resolution spectroscopy for selected stars}

   At the present stage, we have collected high-resolution spectra for 98 stars (Programme IT-2021B-004 and Programme IT-2022B-006, P.I.: M. Bellazzini) with the optical spectrograph PEPSI \citep{pepsi} mounted at the Large Binocluar Telescope in Arizona. The observations have been performed with the CD3 grating for the Blue Arm, that cover the spectral range between 4800 - 5440 \r{A}, and the CD6 grating for the Red Arm with a spectral coverage between 7410 - 9140 \r{A}, both with a spectral resolution of 40,000. The choice of this set-up for the two arms was dictated by the trade-off between the access to measurable lines, allowing to derive the abundance of scientifically relevant chemical elements, and the efficiency of the spectrograph with the goal of achieving S/N $>$ 30 for a reliable measure of the chemical abundances we are interested in. In order to maximize the number of observed stars, we aimed for short exposures. This translates into a magnitude limit of G $<$ 14. %Thus, we opted for the CD3 grating for the Blue Arm due to its superior efficiency compared to CD1 and CD2. As CD3 and CD4 cannot be used simultaneously, we chose CD6 in combination with CD3, since it allows us to recover atomic lines for a broader number of chemical elements compared to CD5.
   CD3 is less efficient than CD6, therefore it drives the exposure times. The short exposures allow us to reach typically S/N $\ga$ 40 for the CD3 grating and S/N $\ga$ 60 for CD6. The spectra have been reduced with the dedicated PEPSI pipeline, which contains bias subtraction, flat-fielding, spectral extraction, wavelength calibration and normalization. The observations of close sky region in the same exposure times of the targets have been executed in order to subtract the sky contribution to the stellar spectra.
   
   We also observed 100 stars (Programme 0109.B-0522, P.I.: A. Mucciarelli) with the optical spectrograph UVES \citep{dekker2000} mounted at the Very Large Telescope of the European Southern Observatory. Observations were performed with UVES in Dichroic mode adopting the standard settings Dic 1 Blue Arm CD2 390 (3600 - 4800 \r{A}) and Dic 1 Red Arm CD3 564 (4800 - 6600 \r{A}) and with the 1"x12" slit, thus yielding a resolution of R=40,000. We are able to obtain on average S/N $\ga$ 20 for the Blue Arm and S/N $\ga$ 40 for the Red Arm. All the observed spectra have been reduced with the ESO pipeline\footnote{\url{https://www.eso.org/sci/software/pipelines/}}.
   
   The spectra of 12 targets have been collected with both instruments as a reference point for the purpose of calibration and direct comparison.
      
   Detailed information on some representative stellar spectra are listed in Table \ref{TabSpectrum}, bracketing the range of exposure times adopted and reporting the reached S/N in several spectral region of interest. As an example of the quality of the spectroscopic dataset, we plot in Figure \ref{FigSpectra} the spectra of four target stars with very similar atmospheric parameters and a difference in [Fe/H] of $\approx$ 1.2 dex.

%-------------------------- Table -------------------------
%
\begin{table*}
\caption{Magnitude, exposure times, airmass and S/N ratios for some of the observed target spectra with the spectrographs UVES and PEPSI.}        
\label{TabSpectrum}      
\centering          
\begin{tabular}{c c c c c c c c}  
\hline      
ID Gaia EDR3 & Instrument & G & Exposure time & Airmass & S/N & S/N & S/N \\ 
 & & (mag) & (s) &  & ($4200$ \r{A}) & ($5200$ \r{A}) & ($8500$ \r{A})   \\ 
\hline 
2503491051919554304 	 & 	UVES	 & 12.28989	& 400	 & 	1.482	 	 & 	34	 & 	60	 & 	    -	  \\
3531611252266342400 	 & 	UVES	 & 12.39454	& 600	 & 	1.028	 	 & 	36	 & 	62	 & 	    -	  \\
4479226310758314496 	 & 	UVES	 & 12.92314	& 1000	 & 	2.388	 	 & 	24	 & 	42	 & 	    -	  \\
5146060253053394176 	 & 	UVES	 & 13.29777	& 1200	 & 	2.055	 	 & 	32	 & 	51	 & 	    -	  \\
\hline 
3857833427353671808 	 & 	PEPSI	 & 11.45480	& 1200	 & 	1.459	 	 & 	-	 & 	85	 & 	   130	  \\
4479226310758314496 	 & 	PEPSI	 & 12.92314	& 1600	 & 	1.144	 	 & 	-	 & 	48	 & 	    68	  \\
4566011038396861440 	 & 	PEPSI	 & 13.67105	& 2400	 & 	1.270	 	 & 	-	 & 	44	 & 	    62	  \\
2565307863375846912 	 & 	PEPSI	 & 13.8165	& 3000	 & 	1.213	 	 & 	-	 & 	63	 & 	    77	  \\

\hline                  
\end{tabular}
\end{table*}
%
%-------------------------------------------------------------
   
%-------------------------------------------------------------   
\section{Orbital parameters}\label{orbital_parameters}
   Position, proper motion and parallax measurements were taken from Gaia EDR3 \citep{GC21}. We corrected each star's parallax for zero-point offsets following the prescriptions described by \citet{lindegren21}.  $V_{\mathrm{los}}$ has been measured from the high-resolution spectra by cross-correlation with template spectra using the \texttt{fxcor} task of \texttt{IRAF}\footnote{IRAF was distributed by the National Optical Astronomy Observatory, which was managed by the Association of Universities for Research in Astronomy (AURA) under a cooperative agreement with the National Science Foundation.}. For each star, as template spectrum we exploited a synthetic spectrum calculated as discussed in Section \ref{ls}, using the correct atmospheric parameters. The $V_{\mathrm{los}}$ is measured at the maximum peak of the cross-correlation function. Uncertainties on the line-of-sight velocity are computed following the method described in \citet{tonry&davies1979}. As shown in the central panel of Fig. \ref{vlos}, they are of the order of $0.1 - 0.9$ km $\mathrm{s^{-1}}$, significantly reducing the uncertainty arising from the Gaia measurement, especially for faint stars. Furthermore, we corrected the measured $V_{\mathrm{los}}$ for the gravitational redshift effect following the prescription described by \citet{zwitter18}. Since most of our stars are dwarfs and the precision of our $V_{\mathrm{los}}$ measures is so high, this correction, albeit small, is non-negligible, ranging from $0.1 - 0.7$ km s$^{-1}$.
   The distribution of the final $V_{\mathrm{los}}$ is displayed in the left panel of Fig. \ref{vlos} and the values are listed in Table \ref{TabOP}. The difference between our measurements and those performed by Gaia DR3 \citep{GC23,katz23}, defined as $\Delta V_{\mathrm{los}} = V_{\mathrm{los, spectrum}} - V_{\mathrm{los, GaiaDR3}}$, is consistent within $3\sigma$ for all except five stars, where $\sigma$ is calculated as the squared sum of the uncertainties arising from high-resolution spectroscopy and Gaia DR3 (see right panel of Fig. \ref{vlos}). The discrepancy observed in these five stars could hint at a binary nature for these objects. This implies the possibility of their having undergone a distinct and unconventional chemical evolution or either having the spectrum contaminated by the companion star. Consequently, we flag these stars with a distinct marking in Table \ref{TabOP}, and they will be omitted from the following chemical analysis. The final $V_{\mathrm{los}}$ have been coupled with the parallax and proper motions provided by the EDR3 to draw the complete $6$-D phase space information for each target.
   
   We then transformed the observed kinematic information to the Galactocentric reference frame. To do so, we assume the Sun to be located $R_{0} = 8.122$ kpc away from the Galactic Centre \citep{gravitycollaboration18} and $z_{\odot} = 20.8$ pc above the midplane \citep{bennetandbovy19}. We obtain the solar velocity by combining the definition of the local standard of rest by \citet{schonrich2010} with the proper motion of Sgr $\mathrm{A^{\ast}}$ \citep{reidandbrunthaler04}. In this frame, the Sun moves with ($U_{\odot}$, $V_{\odot}$, $W_{\odot}$) = (12.9, 245.6, 7.78) km $\mathrm{s^{-1}}$ \citep{drimmelandpoggio18}. \\ We computed stars orbital parameters with the software \texttt{AGAMA} \citep{vasiliev19} using a \citet{mcmillan17} potential for the MW. We ran 100 Monte Carlo simulations of the orbit for each star assuming Gaussian distributions for the uncertainties in distance, proper motion and radial velocity. The final values of the dynamical parameters are computed as the median of their derived distributions, with associated uncertainties at the 16th and 84th percentiles. The typical error in $E$, $L_{\mathrm{z}}$ and $L_{\perp}$ are $\la$ 0.8\%, $\la$ 2.6\% and $\la$ 1.8\%, respectively.
   The distribution of our sample in these dynamical spaces is shown in Section \ref{identification}.
 %------------------------ Figure ---------------------------
   \begin{figure*}
   \centering
   \includegraphics[width=1.0\textwidth]{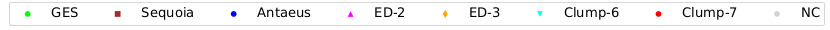}
   \begin{minipage}{0.33\textwidth}
        \centering
        \includegraphics[width=1.0\textwidth]{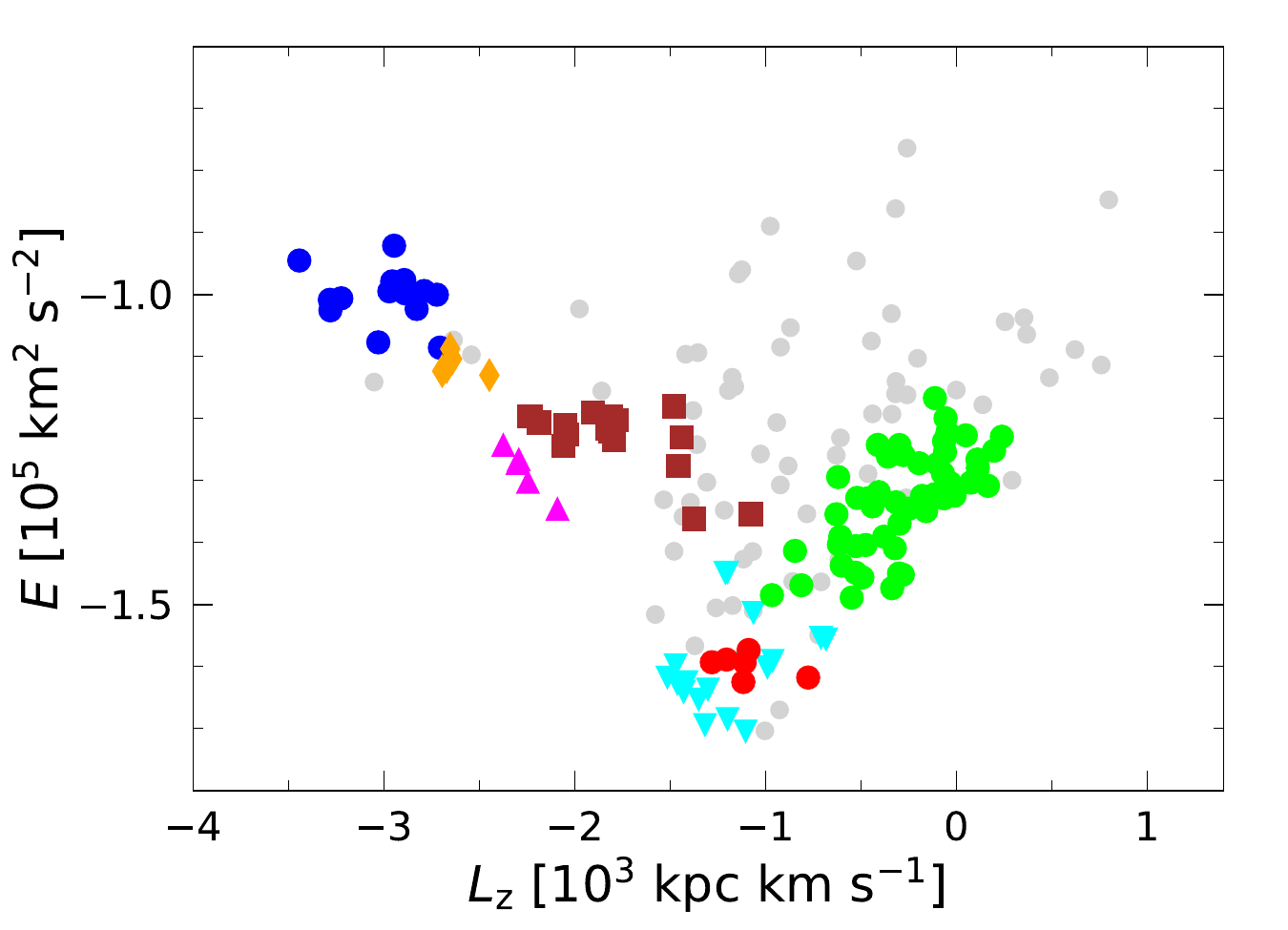} 
        \includegraphics[width=1.0\textwidth]{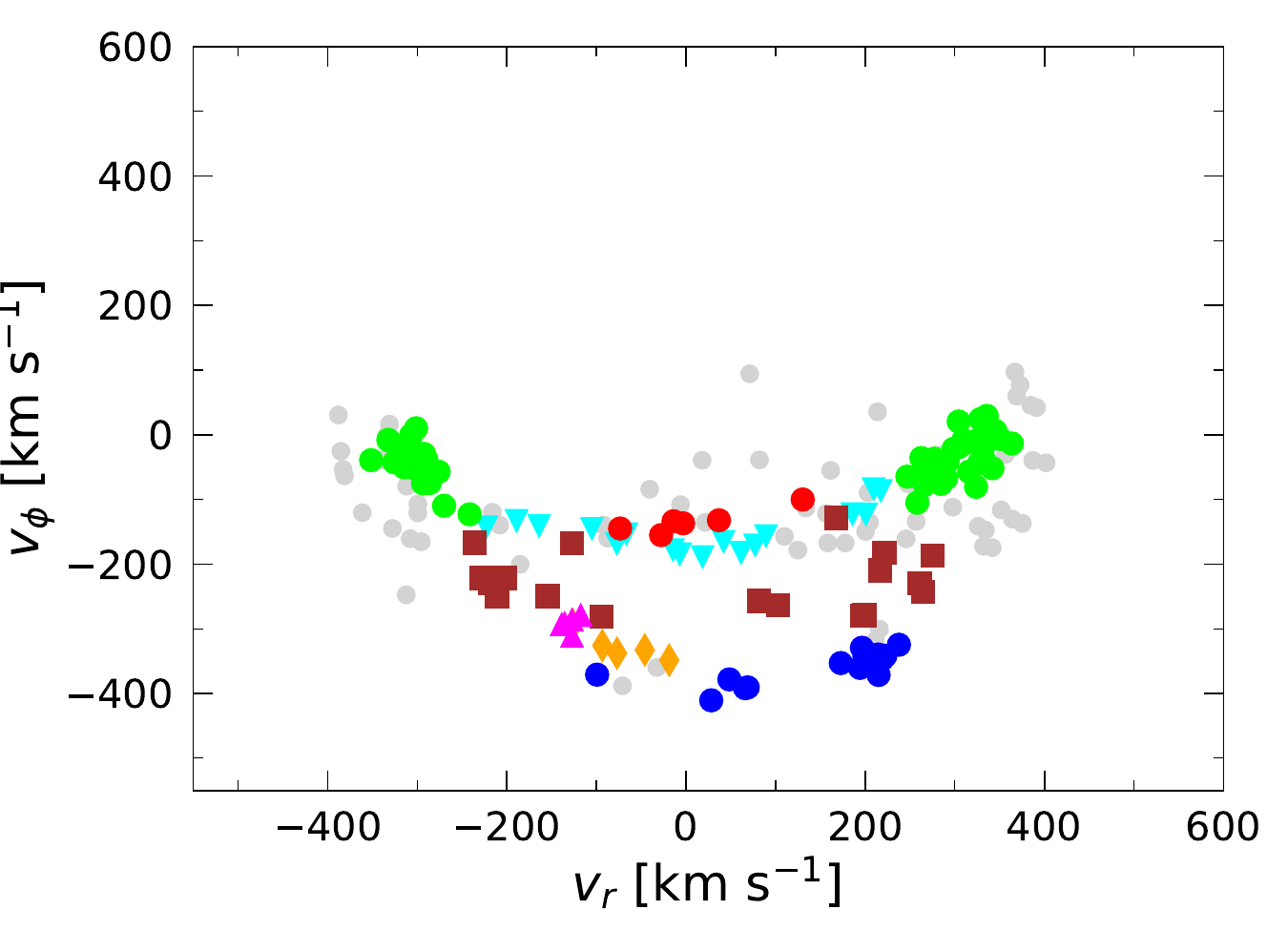}
   \end{minipage}\hfill
   \begin{minipage}{0.33\textwidth}
        \centering
        \includegraphics[width=1.0\textwidth]{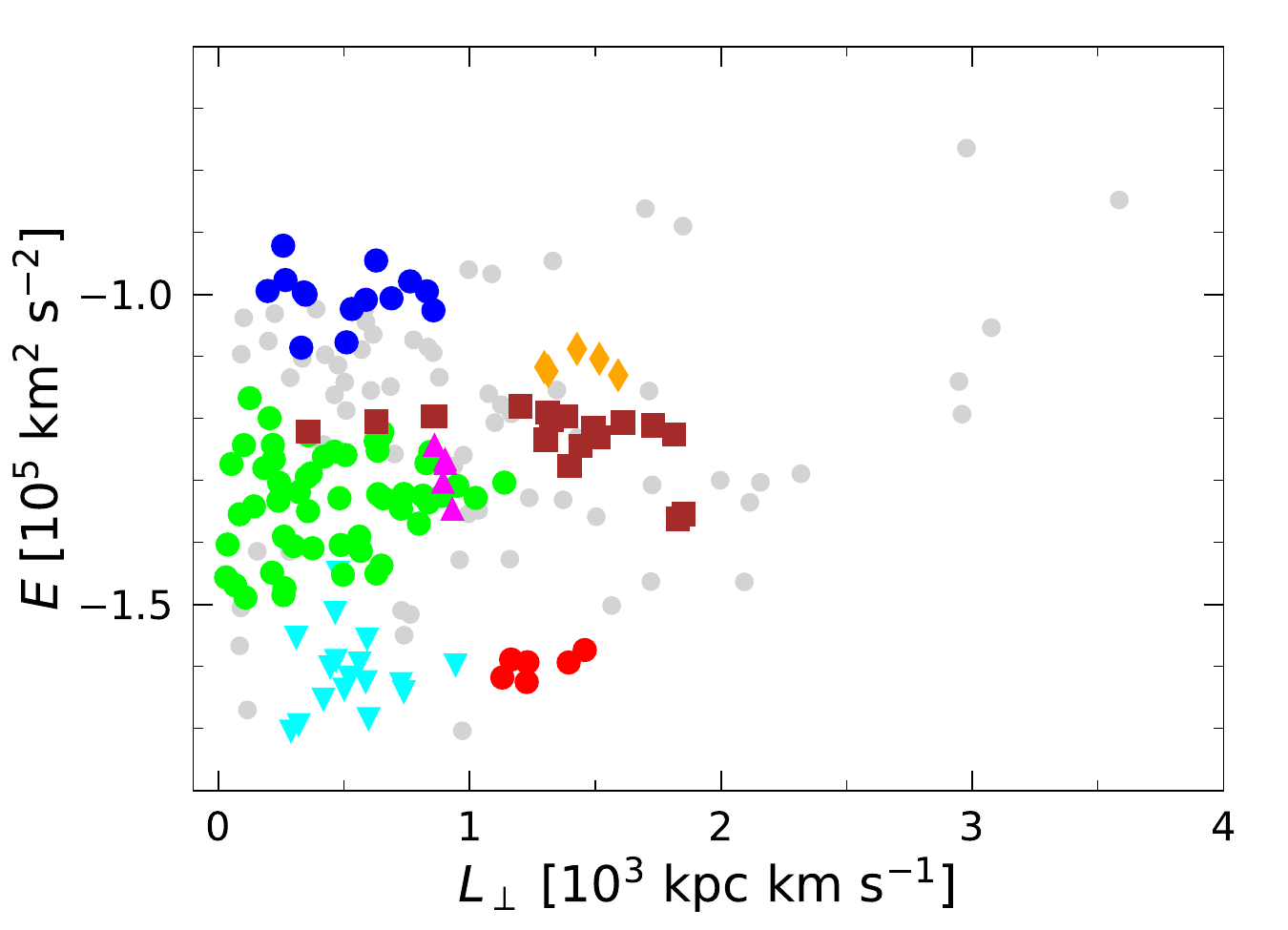}
        \includegraphics[width=1.0\textwidth]{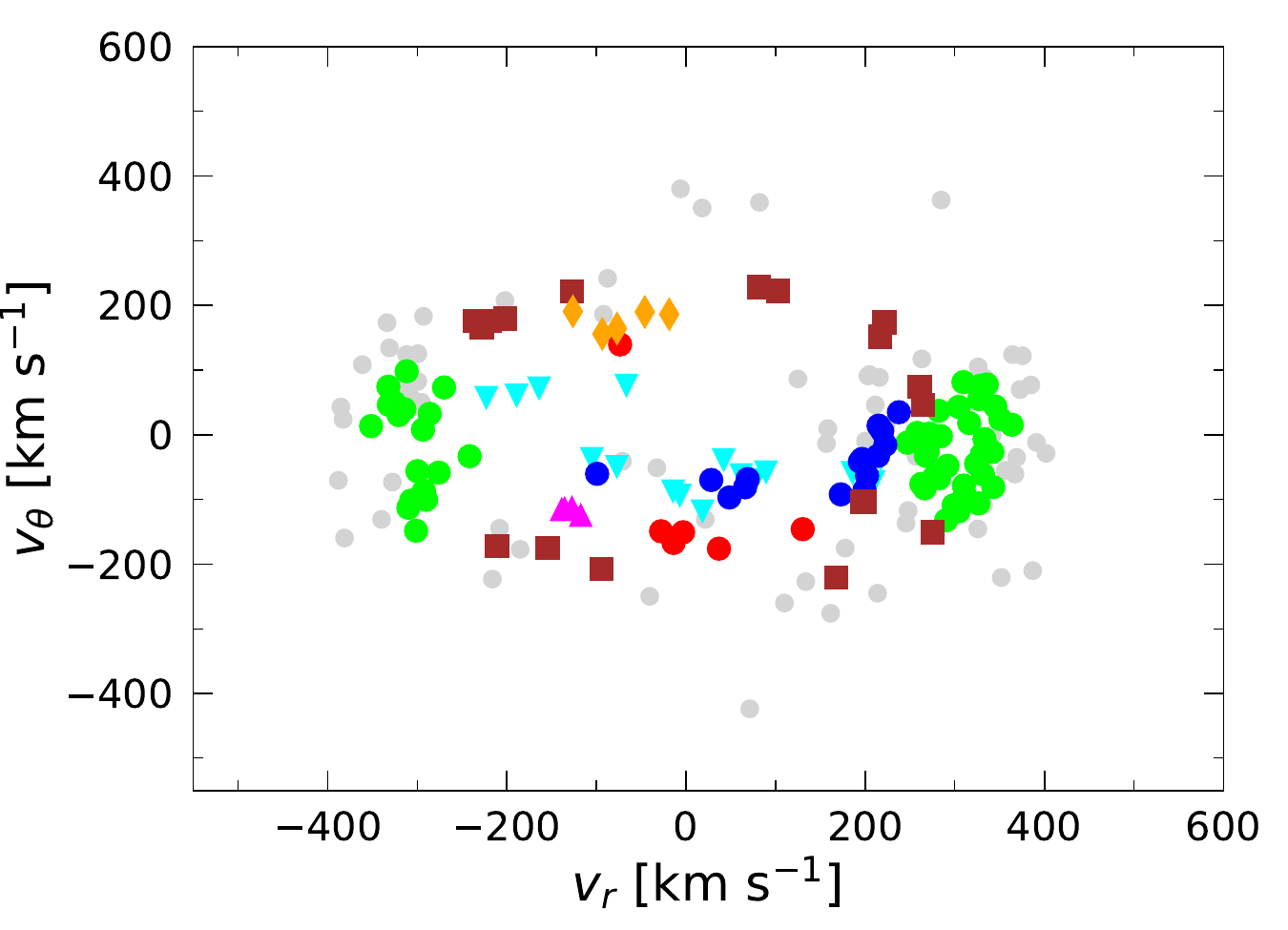}
   \end{minipage}
   \begin{minipage}{0.33\textwidth}
        \centering
        \includegraphics[width=1.0\textwidth]{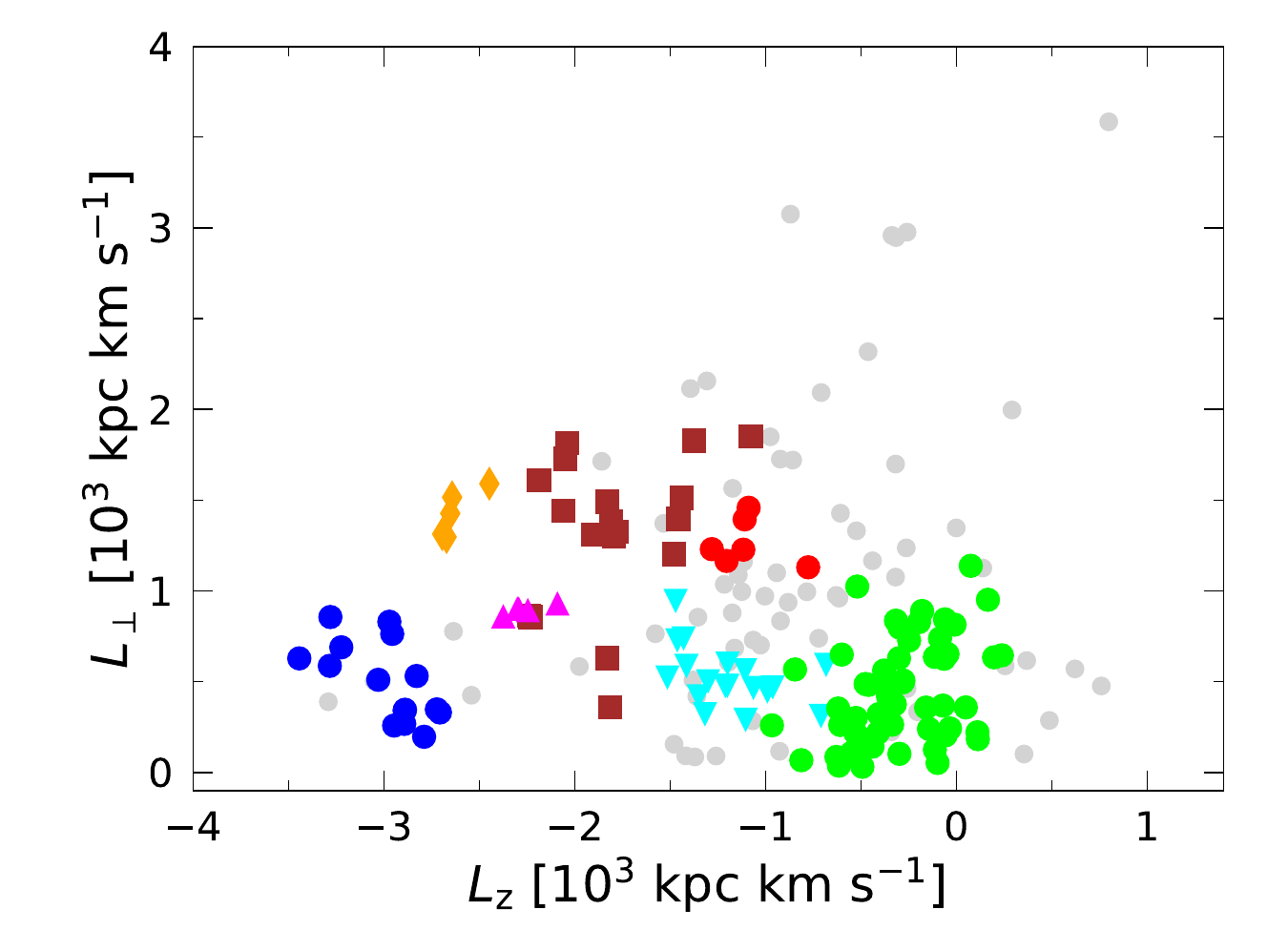}
        \includegraphics[width=1.0\textwidth]{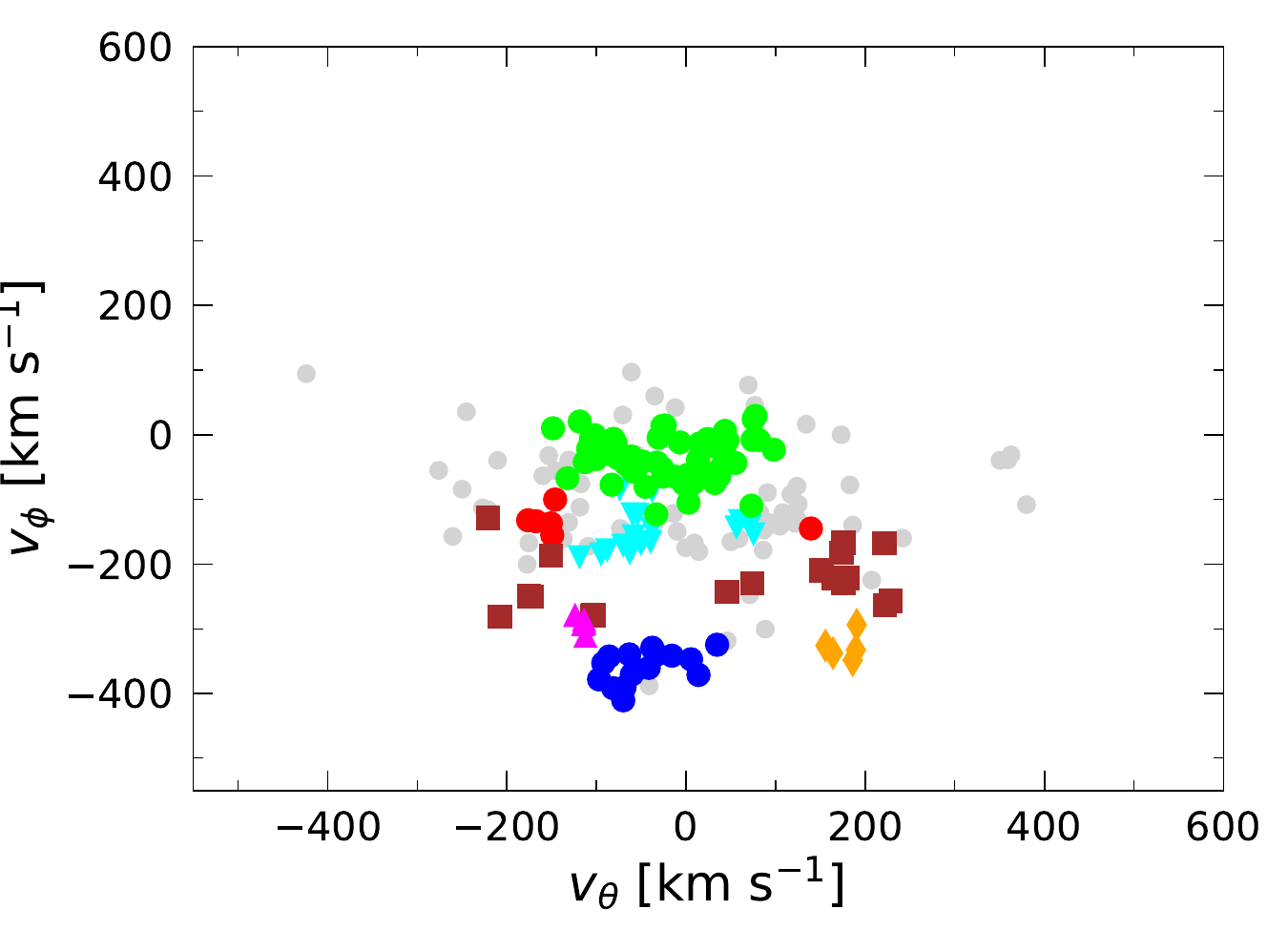}
   \end{minipage}
   \caption{Distribution of the observed stars in IoM and velocity space (top and bottom panels, respectively). We highlight stars associated with various substructures with different colors (green for GES, brown for Sequoia, blue for Antaeus, magenta for ED-2, orange for ED-3, cyan for Clump-6 and red for Clump-7).}
              \label{FigIoM}%
    \end{figure*}    

%---------------------------------------------------------------
%-------------------------- Table -------------------------
%
\begin{table*}
\caption{Kinematic information for the stars: ID from Gaia EDR3, orbital energy, angular momentum along the $z$-axis, perpendicular angular momentum and line-of-sight velocity with uncertainties. The entire table is available in electronic form.}          
\label{TabOP}      
\centering          
\begin{tabular}{c c c c c c c c c c}  
\hline      
ID Gaia EDR3 & E & $\sigma$(E) & $L_{\mathrm{z}}$ & $\sigma$($L_{\mathrm{z}}$) & $L_{\mathrm{perp}}$ & $\sigma$($L_{\mathrm{perp}}$) & $V_{\mathrm{los}}$ & $\sigma$ ($V_{\mathrm{los}}$) & Binary Star\\ 
 & \multicolumn{2}{c}{(x$10^{5}$ $\mathrm{km^{2}}$ $\mathrm{s^{-2}}$)} & \multicolumn{2}{c}{(kpc km $\mathrm{s^{-1}}$)} & \multicolumn{2}{c}{(kpc km $\mathrm{s^{-1}}$)} & \multicolumn{2}{c}{(km $\mathrm{s^{-1}}$)} & \\ 
\hline 
1156158780372565120 	 & 	-1.307	 & 	0.010	 & 	-923	 & 	  20	 & 	1726	 & 	  18	 & 	-89.6	 & 	0.4	     &  no \\
6155896330944952576 	 & 	-1.323	 & 	0.009	 & 	 -86	 & 	  19	 & 	 738	 & 	  10	 & 	-10.2	 & 	0.4	     &  no 	  \\
3028486001397877120 	 & 	-1.236	 & 	0.008	 & 	 -65	 & 	  24	 & 	 626	 & 	   4	 & 	-18.2	 & 	0.5	     &  no 	  \\
2391446689585357568 	 & 	-1.254	 & 	0.011	 & 	 -60	 & 	  17	 & 	 843	 & 	  18	 & 	-6.4	 & 	0.8	     &  no 	  \\
6661345365288720000 	 & 	-1.463	 & 	0.010	 & 	-858	 & 	  24	 & 	1721	 & 	  14	 & 	32.0	 & 	0.3	     &  no 	  \\
6378884813840372864 	 & 	-1.409	 & 	0.006	 & 	-323	 & 	  13	 & 	 375	 & 	  11	 & 	66.4	 & 	0.1	     &  no 	  \\
4479226310758314496 	 & 	-1.302	 & 	0.015	 & 	-2246	 & 	  29	 & 	 894	 & 	   7	 & 	-264.8	 & 	0.6	     &  no 	  \\
4453220730438373504 	 & 	-1.211	 & 	0.017	 & 	-2049	 & 	  33	 & 	1729	 & 	  20	 & 	-83.8	 & 	0.3	     &  no 	  \\
4752251952905666048 	 & 	-1.140 & 	0.008	 & 	-317	 & 	  17	 & 	2947	 & 	  16	 & 	-138.8	 & 	0.2	     &  no 	  \\
4855735169813450624 	 & 	-1.362	 & 	0.009	 & 	-1358	 & 	  19	 & 	1829	 & 	  19	 & 	23.8	 & 	0.2	     &  no 	  \\
... 	 & ...	 & ...	 & ...	 & ...	 & ...	 & ...	 & ...	 & ...	     \\

\hline                  
\end{tabular}
\end{table*}
%---------------------------------------------------------------
\section{Identification of substructures in the RH}\label{identification}

   Recent studies have revealed that the RH exhibits plenty of substructures that significantly contribute  to the total mass budget of the stellar halo \citep{naidu20}. In most cases, they have been identified as overdensities in the space of IoM, such as their actions, the total energy ($E$), the angular momentum along the $z$-direction ($L_{\mathrm{z}}$) and the angular momentum perpendicular to $L_{\mathrm{z}}$ ($L_{\mathrm{\perp}}$). We note that $L_{\mathrm{\perp}}$ has been extensively used to identify stars formed in the same progenitor, even if it is not completely conserved in an axisymmetric potential such as the one we assumed for our galaxy \citep{helmi99,helmidezeeuw00,bonaca2021}.
   %To avoid any potential biases induced by our selection criterion, by the relatively low number of stars in our sample, or by using the chemical abundances as further parameter, we follow the classification provided by \citet{dodd23} (D23 hereafter), 
  
   As a first application of this first bunch of WRS data, we attempt the chemical characterisation of some of the substructures recently listed and discussed by \citet{dodd23} (D23 hereafter), who build up on the findings by \citet{lovdal22} and \citet{ruiz-lara22}. These authors used a specifically designed clustering algorithm based on the IoM computed using Gaia DR3 data to identify groups of stars sharing similar properties in the phase space. 
   
   In the region of the $E$, $L_{\mathrm{z}}$, and $L_{\mathrm{\perp}}$ space covered by our sample, the most largely represented among the D23 groups is GES, of which we sample the most retrograde and high energy slice, due to our selection window. In addition to this, there are two mid-sized slightly retrograde substructures associated to Sequoia \citep{myeong19} and Thamnos \citep{koppelman19}, and three newly-found small, high-energy and extremely retrograde clumps, namely L-RL64, also known as Antaeus \citep{oria2022}, ED-2 and ED-3, plus other minor groups that do not appear to have significant overlap with our sample (see below). Indeed, we find that 179 of our 186 stars are included in the D23 sample. 55 of them are associated with one of the D23 groups. However, there are five groups that have just one member star included in our sample, namely Thamnos, ED-5, ED-6, Group 13 and Group 14. Since no meaningful conclusion can be drawn from an individual star we do not consider these substructures anymore in the following. Still, we will make the chemical abundances available (see Table \ref{TabAbu}) since they might be of interest for upcoming investigations. Of the remaining 50 stars, 12 are associated to GES, 18 to Sequoia, 11 to Antaeus, 4 to ED-2 and 5 to ED-3.
   
   Starting from this 50 "core members", and in order to increase the number of stars associated to a substructure, we performed a clustering analysis with the algorithm \texttt{DBSCAN} \citep{ester1996}. We used $E$, $L_{\mathrm{z}}$, and $L_{\mathrm{\perp}}$ as input parameters for the clustering algorithm, for the purpose of having a fair comparison with D23. We scaled the space of these parameters with the \texttt{RobustScaler} tool in \texttt{scikit-learn} \citep{pedregosa12}. Due to the significant overlap between halo substructures in the IoM spaces, it is difficult for any clustering algorithm to identify each individual overdensity. Furthermore, some of the newly discovered substructures include a low number of stars. For this reasons, the algorithm used in this study was fine-tuned to detect all the small-scale clumps identified by D23. We find the best matching results with D23 by using \texttt{DBSCAN} default hyperparameters and by setting \texttt{eps} = 0.34 and \texttt{min\_samples} = 5. \\ Fig. \ref{FigIoM} shows the position of the final candidate members of each substructure in the IoM space. With our final selection, we are able to relate 119 stars in our parent sample to the known retrograde substructures. In the following sections we discuss in details the sample defined for each substructure. 

   Before starting the discussion of the various substructures it is important to warn the reader that the identification and classification of completely disrupted relics of ancient merging events is still in its infancy and, as it will be clear in the following discussion, it is subject to considerable uncertainty. On one side there is no consensus on the best selection criteria to isolate the purest samples of a given substructure and any criterion will imply some degree of contamination from unrelated sources \citep[see, e.g.,][for references and discussion]{buder22}. On the other side, as said, there are substructures that may be  erroneously considered as an individual entity while being in fact the sum of independent substructures or the unrecognised part of a larger one \citep{koppelman2020,naidu20,belokurov23,dodd23}. Here we are trying to get a clearer insight into the nature of some of the retrograde substructures by adding the chemical dimensions, as done by other authors \citep[like, e.g,][]{feuillet2020,feuillet21,buder22,horta23}. In general this is not an easy task, in the present phase, and it is worth to proceed cautiously, taking care of any little step ahead in our exploratory path.

\subsection{Gaia-Enceladus / Sausage}

  The dominant accreted component in the nearby stellar halo is the remnant of the last major merger event experienced by the MW with an ancient (8-11 Gyr ago) and relatively massive ($ \mathrm{M_{\star}} \sim 10^{8} - 10^{9} \mathrm{M_{\odot}}$) dwarf galaxy, that has been named Gaia-Enceladus / Sausage \citep{belokurov2018,helmi2018,gallart19,mackereth19,massari19,vincenzo2019,kruijssen2020,hasselquist21}. One of the defining characteristics of this population is the stars' highly eccentric orbits and slightly retrograde motion. The metallicity distribution function of GES peaks from [Fe/H] $\simeq -1.5$ to  $-1.1$ dex, depending on the sample selection and the spectroscopic survey \citep{naidu20,feuillet2020,bonifacio21,bellazz23}. Additionally, these stars exhibit lower [$\alpha$/Fe] values for fixed metallicity when compared to in-situ populations, thus suggesting that they originated in an environment characterized by a less efficient star formation \citep{fernandezalvar2018,helmi2018,das2020,feuillet2020,naidu20,limberg22,myeong22,horta23}.
  
  Due to the criterion employed to select retrograde stars (see Section \ref{selection}) that significantly cuts the IoM region typically populated by GES stars, we are able to retain only 12 stars in common with D23 that belong to GES. By using the clustering algorithm, all of the original members of this cluster were retained as GES members and an additional 41 stars were added to it, making the total number of stars in this cluster equal to 53 (see the green colored symbols in Fig. \ref{FigIoM} and thereafter).  
  
\subsection{Sequoia}

   The first evidence of the remnant of an accreted dwarf galaxy in the RH was the identification of a clump of stars moving on highly retrograde orbits, with large orbital energy and forming an arc-like structure in the Toomre diagram, which is now dubbed Sequoia \citep{myeong18,koppelman18,matsuno19,myeong19}. Sequoia has been proposed to be the remnant of an accreted independent small dwarf galaxy due to its position and smaller extent in the IoM space with respect to GES \citep{koppelman19,myeong19}. However, the actual complexity of this substructure is far greater than what was initially perceived. A different scenario has been proposed to explain the Sequoia overdensity, possibly making it tightly related to the GES accretion event. In fact, it has been shown that a GES-like progenitor, depending on its orbital configuration and morphology, is able to display its debris on orbits that reproduce the same arc-like feature associated with Sequoia \citep{koppelman2020,amarante2022,belokurov23}. The chemical composition of Sequoia remains a subject of ongoing debate, as different studies have yielded conflicting results regarding its abundance of $\alpha$-elements compared to the GES substructure. Several works have reported either an enhancement \citep{mackereth19,myeong19} or a depletion \citep{monty2020,matsuno22} of $\alpha$-elements compared to GES abundances. Moreover, \citet{naidu20} found a multiple-peaked metallicity distribution for Sequoia, and claimed that two additional sub-components, namely Arjuna and I'Itoi, occupy the same region in the IoM space \citep[see also,][]{bellazz23}. In addition, \citet{ruiz-lara22} discovered three disconnected clusters in the region of the IoM space previously associated to Sequoia that might be related to different progenitors.
   
   The cross-match provides 18 common members between our targets and the Sequoia sample of stars identified by D23. When using the \texttt{DBSCAN} algorithm, only 8 stars were found to belong to this clump. Additionally, 2 stars from the original sample were associated with ED-2 (see Section \ref{ed2}). This association was primarily based on the alignment of these two stars with the ED-2 clump in the $L_{\mathrm{z}}-L_{\mathrm{\perp}}$ space. However, they were found to move on higher energy orbits with respect to ED-2. Consequently, this distinctive orbital behaviour led us to consider the possibility of these stars being affiliated with the Sequoia substructure. In the end, since all the 8 stars are part of the 18 original members, we decided to stick with the original D23 classification for Sequoia. Thus, our final Sequoia sample contains 18 stars (brown colored symbols in Fig. \ref{FigIoM}).
   
\subsection{Antaeus / L-RL64}
  
   L-RL64 is a newly discovered cluster located in the most retrograde region of the IoM space \citep{lovdal22,ruiz-lara22}. L-RL64 is one of the three clusters associated to the Sequoia merging event and it was initially believed to be its extreme retrograde tail. Subsequently, \citet{oria2022} independently re-discovered this substructure in the action space, giving it the name Antaeus. The peculiar kinematics of this clump, which we will refer to as Antaeus from now on, combined with the mass estimate derived from its mean metallicity, indicate a different origin for this substructure with respect to Sequoia. Indeed, it is now thought that Antaeus is an independent accretion event, not related to Sequoia \citep{oria2022,ruiz-lara22,dodd23}. 
   
   The analysis based on D23 data shows that 11 stars in our entire sample are associated with Antaeus. However, the clustering algorithm identifies two separate clumps composed of 15 stars moving on extremely retrograde orbits. We are going to treat them as one single substructure (blue colored symbols in Fig. \ref{FigIoM}), since their position in the IoM spaces is totally consistent with the one of Antaeus and 13 out of these 15 stars are labeled by \citet{oria2022} as members of this group. Thus, with our final selection we define a fair sample of this newly discovered substructure and we will be able for the first time to provide its detailed chemical characterization. 
   
\subsection{ED-2 and ED-3} \label{ed2}    
   D23 spotted two new small highly retrograde clusters near Sequoia, namely ED-2 and ED-3, that are extremely tight in the IoM and velocity spaces (see their Fig. 3). 
   
   We have a total of 4 and 5 stars in common with the ED-2 and ED-3 clusters, respectively. The results provided by \texttt{DBSCAN} show that we are able to retain all the 5 stars in the ED-3 clump and add one stars to ED-2 substructure (orange and magenta colored symbols in Fig. \ref{FigIoM}, respectively). Notably, as already remarked above, the clustering algorithm spotted as members of ED-2 two additional stars that were previously associated with Sequoia. 

\subsection{Thamnos, Clump-6 and Clump-7}

   \citet{koppelman19} revealed the presence of a less prominent substructure with typically lower orbital energy in the retrograde region of the $E-L_{\mathrm{z}}$ space with respect to GES, which they named Thamnos. Given its small extent in the $E-L_{\mathrm{z}}$ plane, they estimate a stellar mass $<$ $5\times10^{6}\mathrm{M_{\odot}}$. Thamnos has been suggested to be the remnant of an old merging event with the MW, due to its low mass and peculiar location in the IoM \citep{kruijssen2020}, but it is still unclear if the stars associated to this group are from a unique remnant or two \citep{koppelman19,bellazz23}.
   
   In our entire sample, only one star is linked to Thamnos according to D23. As it is in the case of GES, the selection method used does not allow us to include stars with orbital energy as low as Thamnos, which means we are unable to sample this particular substructure. Interestingly, \texttt{DBSCAN} spots two clusters composed of 17 and 6 stars in the low energy regime of our sample at slightly higher orbital energy with respect to Thamnos. These two groups have very similar $E$ and $L_{\mathrm{z}}$, but they differ in $L_{\perp}$. It is worth noting that these clumps may be artificial overdensities identified by the algorithm and produced by our selection cut. In the following we will refer to them as Clump-6 and Clump-7 (cyan and red colored symbols in Fig. \ref{FigIoM}).
   
\subsection{Not associated}   
   
    A significant fraction of our sample, 67 out of 186 stars, does not appear to be associated with any known structure according to D23 and the clustering algorithm. In the rest of the paper, we will not focus on the chemical features of these unassociated stars, as they will be part of future contributions to the WRS project, exploiting their chemistry in order to determine whether they are accreted or not. 
%---------------------------------------------------
\section{Abundance analysis}\label{abu}
\subsection{Stellar parameters}\label{sp}

   Establishing precise atmospheric parameters (namely the effective temperature, $T_{\mathrm{eff}}$, the surface gravity, log $g$, and the microturbulent velocity, $v_{\mathrm{t}}$) is a crucial step to obtain reliable chemical abundances.
   
   We derived both $T_{\mathrm{eff}}$ and log $g$ relying on the EDR3 photometry (see Table \ref{TabMain}). $T_{\mathrm{eff}}$ have been calculated with the $\mathrm{(BP-RP)_{0}}$ - $T_{\mathrm{eff}}$ relation provided by \citet{mucciarelli21}. The color (BP-RP) has been corrected for extinction following the iterative procedure described in \citet{GC18_extinction}. The color excess E(B-V) adopted for each star is provided by the online tool \texttt{EXPLORE}\footnote{\url{https://explore-platform.eu/}}. Since the adopted color - $T_{\mathrm{eff}}$ relation depends on the star metallicity ([Fe/H]), that is not known a priori, we calculated $T_{\mathrm{eff}}$ assuming a fixed value of [Fe/H] = -1.5 dex for any star. The uncertainty on $T_{\mathrm{eff}}$ is due to the errors in photometric data, parallax, color excess and $\mathrm{(BP-RP)_{0}}$ - $T_{\mathrm{eff}}$ relation. Given that the observed targets are bright, very nearby and affected by minimal extinction, the main source of error in the effective temperature arises from the color - $T_{\mathrm{eff}}$ relation. Thus, we assume the dispersion of the $\mathrm{(BP-RP)_{0}}$ - $T_{\mathrm{eff}}$ relation, which is 61 K \citep{mucciarelli21}, as the typical internal error on $T_{\mathrm{eff}}$. Surface gravities have been estimated assuming the photometric $T_{\mathrm{eff}}$, a stellar mass derived from a BaSTI isochrone \citep{pietrinferni2021} with 12 Gyr, [Fe/H] = -1.5 dex and [$\alpha$/Fe] = +0.4 dex (according to the typical values for the accreted component of the stellar halo derived in previous studies, \citealt{nissen&schuster2010}, \citealt{kruijssen2020}, \citealt{montalban2021}, \citealt{horta23}), and the G-band bolometric corrections described by \citet{andrae18}. We note that employing an alternative isochrone varying the parameters has a minimal impact on the resulting surface gravity of the star. For instance, when using an isochrone with 10 Gyr age, [Fe/H] = -1.1 dex and [$\alpha$/Fe] = +0.4 dex, the discrepancy with the first results remains consistently below 0.04 dex. The uncertainties on log $g$ are derived through the propagation of the errors on the effective temperature, photometry and distance. The obtained values are systematically lower than 0.1 dex. Finally, microturbulent velocities have been calculated by requiring no trend between iron abundances and reduced equivalent widths, defined as $\mathrm{log_{10}}$(EW/$\lambda$). The uncertainties on $v_{\mathrm{t}}$ are estimated as described in \citet{gala} and they are of the order of 0.2 km $\mathrm{s^{-1}}$. 
   
\begin{figure}
   \centering
   \includegraphics[width=\hsize]{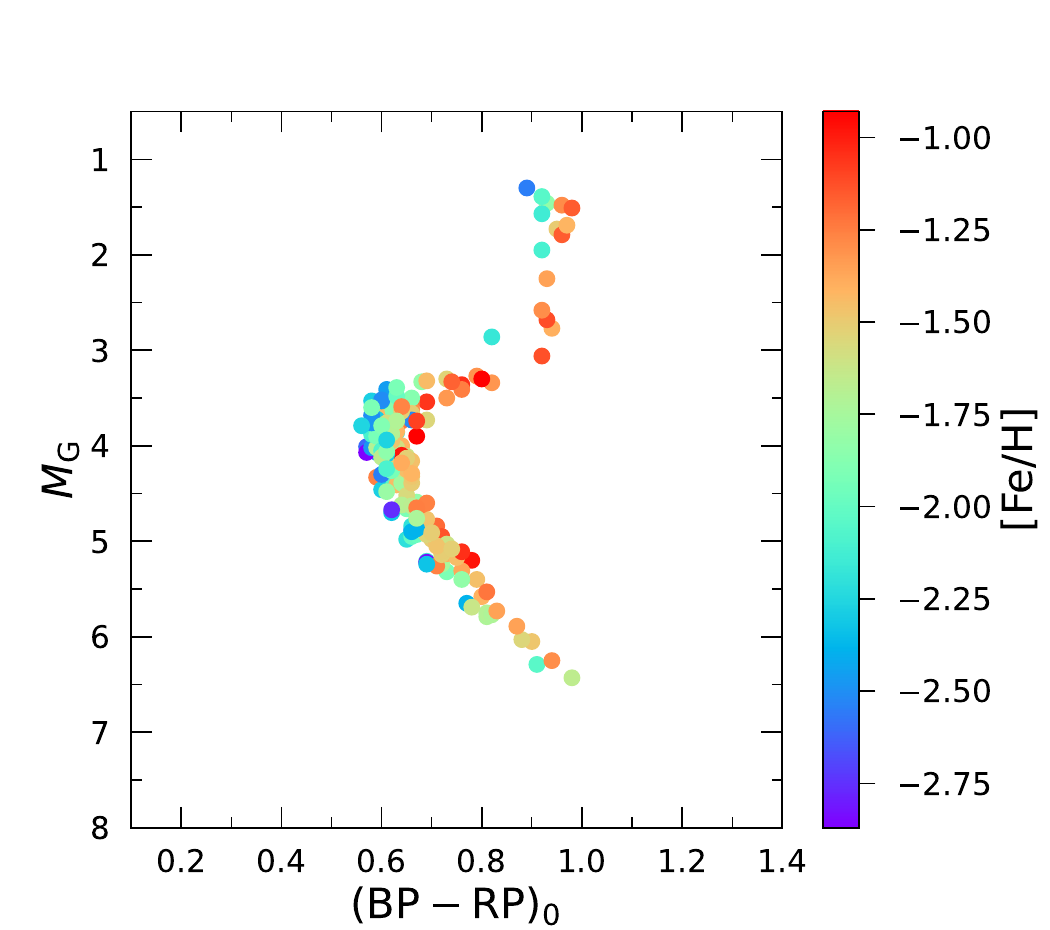}
      \caption{Gaia EDR3 distance-corrected color-magnitude diagram of the 175 analyzed stars. Targets are color coded by the metallicity.}
         \label{CMD_fe}
   \end{figure}

   Afterwards, we improved the estimate of the stellar parameters by using the correct [Fe/H] value derived from the chemical analysis. The final effective temperature value deviates from the initial estimate by a margin persistently lower than 60 K. We excluded from the chemical analysis the 3 hottest stars of the sample ($T_{\mathrm{eff}} >$ 6650 K). This decision stems from the fact that the combination of their atmospheric parameters leads to the formation of extremely weak absorption lines. As a consequence, this renders the measurement of abundances unreliable. In Fig. \ref{CMD_fe} we show the color-magnitude diagram (CMD) for all the analyzed stars, color coded by the metallicity. The final values of the atmospheric parameters are listed in Table \ref{TabSP}.
%-------------------------------------- Figure
%   \begin{figure}
%   \centering
%   \includegraphics[width=\hsize]{figures/CMD/CMD_Teff.pdf}
%      \caption{Gaia EDR3 distance-corrected color-magnitude diagram of all the targets. Stars are color coded by the temperature.}
%         \label{CMD_Teff}
%   \end{figure}
%--------------------------------------- 
%-------------------------------------- Figure
%--------------------------------------- 
%-------------------------- Table -------------------------
%
\begin{table*}
\caption{Stellar parameters for the selected targets: ID from Gaia EDR3, derived effective temperature, surface gravity, microturbulent velocity and iron abundance ratio. The entire  table is available in electronic form.}          
\label{TabSP}      
\centering          
\begin{tabular}{c c c c c c}  
\hline      
ID Gaia EDR3 & $T_{\mathrm{eff}}$ & log $g$ & $v_{\mathrm{t}}$ & [Fe/H] & $\sigma$[Fe/H]  \\ 
 & K & dex & km $\mathrm{s^{-1}}$ & dex & dex \\ 
\hline 
1156158780372565120 	 & 	6426	 & 	4.20	 & 	1.0	 & 	-2.06	 & 	0.05	  \\
6155896330944952576 	 & 	6137	 & 	4.47	 & 	0.9	 & 	-2.21	 & 	0.05	  \\
3028486001397877120 	 & 	6412	 & 	4.36	 & 	1.3	 & 	-1.26	 & 	0.04	  \\
2391446689585357568 	 & 	6140	 & 	4.07	 & 	0.1	 & 	-2.54	 & 	0.05	  \\
6661345365288720000 	 & 	6418	 & 	4.20	 & 	1.0	 & 	-2.16	 & 	0.05	  \\
6378884813840372864 	 & 	6301	 & 	4.19	 & 	1.0	 & 	-1.33	 & 	0.05	  \\
4479226310758314496 	 & 	6152	 & 	4.49	 & 	1.0	 & 	-2.41	 & 	0.05      \\
4453220730438373504 	 & 	6246	 & 	4.34	 & 	1.2	 & 	-1.45	 & 	0.05	  \\
4752251952905666048 	 & 	6469	 & 	4.13	 & 	1.5	 & 	-1.44	 & 	0.04	  \\
4855735169813450624 	 & 	5999	 & 	4.57	 & 	0.2	 & 	-2.34	 & 	0.05	  \\
... 	 & ...	 & ...	 & ...	 & ...	 & ...	     \\
\hline                  
\end{tabular}
\end{table*}
%
%-------------------------------------------------------------
\subsection{Line selection}\label{ls}

   In order to obtain the chemical abundances of each star, we made a list of unblended lines to analyze, automatically pre-selected with our own code \texttt{AUTOKUR} and consequently refined with a visual comparison between the observed and the synthetic spectra. The latter have been calculated with the code \texttt{SYNTHE} \citep{kurucz}, adopting stellar parameters of the observed targets (see Section \ref{sp}), assuming \texttt{ATLAS9} \citep{kurucz} model atmospheres and using the linelists containing atomic and molecular data from the Kurucz/Castelli\footnote{\url{http://wwwuser.oat.ts.astro.it/castelli/linelists.html}} database. The synthetic spectra have been convolved with a Gaussian profile to reproduce the observed resolution.
   
   To address the fact that the degree of blending for a particular transition is dependent on the metallicity, which is not known a priori in this case, we utilized an iterative approach to establish the list of spectral lines to analyze for each star. The synthetic spectra have been calculated adopting an $\alpha$-enhanced mixture ([$\alpha$/Fe] = +0.4 dex) with an initial metallicity of [Fe/H] = -1.5 dex. Following the initial chemical analysis, a revised set of unblended spectral lines has been identified for each star using a synthetic spectrum that was computed using the appropriate chemical composition.
   
   During the visual inspection, we noticed that the spectra of 3 low main sequence stars ($T_{\mathrm{eff}} <$ 5250 K and log $g$ $>$ 4.60) were highly contaminated by molecular transitions from MgH. Thus, we opted to exclude them from the abundance analysis. In the end, when considering also the exclusion of the 5 binary stars, a total of 11 stars were removed from the initial sample. We will exclude them from any further consideration in our study. Among these excluded stars, four were associated with GES, one with ED-2, one with Sequoia, one with Clump-6 and four star remained unassociated with any specific substructure.  

%---------------------------------------------------------------
\subsection{Chemical analysis}

   The chemical abundances of Na, Ca, Ti, Fe, Ni and Zn have been derived through a comparison between the observed equivalent widths (EWs) and the theoretical line strengths using the code \texttt{GALA} \citep{gala}. EWs have been measured with the code \texttt{DAOSPEC} \citep{daospec} through the automatic tool \texttt{4DAO} \citep{4dao}. 
   
   The chemical abundances of elements which show saturated core lines but with wings still sensitive to the abundance or hyperfine/isotopic splitting transitions (Mg, Al, Sc, Mn, and Y) have been derived through spectral synthesis with the proprietary code \texttt{SALVADOR}. This program runs a $\chi^{2}$-minimization between the observed line and a grid of synthetic spectra computed on-the-fly by the code \texttt{SYNTHE} keeping the stellar parameters fixed and varying only the abundance of the matching element. 
   
   All the abundances are scaled to the solar composition, taking as reference the values from \citet{grevesse1998}. We adopted these solar abundances since the \texttt{ATLAS9} model atmospheres used in this study are computed based on a chemical mixture consistent with the one outlined in \citet{grevesse1998} (see \citealt{castelli&kurucz2003}). For all the elements that were examined, in cases where the spectral lines were not clearly detectable, we determined upper limits by comparing the observed spectra to the synthetic spectra.
   
   In order to put the chemical abundances measured for the stars in the UVES and PEPSI samples onto the same scale, we made a direct comparison between the final abundances ratios of the 12 stars that have been observed with both instruments. For each element, we determined a median abundance difference ($\Delta_{\mathrm{abu}} = \mathrm{[X/Fe]_{PEPSI} - [X/Fe]_{UVES}}$), and we applied an offset correction by adopting UVES as the reference scale (due to the largest number of available lines) in the cases where:
   \begin{equation} \label{delta_abu}
       \Delta_{\mathrm{abu}} > \displaystyle\frac{\sigma}{\sqrt{N_{\star} -1}}
   \end{equation}
   where $\sigma$ is the standard deviation of the abundance differences for a given element and $N_{\star}$ is the number of stars with a measured abundance with both instruments for that specific element. For Fe, Ca, Sc, Ni, Zn and Y, we corrected the abundances for the PEPSI sample by subtracting the offsets listed in Table \ref{TabOffset}, thus bringing them onto the UVES scale.
   %In particular, the offset in Ca is due to the peculiar lines available in the spectral range of PEPSI. In fact, only the five Ca lines, at 5260.4, 5261.7, 5262.2, 5265.6 and 5349.5 \r{A}, are observed with the PEPSI CD3 grating. On the other hand, for the UVES spectra, more than 15 lines have been used to derive the Ca abundance. In the UVES sample, we found a median difference of 0.09~dex between the abundances derived using only those five lines and the abundances measured from the entire list of Ca lines. \\
   
   We focus here on the methodology we employed specifically to derive chemical abundances for problematic species: 

   \begin{enumerate}
      %\item \textit{Oxygen}. The only available oxygen lines are the permitted OI triplet at 7771 - 75 \r{A}. Since this feature is affected by non-LTE effects, we applied the corrections from \citet{amarsi15}. 
      \item \textit{Sodium}. The available two lines of Na in the spectral range of PEPSI are centered at 8183.3 and 8194.8 \r{A}. We highlight that these atomic lines are situated within a spectral region significantly affected by telluric absorption lines. As a precautionary measure to ensure the robustness of our analysis, we visually inspected the possibile telluric contamination of the Na lines using appropriate synthetic spectra of the Earth atmosphere calculated with \texttt{TAPAS} \citep{bertaux2014}. Subseqeuntely, we excluded any Na line exhibiting signs of contamination attributable to telluric features. We applied the correction for departures from LTE conditions from \citet{lind2011}. All the corrections for Na are available through the INSPECT database\footnote{\url{http://www.inspect-stars.com/}}. The values listed in Table \ref{TabAbu} are corrected for NLTE.
      \item \textit{Aluminum}. Within the UVES spectral range, the presence of Al is characterized by solely two strong lines located at 3944.0 and 3961.5 \r{A}. However,  these lines become saturated at [Fe/H] $> -1.3$ dex. In contrast, the PEPSI sample features only three weak aluminum lines suitable for the chemical analysis: one line at 7836.1 Å and the doublet at 8772 - 73 Å. Unfortunately, these weak lines are not detected at metallicities lower than $\sim -1.5$ dex. The corrections for non-LTE effects for each line are taken from \citet{lind22}.
      %\item \textit{Barium}. There are four Ba II lines inside the spectral coverage of UVES located at 4934.1, 5853.7, 6141.7 and 6496.9 \r{A}, while only the first one is available in the spectra of the stars observed with PEPSI. As shown in Fig. \ref{Ba}, the first line saturates at [Fe/H] > -1.5, thus becoming insensitive to variations in Ba abundance. We made the decision to exclude this line from the analysis. Moreover, the line at 6496.9 \r{A} consistently yields higher abundances compared to the other two lines for all the target samples. Therefore, we also excluded this transition from the chemical analysis. In this way, we are able to obtain a reliable abundance measurement only for the stars in the UVES sample.  
   \end{enumerate}

   %-------------------------- Table -------------------------
\begin{table}
\caption{Differences in chemical abundances obtained from the UVES and PEPSI samples, including the calculated offsets, corresponding standard deviations, and the number of stars used for each element. These values were employed to ensure a consistent scale alignment between the two sets of measurements.}          
\label{TabOffset}      
\centering          
\begin{tabular}{c c c c}  
\hline     
Element & $\Delta_{\mathrm{abu}}$ & $\sigma$ & $N_{\star}$  \\ 
\hline 
Fe   	 & 	  0.02	 & 	  0.05	 & 	     12	  \\
Mg   	 & 	 -0.01	 & 	  0.06	 & 	     12	  \\
%Si   	 & 	 -0.03	 & 	  0.05	 & 	      3	  \\
Ca   	 & 	  0.09	 & 	  0.06	 & 	      9	  \\
Sc   	 & 	 -0.18	 & 	  0.22	 & 	     10	  \\
Ti I  	 & 	  0.00	 & 	  0.06	 & 	     12	  \\
Ti II 	 & 	  0.00	 & 	  0.10	 & 	     11	  \\
%Cr I  	 & 	  0.02	 & 	  0.06	 & 	     12	  \\
Mn   	 & 	 -0.05	 & 	  0.14	 & 	      8	  \\
Ni   	 & 	  0.05	 & 	  0.08	 & 	      8	  \\
Zn   	 & 	  0.11	 & 	  0.07	 & 	      5	  \\
% Co   	 & 	  0.365	 & 	  0.305	 & 	      2	  \\
Y    	 & 	  0.12	 & 	  0.11	 & 	      4	  \\
% Ba   	 & 	  0.389	 & 	  0.175	 & 	      8	  \\

\hline                  
\end{tabular}
\end{table}
%
%-------------------------- Figure -----------------------------
%   \begin{figure*}
%   \centering
%   \includegraphics[width=1.0\textwidth]{figures/Chemical Analysis/differenze_Ba.pdf}
%   \caption{Synthetic spectra calculated for a representative dwarf star with $T_{\mathrm{eff}}$ = 5700 K, log $g$ = 4.50 and $v_{\mathrm{t}}$ = 1.40 km $\mathrm{s^{-1}}$ at four different metallicities ([Fe/H] = -2.0, -1.5, -1.0 and -0.5; from bottom to top panel, respectively), around the Ba II lines at 4934.1, 5853.7, 6141.7 and 6496.9 \r{A} (from left to right panels, respectively). For each metallicity, synthetic spectra have been computed with different Ba abundances, namely, [Ba/Fe] $= -0.2$ (red lines), 0.0 (black lines), +0.2 (blue lines), +0.4 (green lines) and +0.6 (orange lines).}
%              \label{Ba}%
%    \end{figure*}
%-------------------------------------------------------------- 
\subsection{Abundance uncertainties}
\label{abu_uncertainties}
   In inferring the uncertainties on the abundance ratios, it is essential to consider two primary sources of error: internal errors related to EW measurements and errors arising from the adopted atmospheric parameters.
   
   Uncertainties due to EW measurements have been estimated as the dispersion around the mean of the individual line measurements, divided by the root mean square of the number of used lines. For the elements whose abundance has been measured using spectral synthesis, we exploit Monte Carlo simulations to estimate the internal error. Specifically, we generated a sample of 500 synthetic spectra with Poissonian noise to reproduce the S/N of the observed spectra and then we repeated the analysis to derive the abundance. The internal error is estimated as the standard deviation of the abundance distribution of the 500 noisy synthetic spectra. 
   
   To determine the uncertainties arising from atmospheric parameters, we replicated the chemical analysis by varying only one stellar parameter at a time and keeping the others fixed (see Section \ref{sp} for the details about the derivation of the uncertainties on the atmospheric parameters). %The main source of error in the effective temperature arises from the assumed color-$T_{\mathrm{eff}}$ relation. Thus, we applied variation on the effective temperature equal to the dispersion of the assumed relation, which is 61 K \citep{mucciarelli21}. For log $g$ and $v_{\mathrm{t}}$, we chose \textbf{conservative estimates of the} uncertainties as variations for the analysis, that equal to 0.1 dex and 0.2 km $\mathrm{s^{-1}}$, respectively.
   Since we refer to chemical abundances with abundance ratios ([X/Fe]), the uncertainties on the iron abundance [Fe/H] have been considered as well.
   
   The intrinsic and systematic components have been summed in quadrature to compute the total uncertainty. Therefore, the final errors have been estimated as:
   
   \begin{flalign}
      &\sigma_{[Fe/H]}  = 
      \sqrt{\frac{\sigma_{Fe}^{2}}{N_{Fe}}  +
      (\delta_{Fe}^{T_{\mathrm{eff}}})^{2}  +
      (\delta_{Fe}^{\mathrm{log} \; g})^{2} + 
      (\delta_{Fe}^{v_{\mathrm{t}}})^{2}}  \\ \nonumber
      & \\ \nonumber
      &\sigma_{[X/Fe]} = \\ 
      & \sqrt{\frac{\sigma_{X}^{2}}{N_{X}} + \frac{\sigma_{Fe}^{2}}{N_{Fe}} + (\delta_{X}^{T_{\mathrm{eff}}} - \delta_{Fe}^{T_{\mathrm{eff}}})^{2} + (\delta_{X}^{\mathrm{log} \; g} - \delta_{Fe}^{\mathrm{log} \; g})^{2} +(\delta_{X}^{v_{\mathrm{t}}} - \delta_{Fe}^{v_{\mathrm{t}}})^{2}}
   \end{flalign}
   where $\sigma_{X,Fe}$ is the dispersion around the mean of chemical abundances, $N_{X,Fe}$ is the number of used lines and $\delta_{X,Fe}^{i}$ are the abundance differences obtained by varying the parameter $i$. 

   All the abundance ratios for individual stars are listed in Table \ref{TabAbuAlpha} and \ref{TabAbu} together with the corresponding uncertainties, as described in Section \ref{abu_uncertainties}.
%----------------------------------------------
%-----------------------------------------------------
%\begin{landscape}% Landscape page
% \topskip0pt
% \vspace*{\fill}
% \begin{table}[htb!]
\begin{table*}
  \caption{Chemical abundances for the $\alpha$-elements for the stars associated to known substructures. In the last column we report our association with the former progenitor and the result provided by D23 (inside the brackets).  The entire table is available in electronic form.}\label{TabAbuAlpha}
  \centering
  \begin{tabular}{ccccccc} 
   \hline\hline             
ID Gaia EDR3 & [Fe/H] & [Mg/Fe] & [Ca/Fe] & [TiI/Fe] & [TiII/Fe] & Substructure (D23) \\ 
\hline 
6155896330944952576 	 & 	-2.21 $\pm$ 0.05 & 	0.23 $\pm$ 0.03 & 0.26 $\pm$ 0.04 & 0.29 $\pm$ 0.02	 & 0.58	$\pm$ 0.32  &                  GES (GES) \\
3028486001397877120 	 & 	-1.26 $\pm$ 0.04 & 	0.25 $\pm$ 0.03 & 0.35 $\pm$ 0.04 & 0.29 $\pm$ 0.02	 & 0.28	$\pm$ 0.02  &                  GES (N/A)	  \\
2391446689585357568 	 & 	-2.54 $\pm$ 0.05 & 	0.24 $\pm$ 0.03 & 0.43 $\pm$ 0.03 & 0.43 $\pm$ 0.02	 & 0.50	$\pm$ 0.10  &                  GES (Group 14)	  \\
6378884813840372864 	 & 	-1.33 $\pm$ 0.05 & 	-0.02 $\pm$ 0.04 & 0.26	$\pm$ 0.03 & 0.26 $\pm$ 0.03	 & 0.23	$\pm$ 0.03  &                  GES (N/A)	  \\
4479226310758314496 	 & 	-2.41 $\pm$ 0.05 & 	-0.03 $\pm$ 0.03 & 0.22	$\pm$ 0.04 & 0.59 $\pm$ 0.13	 & -	  &                  ED-2 (ED-2)	  \\
4453220730438373504 	 & 	-1.45 $\pm$ 0.05 & 	0.08 $\pm$ 0.04 & 0.19 $\pm$ 0.04 & 0.28 $\pm$ 0.03	 & 0.25	$\pm$ 0.05  &              Sequoia (Sequoia)	  \\
4855735169813450624 	 & 	-2.34 $\pm$ 0.05 & 	0.14 $\pm$ 0.03 & 0.16 $\pm$ 0.03 & 0.20 $\pm$ 0.04	 & 0.20	$\pm$ 0.06  &              Sequoia (Sequoia)	  \\
3499616700847886592 	 & 	-0.97 $\pm$ 0.05 & 	0.16 $\pm$ 0.04 & 0.19 $\pm$ 0.04 & 0.09 $\pm$ 0.02	 & 0.17	$\pm$ 0.06  &                  GES (N/A)	  \\
2458265767848052992 	 & 	-1.64 $\pm$ 0.04 & 	0.23 $\pm$ 0.03 & 0.41 $\pm$ 0.03 & 0.38 $\pm$ 0.02	 & 0.40	$\pm$ 0.04  &                  GES (N/A)	  \\
1157100026750416640 	 & 	-1.66 $\pm$ 0.05 & 	0.31 $\pm$ 0.04 & 0.20 $\pm$ 0.05 & 0.25 $\pm$ 0.04	 & 0.30	$\pm$ 0.08  &              Sequoia (Sequoia)	  \\
... 	 & ...	 & ...	 & ...	 & ...	 & ...	 & ...	      \\
   \hline
  \end{tabular}
\end{table*}

\begin{table*}
  \caption{Chemical abundances for the odd-Z, iron-peak and neutron capture elements for the stars associated to known substructures. The values of [Na/Fe] are corrected for NLTE effects. The entire table is available in electronic form.}\label{TabAbu}
  \centering
  \begin{tabular}{ccccccccc} 
   \hline\hline             
ID Gaia EDR3 &  [Na/Fe]  & [Al/Fe]  & [ScII/Fe]  & [Mn/Fe] & [Ni/Fe] & [Zn/Fe] & [Y/Fe]  \\ 
\hline 
6155896330944952576 	 	 & 	-	 	 & 	-0.15 $\pm$ 0.11	 & 0.07	 $\pm$ 0.07	 	 & -0.46 $\pm$ 0.10	 & -0.06 $\pm$ 0.02	 & -    	 & -0.04 $\pm$ 0.08	  \\
3028486001397877120 	 	 & 	-	 	 & 	-0.37 $\pm$ 0.09	 & 0.12	 $\pm$ 0.06	 	 & -0.41 $\pm$ 0.11	 & -0.07 $\pm$ 0.03	 & -0.01 $\pm$ 0.02	 & -0.18 $\pm$ 0.09	 	  \\
2391446689585357568 	 	 & 	-	 	 & 	-0.44 $\pm$ 0.11	 & 0.22	 $\pm$ 0.07	 	 & -0.51 $\pm$ 0.10	 & 0.17 $\pm$ 0.02	     & -    	 & -0.01 $\pm$ 0.08	 	  \\
6378884813840372864 	 	 & 	-	 	 & 	-                    & 0.02	 $\pm$ 0.08	 	 & -0.38 $\pm$ 0.16	 & -0.20 $\pm$ 0.03	 & -0.21 $\pm$ 0.05	 & -0.12 $\pm$ 0.15	 	  \\
4479226310758314496 	 	 & 	-	 	 & 	-0.23 $\pm$ 0.11	 & 0.12	 $\pm$ 0.07	 	 & -0.48 $\pm$ 0.10	 & -0.08 $\pm$ 0.04	 & -    	 & 0.09	$\pm$ 0.08     	  \\
4453220730438373504 	 	 & 	-	 	 & 	-0.27 $\pm$ 0.16	 & 0.14	 $\pm$ 0.08	 	 & -0.32 $\pm$ 0.16	 & -0.15 $\pm$ 0.04	 & -0.15 $\pm$ 0.05	 & -0.16 $\pm$ 0.15	 	  \\
4855735169813450624 	 	 & 	-	 	 & 	-0.13 $\pm$ 0.11	 & -0.09 $\pm$ 0.07	 	 & -0.48 $\pm$ 0.10	 & -0.07 $\pm$ 0.03	 & -    	 & -    	 	  \\
3499616700847886592 	 	 & 	-	 	 & 	-               	 & -0.04 $\pm$ 0.08	 	 & -0.33 $\pm$ 0.16	 & -0.17 $\pm$ 0.02	 & -0.12 $\pm$ 0.06	 & -0.13 $\pm$ 0.15	 	  \\
2458265767848052992 	 	 & 	-	 	 & 	-0.19 $\pm$ 0.09	 & 0.36	 $\pm$ 0.06	 	 & -0.28 $\pm$ 0.11	 & -0.08 $\pm$ 0.02	 & -0.06 $\pm$ 0.08	 & 0.14	$\pm$ 0.09     	  \\
1157100026750416640 	 	 & 	-	 	 & 	-               	 & -0.12 $\pm$ 0.08	 	 & -0.21 $\pm$ 0.16	 & -0.05 $\pm$ 0.04	 & -    	 & 0.09	$\pm$ 0.15     	  \\
... 	 	 & ...		 & ...	 & ...	 & ...	 & ...	 & ...	 & ...	 	     \\
   \hline
  \end{tabular}
\end{table*}
% \end{table}
% \vspace*{\fill}
%\end{landscape}
%-----------------------------------------------------
\section{Comparison with the literature} \label{comparison_literature}  
%-------------------------------------- Figure
   \begin{figure}
   \centering
   \includegraphics[width=\hsize]{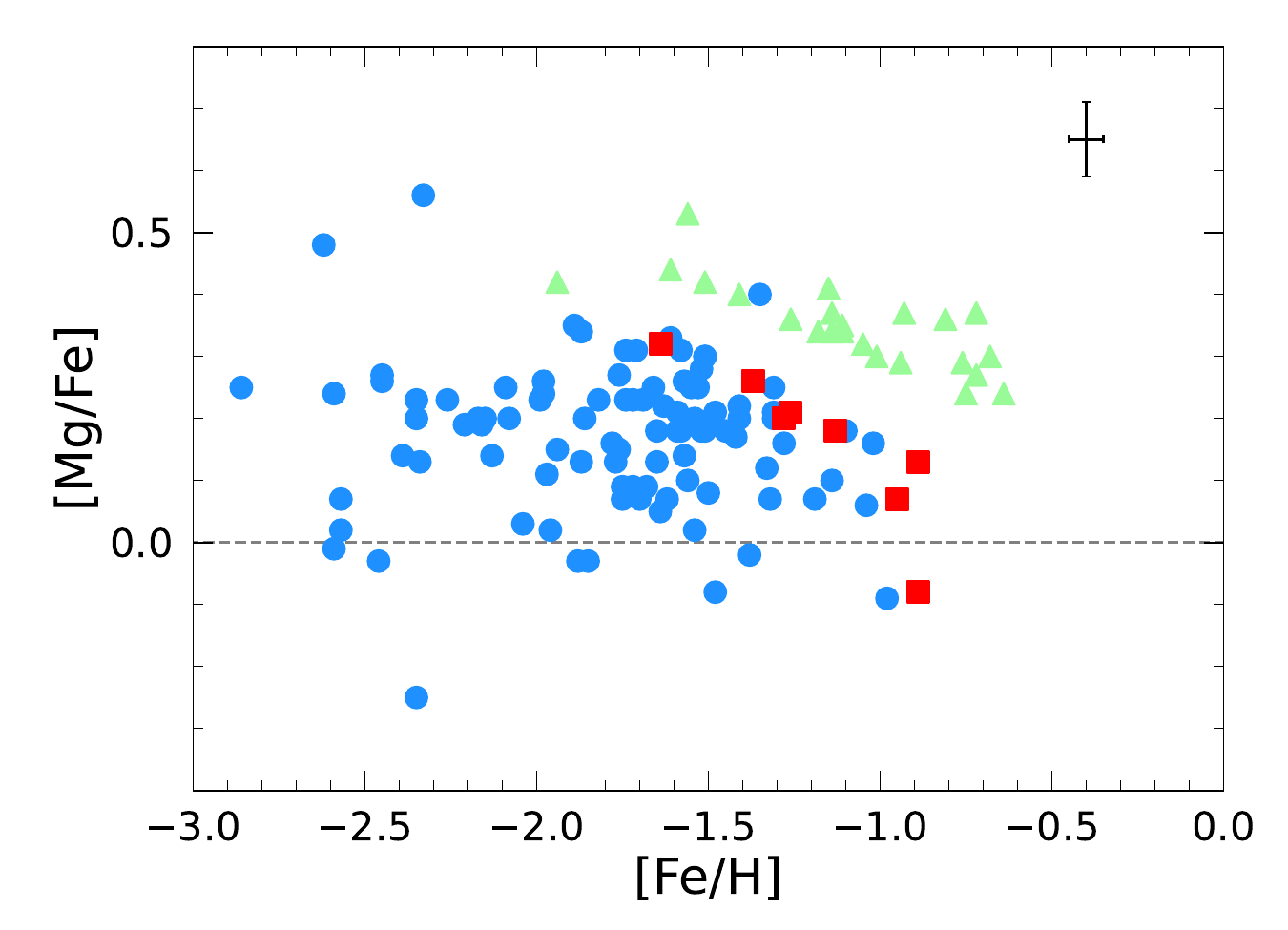}
      \caption{Comparison of Mg abundance ratios of the stars from 
               our selection observed with UVES (blue filled circles) and the NS10 reference sub-sample (red filled squares and green filled triangles for low-$\alpha$ and high-$\alpha$ sequences, respectively). The errorbar in the top-right corner indicate the typical uncertainties.}
         \label{Fig_NS10}
   \end{figure}
%--------------------------------------- 
   \begin{figure*}
   \centering
   \begin{minipage}{0.49\textwidth}
        \centering
        \includegraphics[width=1.0\textwidth]{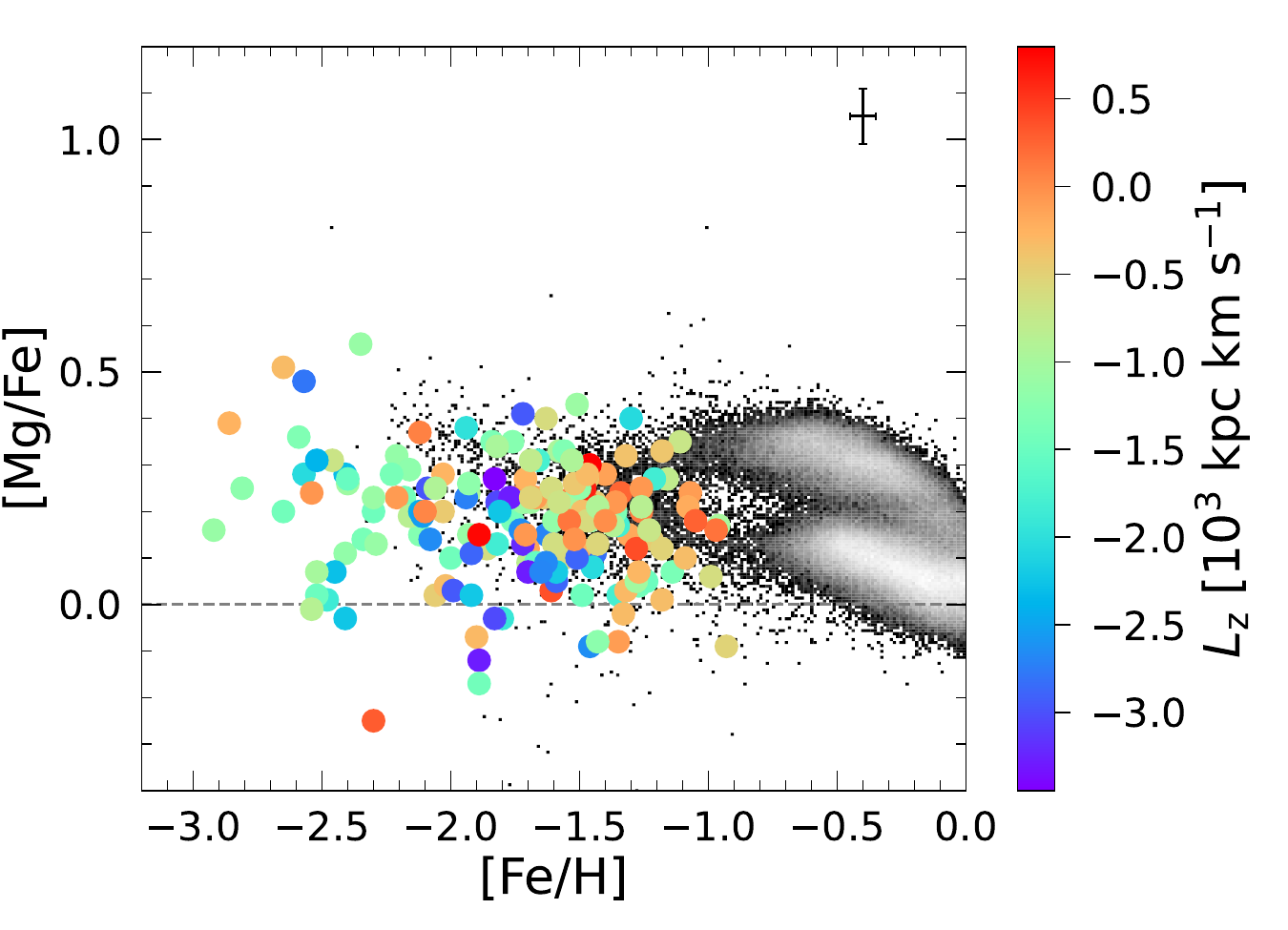} 
   \end{minipage}
   \begin{minipage}{0.49\textwidth}
        \centering
        \includegraphics[width=1.0\textwidth]{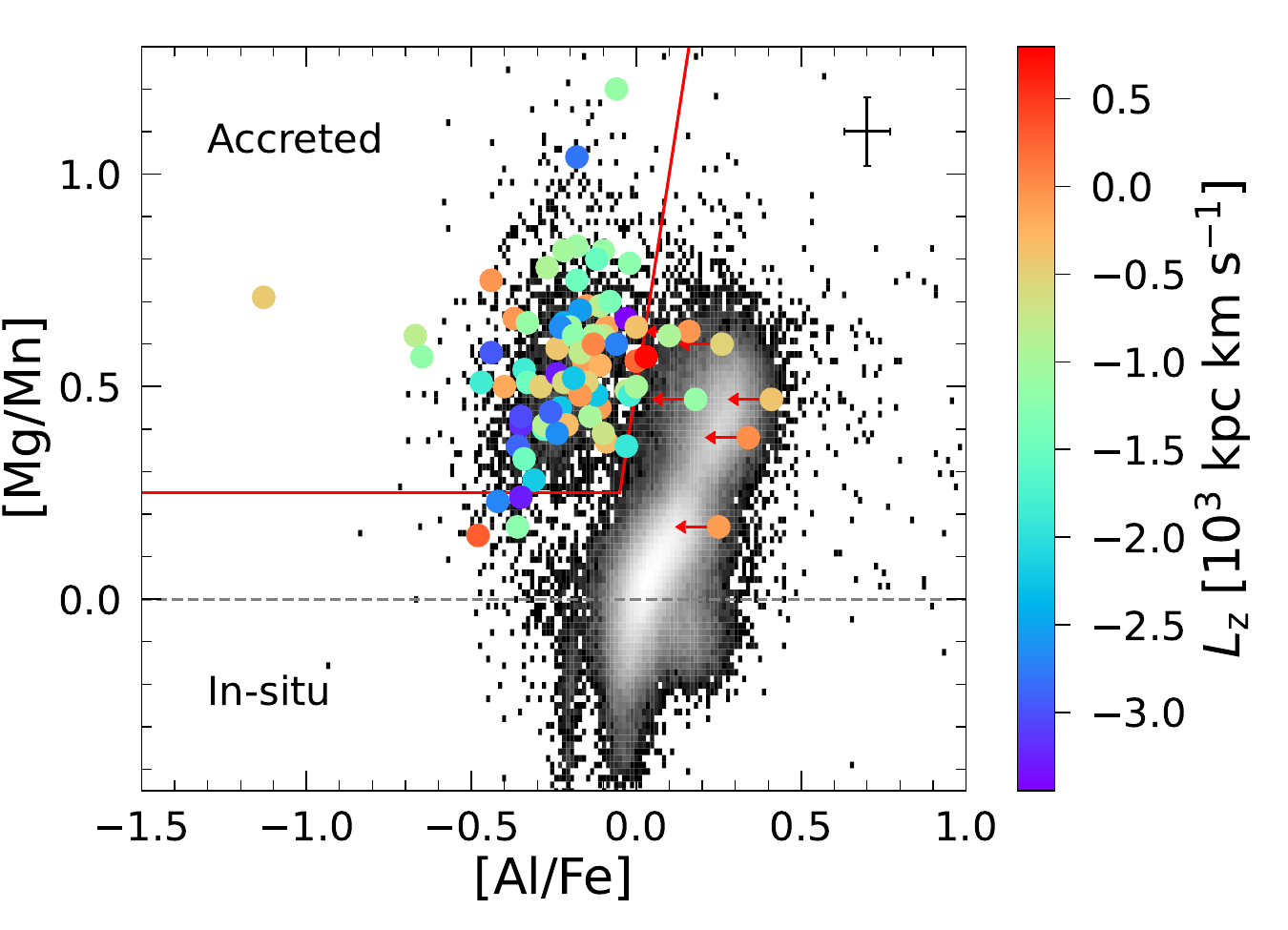}
   \end{minipage}
      \caption{Left panel: [Mg/Fe] abundance ratios as a function of [Fe/H] for our sample (color coded by the angular momentum along the $z$-direction, $L_{\mathrm{z}}$) and for the APOGEE control sample (grey scale 2D histogram). The errorbar in the top-right corner displays the typical uncertainties. Right panel: [Mg/Mn] abundance ratios as a function of [Al/Fe] for the same samples, with the same arrangement. Red arrows indicate upper limits.}
         \label{Fig_APOGEE}
   \end{figure*}
%---------------------------------------
%------------------------ Figure ---------------------------
   \begin{figure*}
   \centering
   \begin{minipage}{0.33\textwidth}
        \centering
        \includegraphics[width=1.0\textwidth]{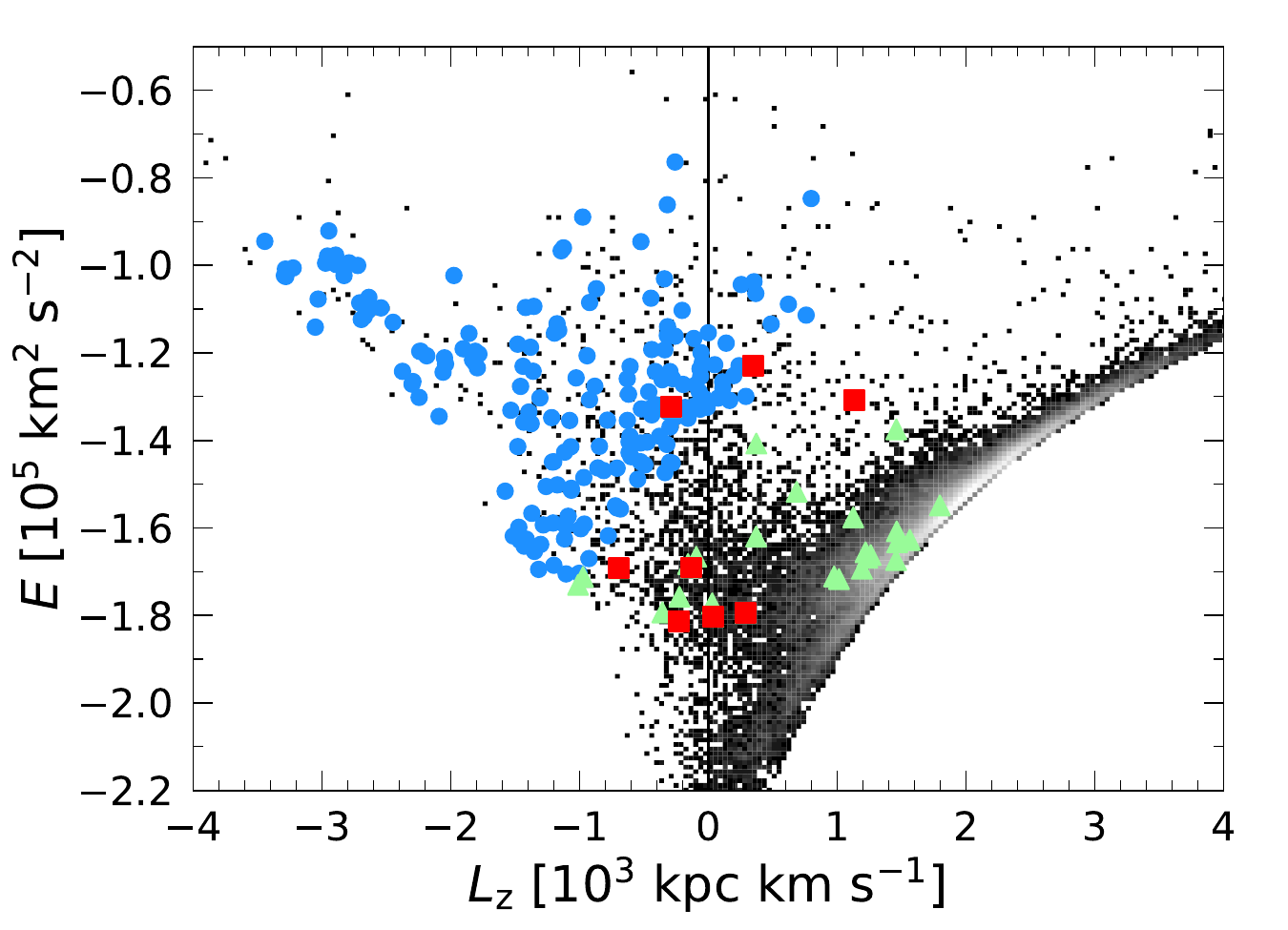} 
   \end{minipage}
   \begin{minipage}{0.33\textwidth}
        \centering
        \includegraphics[width=1.0\textwidth]{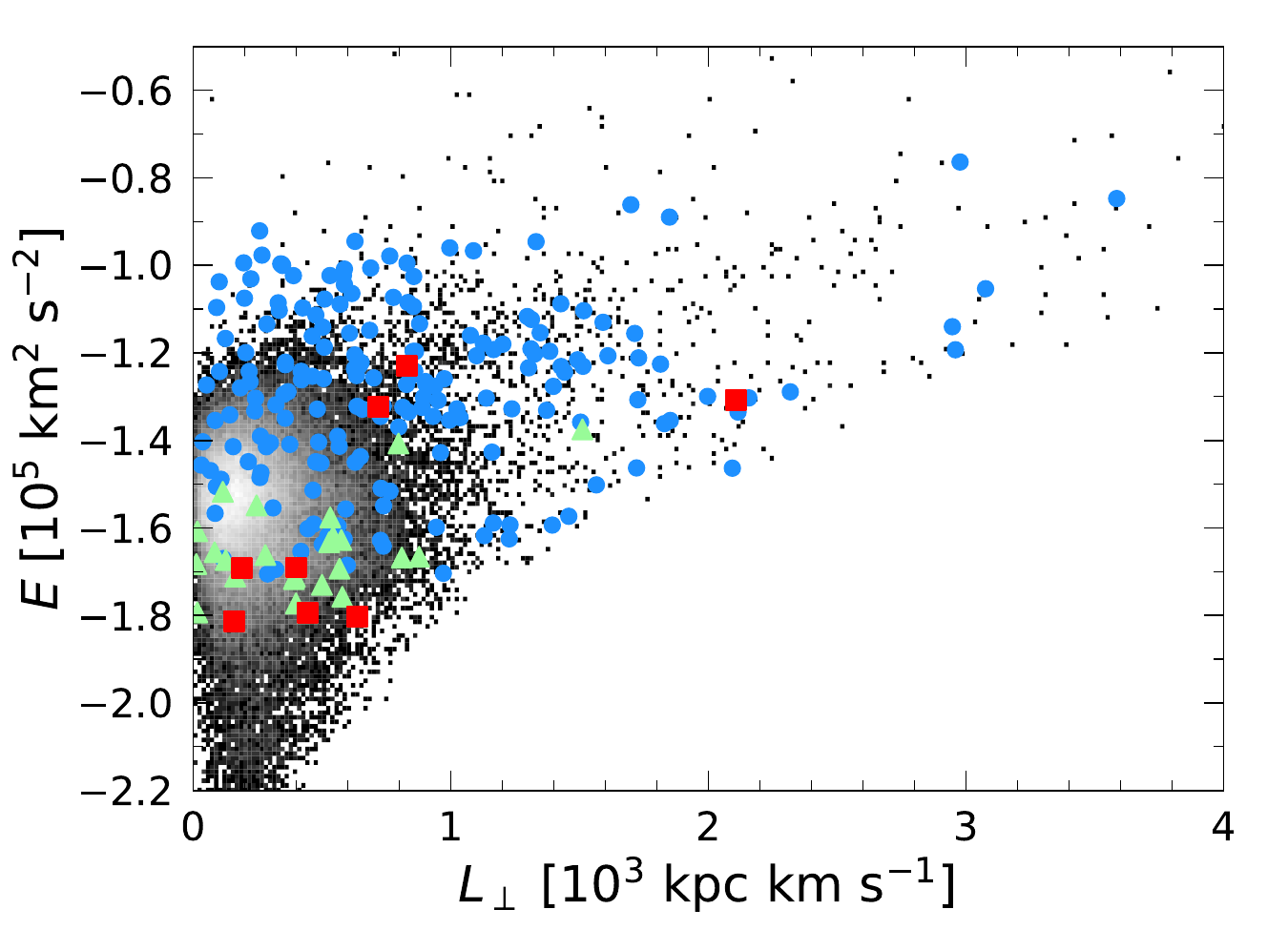}
   \end{minipage}
   \begin{minipage}{0.33\textwidth}
        \centering
        \includegraphics[width=1.0\textwidth]{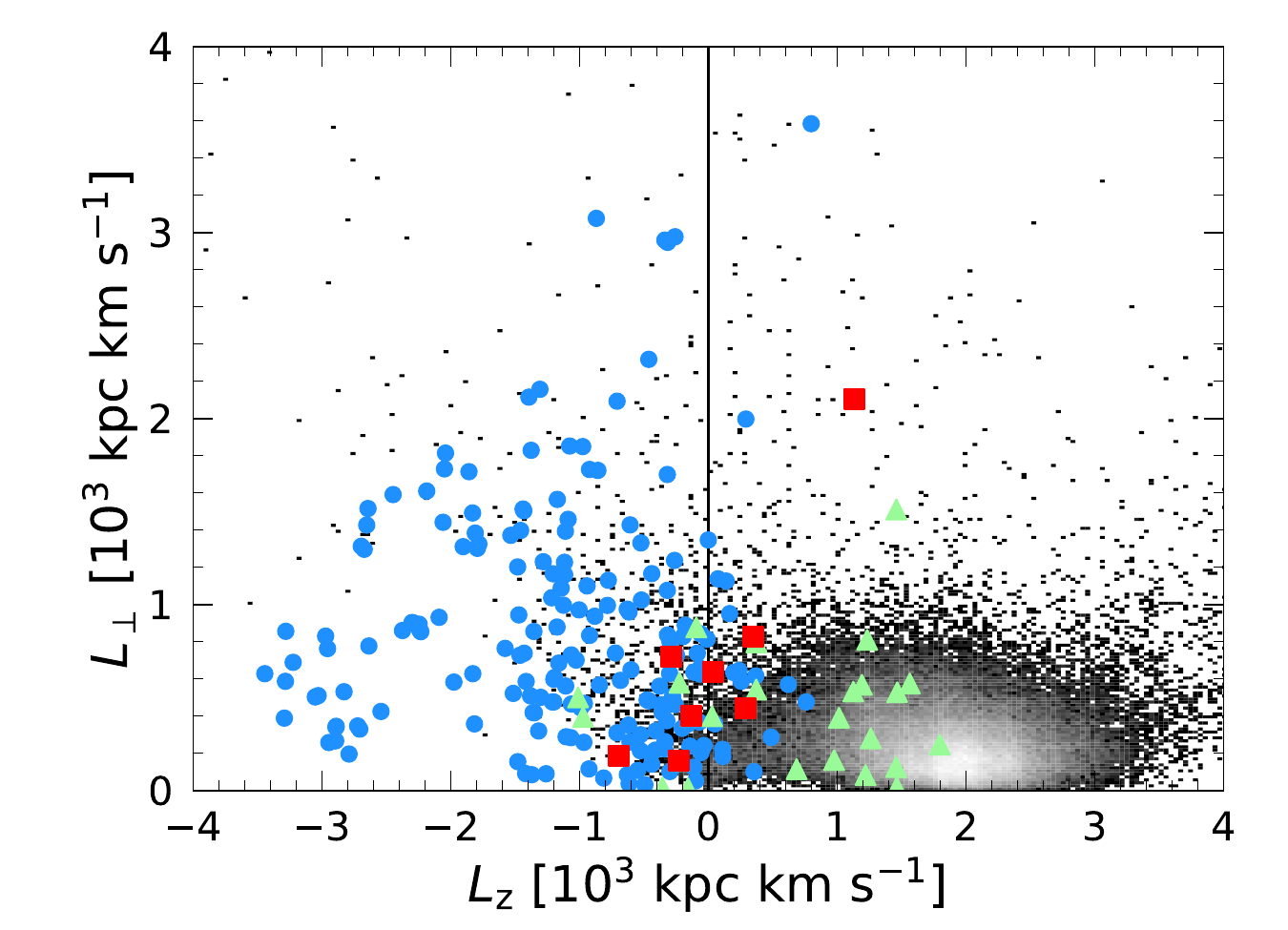}
   \end{minipage}
   \caption{Distribution of the targets stars in the IoM space compared with the NS10 stars and the APOGEE sample. The color coding is the same as in Fig. \ref{Fig_NS10}.}
              \label{FigIoM_all}%
    \end{figure*}    
%-------------------------------------------------

   To put our results in a more general context, in this section we compare chemical abundances of our sample with external datasets, starting with the sample of metal-poor dwarf and subgiant stars associated with the halo population analyzed by \citet{nissen&schuster2010} (NS10 hereafter). This choice was guided by the extremely high-quality stellar spectra used by NS10 (S/N ranging from 250 to 500) that guarantees accurate measurements of chemical abundances. By using a sample of stars observed with UVES at VLT and FIES at NOT, these authors revealed for the first time the existence of two distinct populations within the nearby MW halo, composed of stars having different kinematics and chemical patterns. NS10 proposed that the Galactic halo population ([Fe/H] $\la -0.8$ dex) following the low-$\alpha$ sequence in the classical [Fe/H] vs. [$\alpha$/Fe] plane (see the red squares in Fig. \ref{Fig_NS10}) is composed of stars accreted from disrupted dwarf galaxies, whereas the high-$\alpha$ population (green triangles in Fig. \ref{Fig_NS10}) is made of stars that have formed in-situ, a view that nowadays is generally accepted \citep{bonaca17, hayes2018, haywood2018, helmi2018, mackereth19, helmi2020}. To avoid the introduction of systematic errors due to the different chemical analysis and ensure a fair comparison between homogeneous results, we defined as reference targets a sub-sample of 41 NS10 stars that have been observed with UVES using the same instrument setup as our targets and covering a range of atmospheric parameters similar to our sample. We re-analyzed the NS10 spectra following the same procedure described in Section \ref{abu} and using the same linelist. We find a median discrepancy of [Fe/H] $= -0.05$ dex and $\sigma_{\mathrm{[Fe/H]}} = 0.08$ and given that these values adhere to the criteria described in Eq. \ref{delta_abu}, we made the decision to adjust the abundances to ensure congruence within the NS10 reference frame. The estimates of [Mg/Fe] display a median difference of $0.01$ dex and $\sigma_{\mathrm{[Mg/Fe]}} = 0.08$.
   
   We show the final comparison in Fig. \ref{Fig_NS10}: as expected from the selection we imposed to our targets, which should predominantly include retrograde, accreted objects, stars from our sample (blue symbols) are located in the same region of the $\alpha$-elements chemical planes as the low-$\alpha$ population from NS10, with abundances on average 0.2 dex systematically lower than the high-$\alpha$ population composed of in-situ stars. It is also interesting to note that the low-$\alpha$ sequence of NS10 targets basically defines the upper limit of the distribution of our targets, which are mostly at lower [$\alpha$/Fe] abundance. This might reflect the fact that we are sampling regions of the RH populated also by several different low-mass mergers, for which theoretical models predict low (and different) [$\alpha$/Fe] levels already at fairly low metallicity \citep{matsuno22}. On the other hand, the accreted population of NS10 likely belongs to the most-massive merger, GES. 
%-------------------------- Figure -----------------------------
   \begin{figure*}
   \centering
   \includegraphics[width=1.0\textwidth]{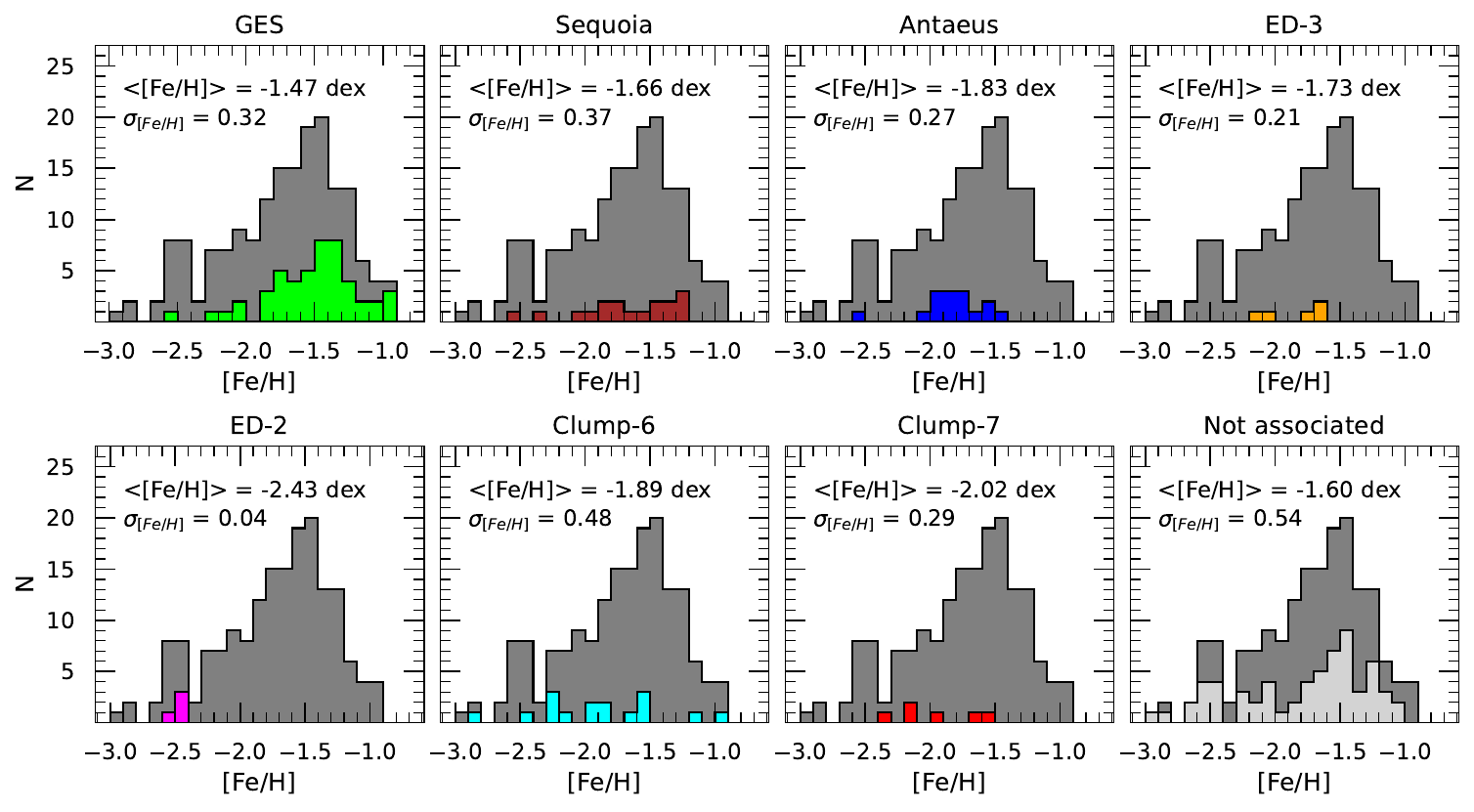}
   \caption{Metallicity distribution functions for the RH substructures labelled at the top of each panel compared to the MDF of the entire sample of stars. The median metallicity value and the standard deviation of the distribution are also reported. The color coding is the same as in Fig. \ref{FigIoM}.}
              \label{MDF}%
    \end{figure*}
%--------------------------------------------------------------   

   As a further and independent check of our results we also compare with chemical abundances from the large and widely used APOGEE DR17 \citep{apogee22} dataset. We limit the comparison to well-measured APOGEE abundances by selecting only stars having \texttt{ASPCAPFLAG} $= 0$, \texttt{STARFLAG} = 0, \texttt{EXTRATARG} = 0, \texttt{SNR} $>$ 70, \texttt{FE\_H\_ERR} $<$ 0.1 and \texttt{ALPHA\_M\_ERR} $<$ 0.2\footnote{See \url{https://data.sdss.org/datamodel/files/APOGEE_ASPCAP/APRED_VERS/ASPCAP_VERS/allStarLite.html} for definitions.}. The final APOGEE sample (APOGEE control sample hereafter), that we will use as a reference throughout the paper, is composed of 192225 stars, with 98.9\% of them lying at $D_{\sun}\le 2.0$~kpc. The left panel of Fig.~\ref{Fig_APOGEE} reveals that, in the [Fe/H] vs. [Mg/Fe] plane, our targets show a good agreement in the common metallicity range of $-2.4 \le$ [Fe/H] $\le -0.8$ dex with the low-$\alpha$ population stars from the APOGEE control sample. Interestingly, the stars from our sample that move on more extreme retrograde orbits ($L_{\mathrm{z}} < -2500$ kpc km $\mathrm{s^{-1}}$) seem to lay on a different pattern with respect to the low-$\alpha$ population at [Fe/H] $< -1.5$ dex, with lower [Mg/Fe] at fixed metallicity. In the right panel of Fig.~\ref{Fig_APOGEE} we show the [Mg/Mn] vs. [Al/Fe] chemical space, which serves as a diagnostic to distinguish between accreted and in-situ populations \citep{das2020, horta2021}. Contrary to the APOGEE control sample, nearly all of our targets are located in the area of the plot where we expect to find accreted stars \citep{horta2021}, emphasizing the fact that we are observing a poorly explored area of the Solar Neighbourhood. Indeed, only less than 1\% of the stars within the APOGEE control sample are located at $D_{\odot} < 1$ kpc and move on retrograde orbits. It is worth noting that only 11 stars among our sample are listed in the entire APOGEE DR17 catalogue, with only one of them aligning with the selection criteria expounded above. Moreover, through our program we have the capability to provide detailed chemical abundance data for elements that are notably absent in APOGEE (e.g, Sc, Zn and Y) and for elements whose measurements in APOGEE may diverge from theoretical expectations (Na and Ti).

   The comparisons presented above show that the selection we adopted was very effective in selecting stars that very likely were originated in ancient and now dissolved MW satellites. In addition to the location of our stars in the [Mg/Mn] vs. [Al/Fe] plane, it is worth noting that (a) our kinematic selection picked up only metal-poor stars ([Fe/H] $\la -0.8$ dex), the majority of which (65\%) having [Fe/H] $\le -1.5$ dex, with a significant very metal-poor tail (47 stars with  $-2.0 \le$ [Fe/H] $< -3.0$ dex, 26\% of the sample), a component that is poorly probed by the APOGEE control sample, and (b) despite the higher resolution, our stars show a larger scatter in [Mg/Fe] with respect to APOGEE, increasing with decreasing metallicity, suggesting again that we are more effective in sampling stars from a variety of disrupted satellites, with different chemical evolutionary paths. In summary, the results of the adopted selection are fully compliant with the main rationale of WRS.
  
   In Fig. \ref{FigIoM_all} we compare the IoM of our targets with those of the NS10 sample and APOGEE. We computed the IoM for the NS10 stars using the $6$-D phase space information provided by the Gaia EDR3. Instead, for the APOGEE control sample we used the phase space values from Gaia listed in the APOGEE DR17 catalogue. First, Fig. \ref{FigIoM_all} provides a direct view of the efficiency of our selection criterion in picking up halo stars in retrograde motion, as discussed in Section \ref{selection}. Though, it introduces a clear bias, as it prevents us from picking stars with low orbital energy ($E$ $\le 1.5 \times 10^{5}$ $\mathrm{km^{2}}$ $\mathrm{s^{-2}}$) moving on slightly retrograde orbits ($L_{\mathrm{z}}$ $\ge -0.5 \times 10^{3}$ kpc km $\mathrm{s^{-1}}$). On the contrary, it is clear that high-quality APOGEE measures primarily targets disc stars with a quite sparse sampling of the retrograde component of the halo. For what it concerns the NS10 sample, although some high-$\alpha$ stars are interestingly located in the retrograde region (they might be part of the dynamically heated disc, sometimes referred to as the Splash, \citealt{belokurov2020}), the bulk of this population moves on disc-like orbits whereas the low-$\alpha$ moves on high-eccentricity orbits as expected for GES stars.

%--------------------------------------------------------------
\section{Chemical abundances of RH substructures}\label{chemical_discussion}
%-------------------------- Figure -----------------------------
   \begin{figure*}
   \centering
   \includegraphics[width=1.0\textwidth]{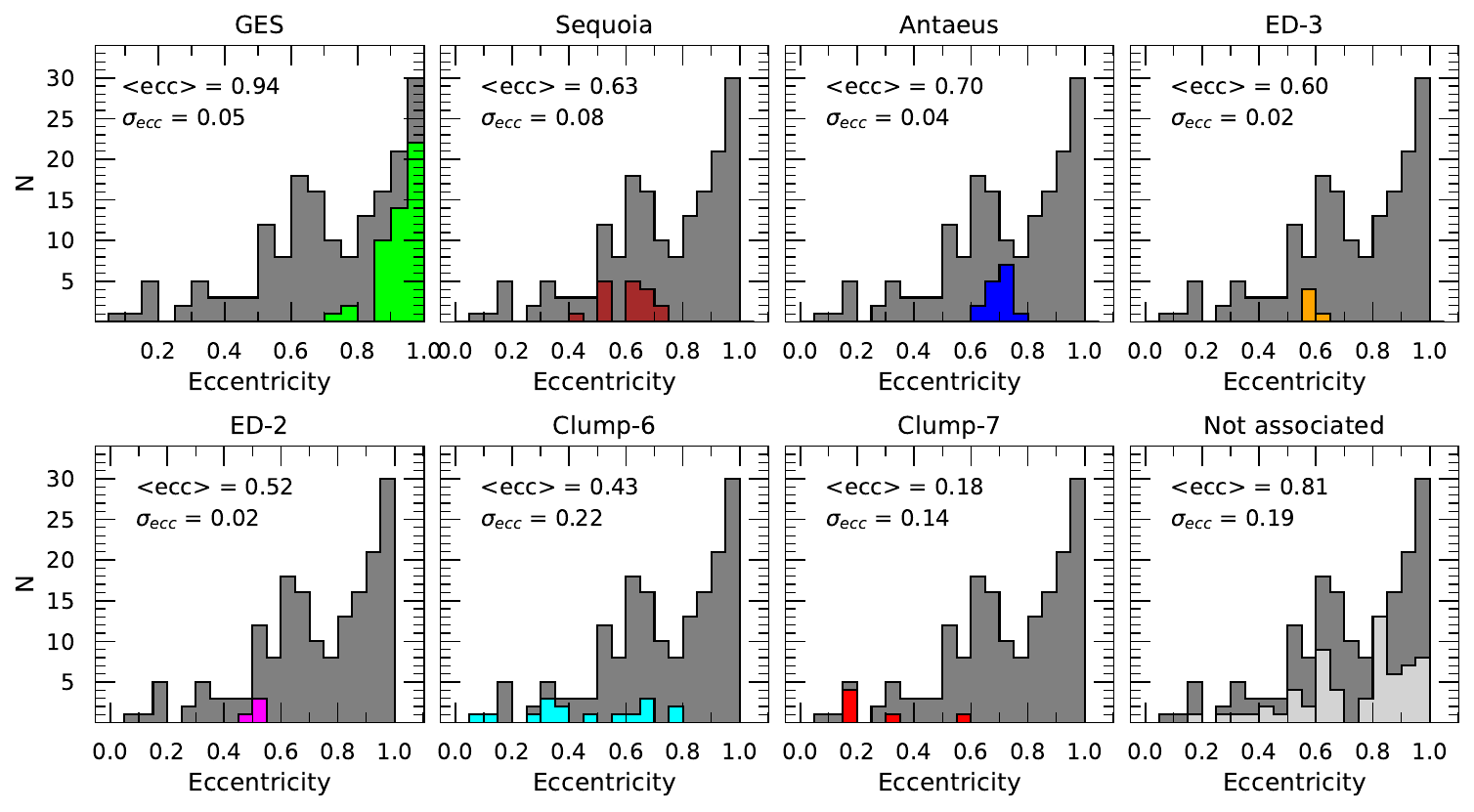}
   \caption{Eccentricity distribution functions for the RH substructures labelled at the top of each panel compared to the distribution of the entire sample of stars. The median eccentricity value and the standard deviation of the distribution are also reported. The color coding is the same as in Fig. \ref{FigIoM}.}
              \label{eccDF}%
    \end{figure*}
%--------------------------------------------------------------
%-------------------------- Figure -----------------------------
   \begin{figure*}
   \centering
   \includegraphics[width=\hsize]{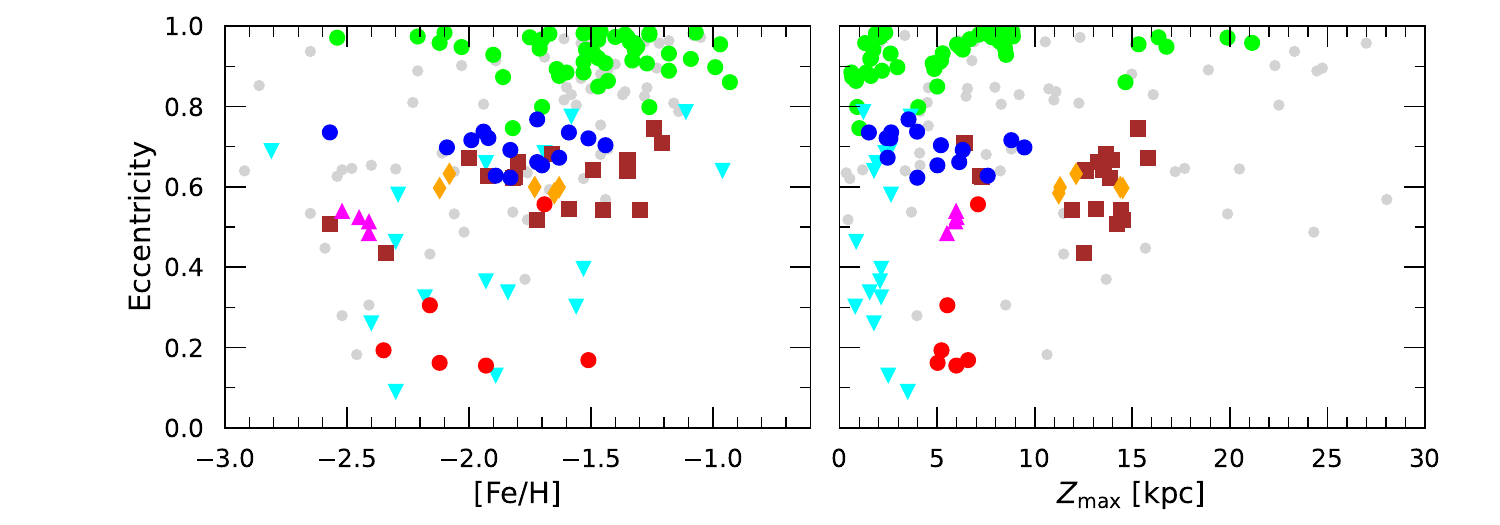}
   \caption{Distribution of the RH substructures in the eccentricity vs. [Fe/H] space (left panel) and in the eccentricity vs. $Z_{\mathrm{max}}$ (right panel). The color coding is the same as in Fig. \ref{FigIoM}.}
              \label{ecc}%
    \end{figure*}
%--------------------------------------------------------------
%-------------------------------------------------------------
   \begin{figure*}
   \centering
   \begin{minipage}{0.49\textwidth}
        \centering
        \includegraphics[width=1.0\textwidth]{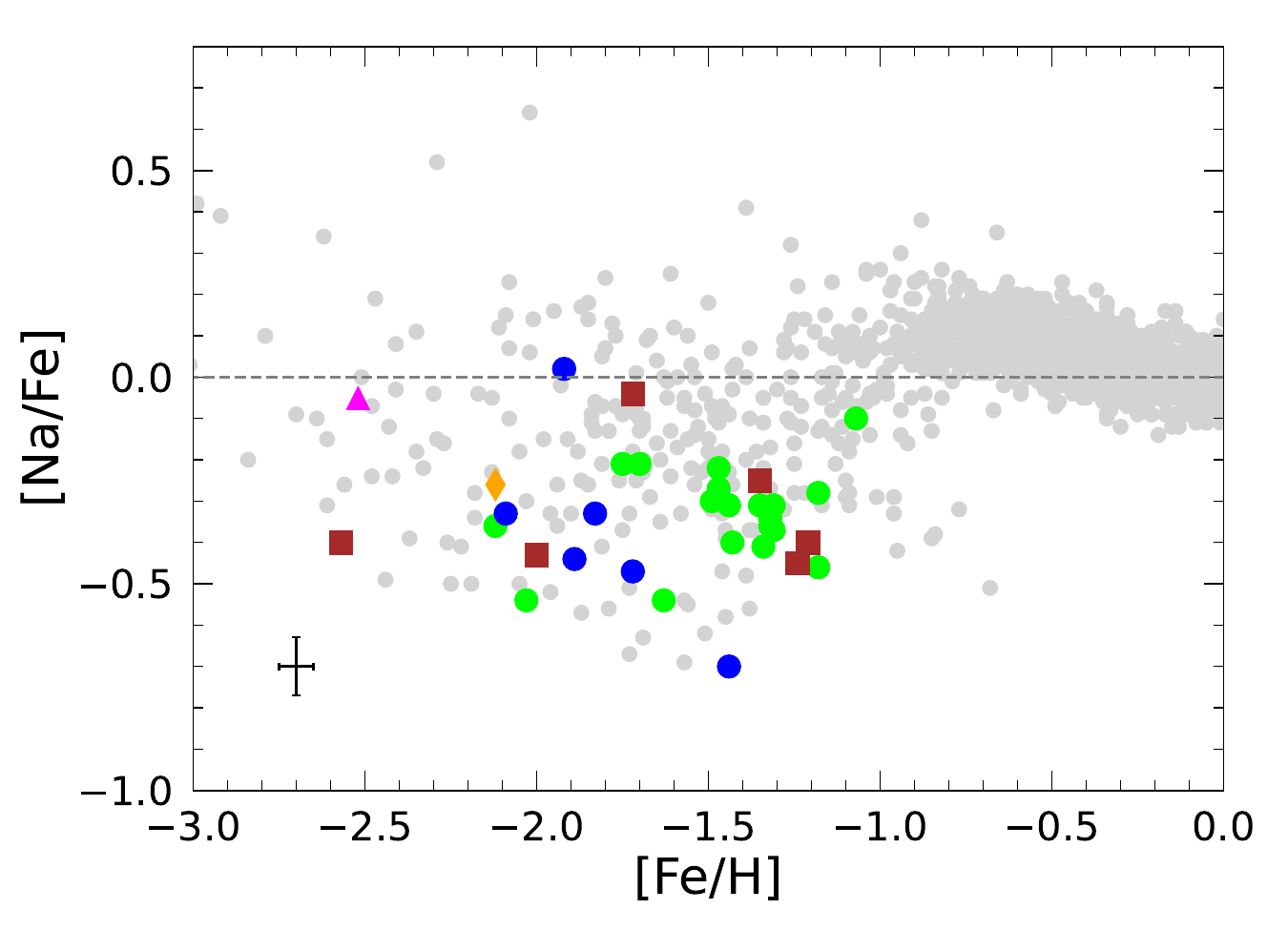}
   \end{minipage}\hfill
   \begin{minipage}{0.49\textwidth}
        \centering
        \includegraphics[width=1.0\textwidth]{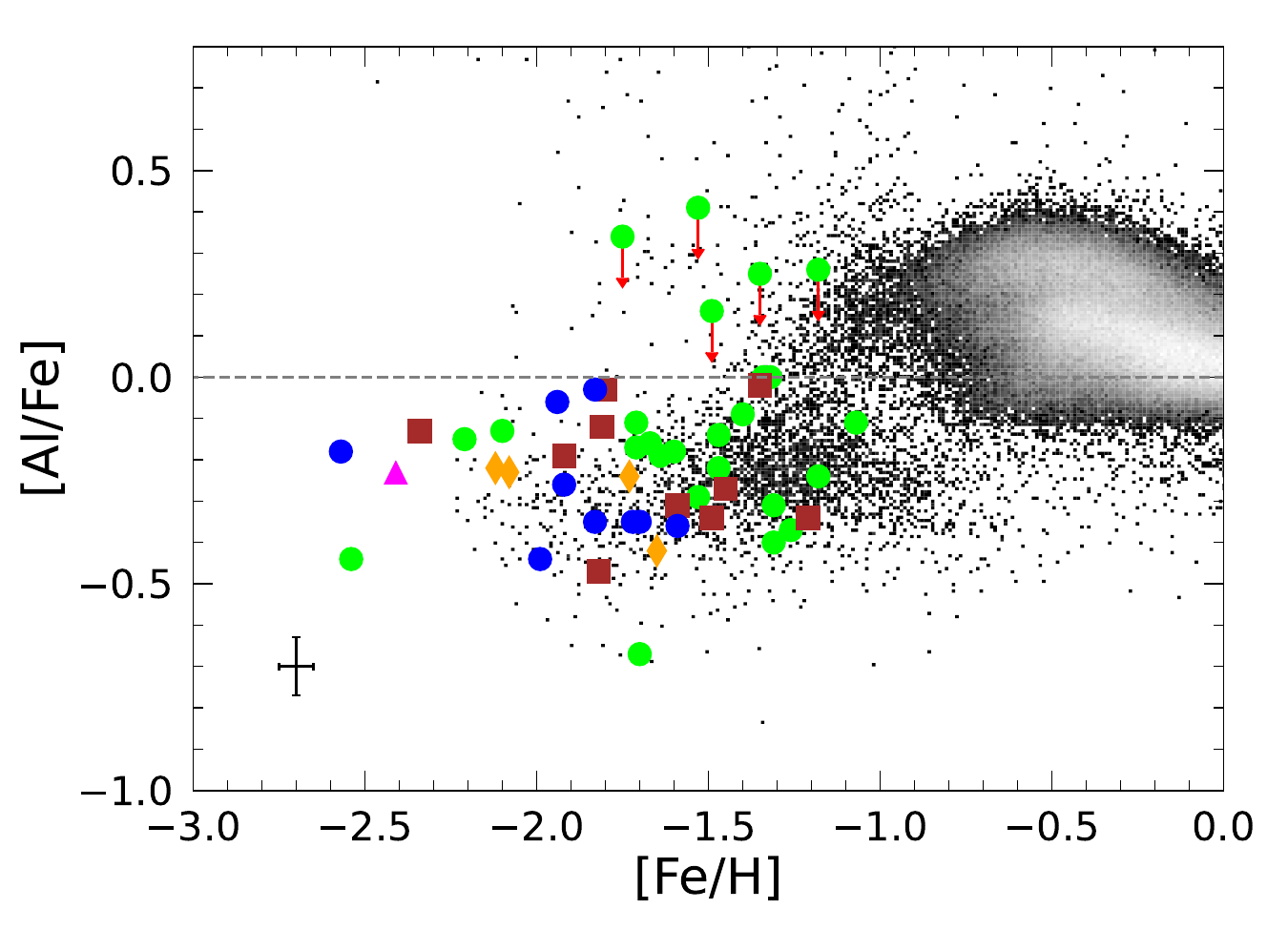} 
   \end{minipage}
   \caption{Behaviour of the odd-Z elements [Na/Fe] and [Al/Fe] abundance ratios as a function of [Fe/H] for GES, Sequoia, Antaeus, ED-2 and ED-3 stars (green, brown, blue, magenta and orange, respectively). Red arrows indicate upper limits. The errorbars in the lower-left corner indicate the typical uncertainties. The MW literature data for Na abundances are taken from \citet{edvardsson93,fulbright2000,stephens2002,gratton03,reddy2003,reddy06,bensby05,bensby14,roederer2014} and \citet{reggiani2017}. The APOGEE control sample, as defined in Section \ref{comparison_literature}, is plotted as 2-D histogram, where black/white colors stand for low/high density regions.}
              \label{FigOddZ}%
    \end{figure*}

%-------------------------------------------------------------
   \begin{figure*}
   \centering
   \begin{minipage}{0.49\textwidth}
        \centering
        \includegraphics[width=1.0\textwidth]{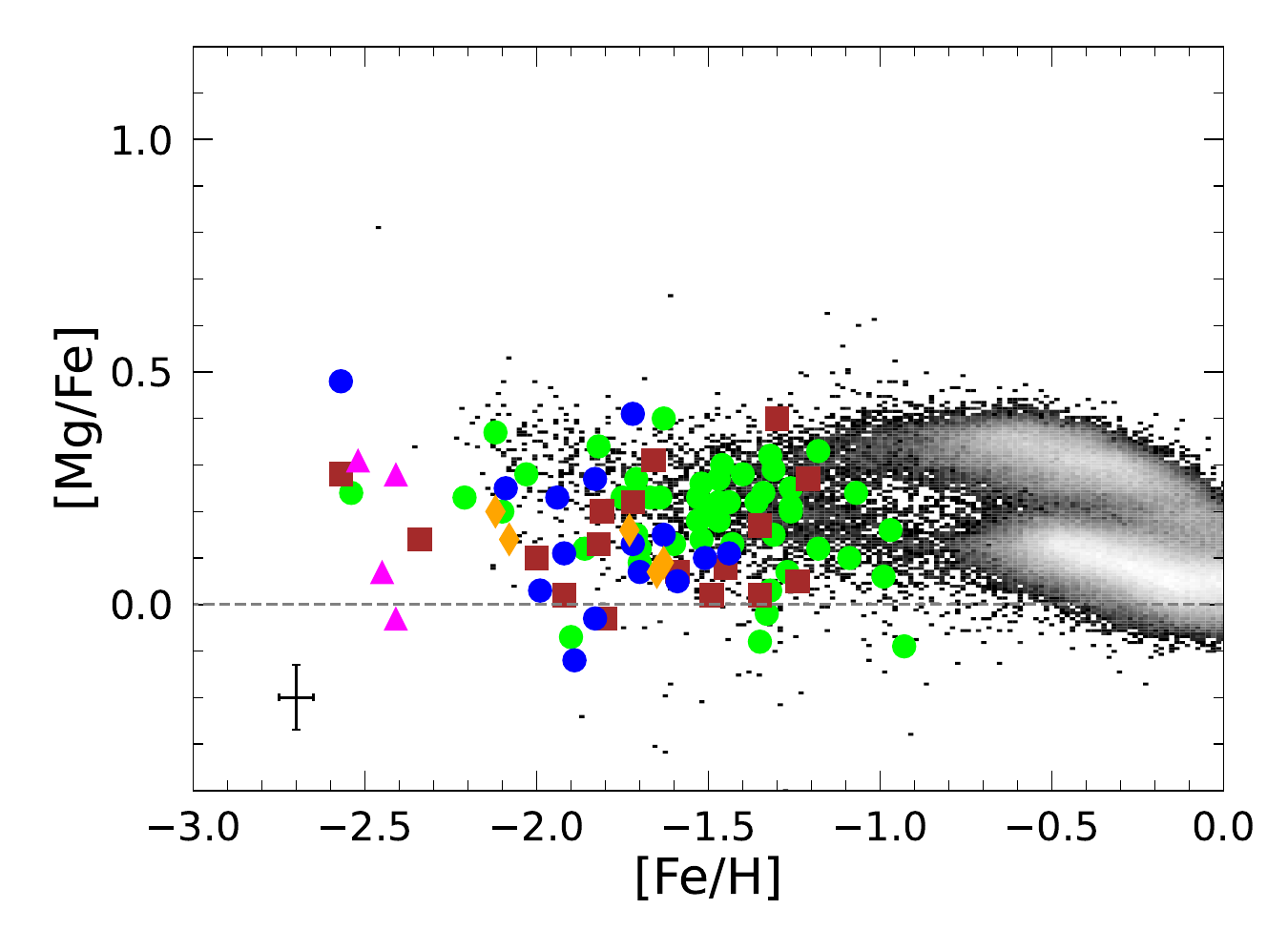}
        \includegraphics[width=1.0\textwidth]{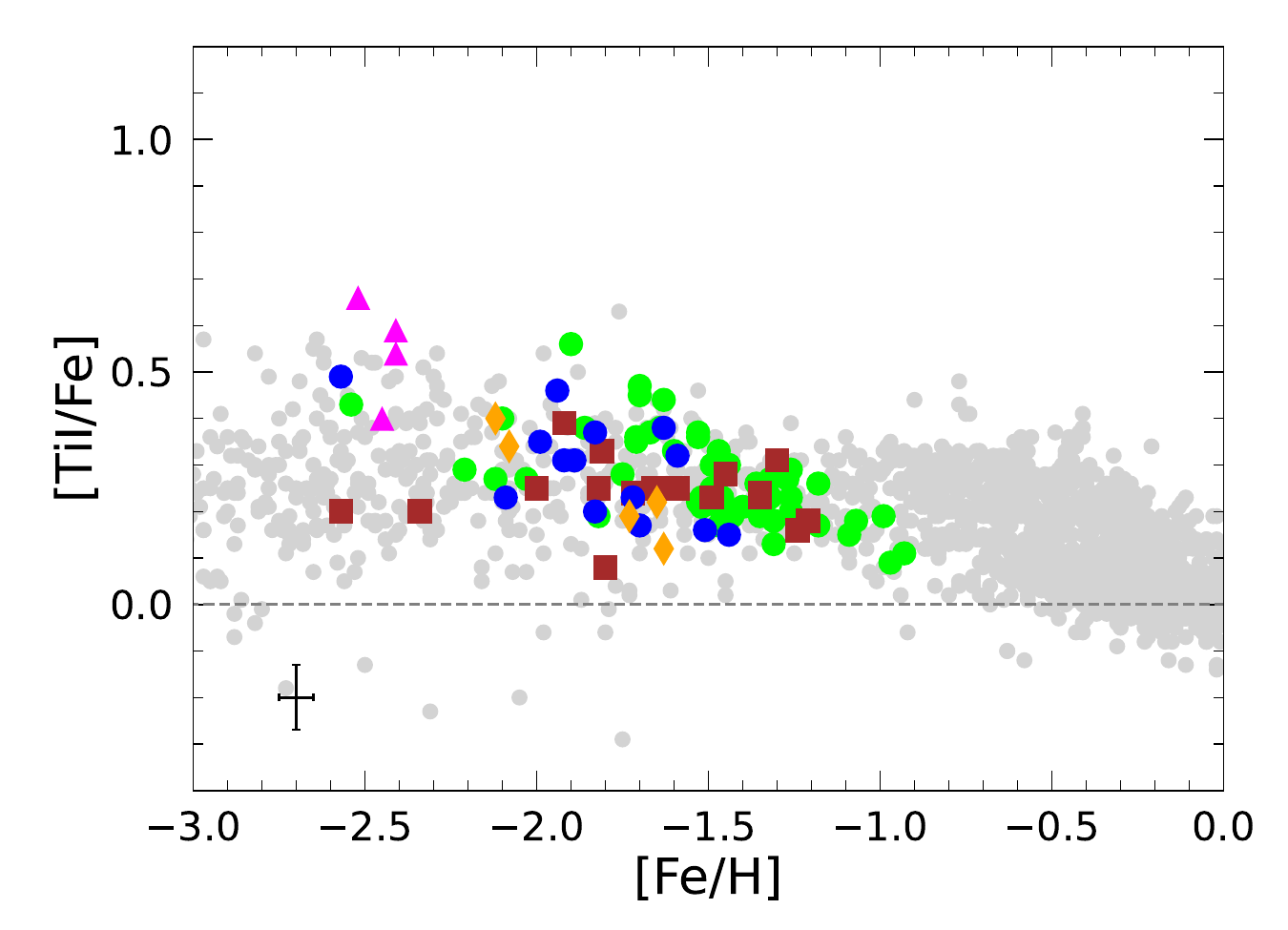}
   \end{minipage}\hfill
   \begin{minipage}{0.49\textwidth}
        \centering
        \includegraphics[width=1.0\textwidth]{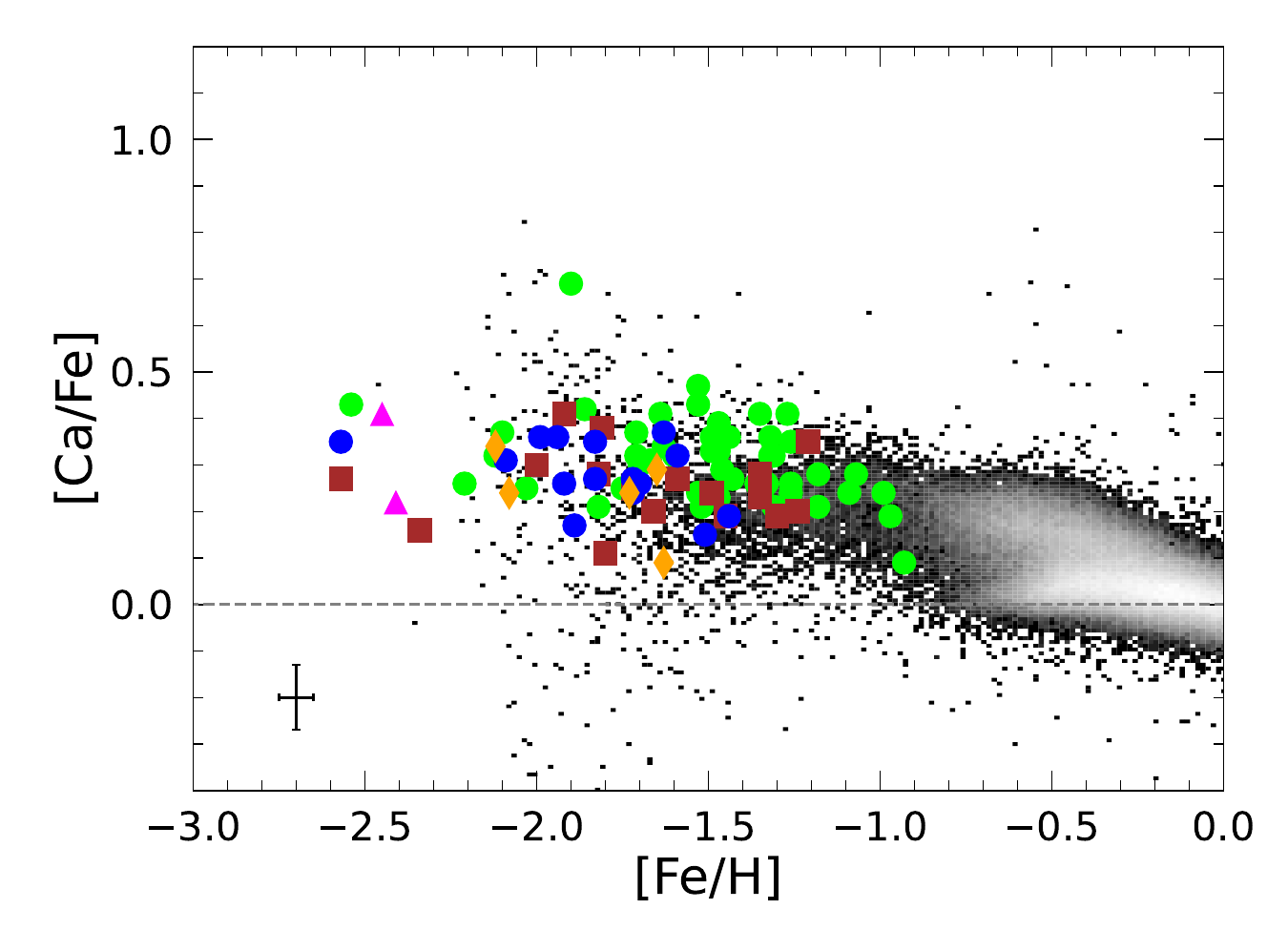}
        \includegraphics[width=1.0\textwidth]{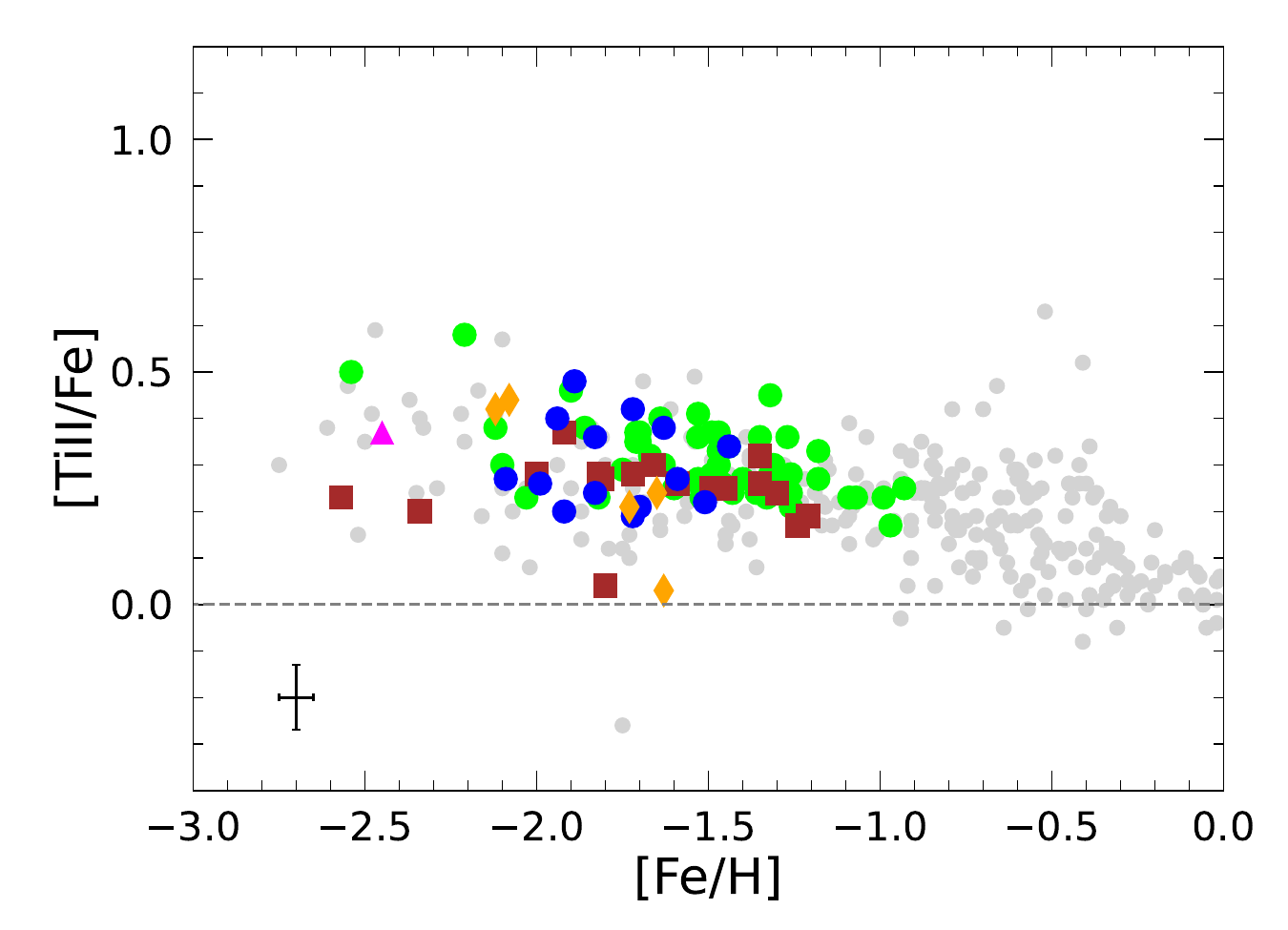}
   \end{minipage}
   \caption{Same illustration as Fig. \ref{FigOddZ} for the $\alpha$- 
           elements [Mg/Fe], [Ca/Fe], [TiI/Fe] and [TiII/Fe] abundance ratios. The MW literature data for TiI and TiII abundances are taken from \citet{edvardsson93,fulbright2000,stephens2002,gratton03,reddy2003,reddy06,barklem05,bensby05,bensby14,roederer2014} and \citet{reggiani2017}.}
              \label{FigAlpha}%
    \end{figure*}
%-------------------------------------------------------------
   \begin{figure*}
   \centering
   \begin{minipage}{0.49\textwidth}
        \centering
        \includegraphics[width=1.0\textwidth]{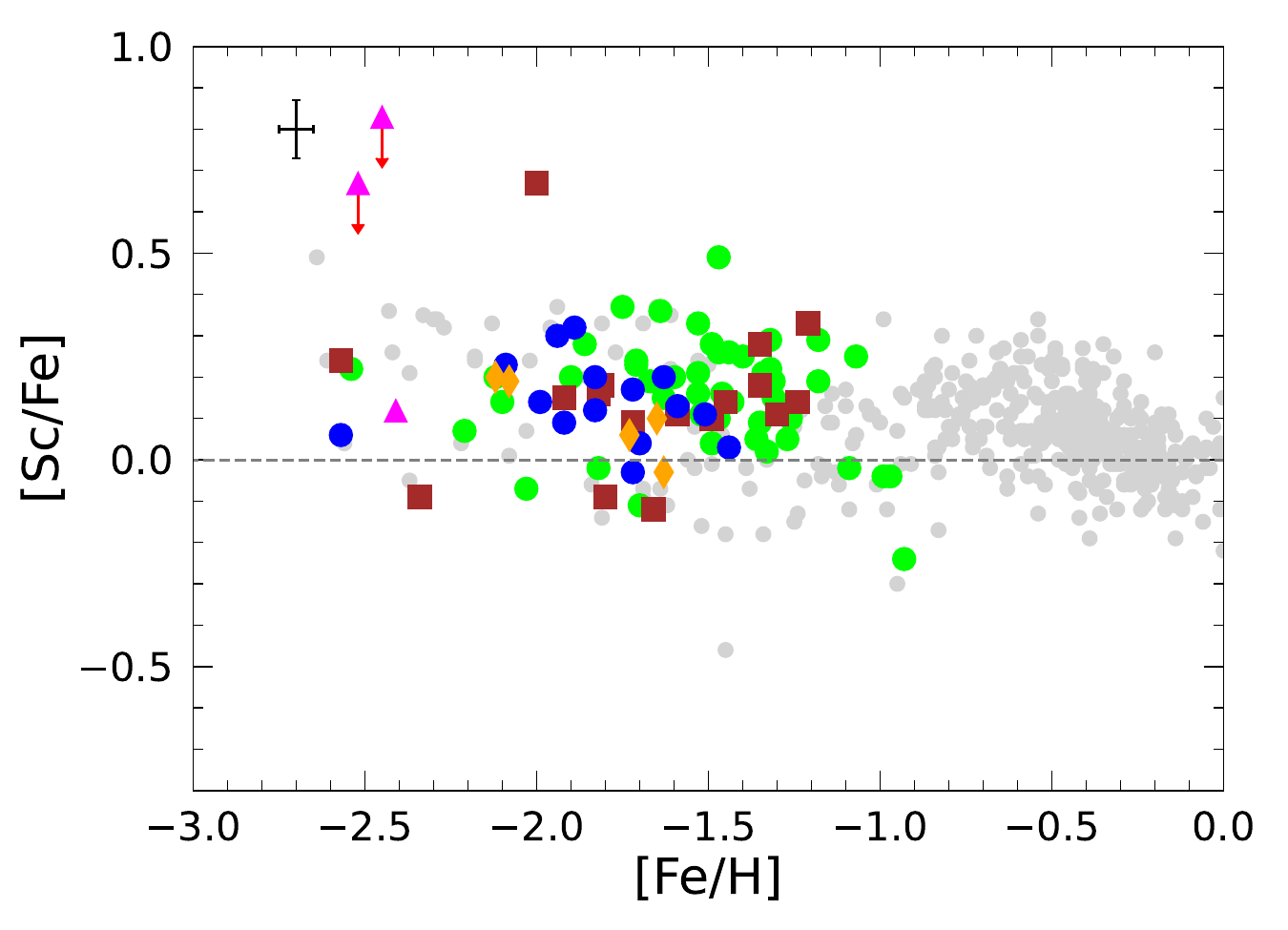}
        \includegraphics[width=1.0\textwidth]{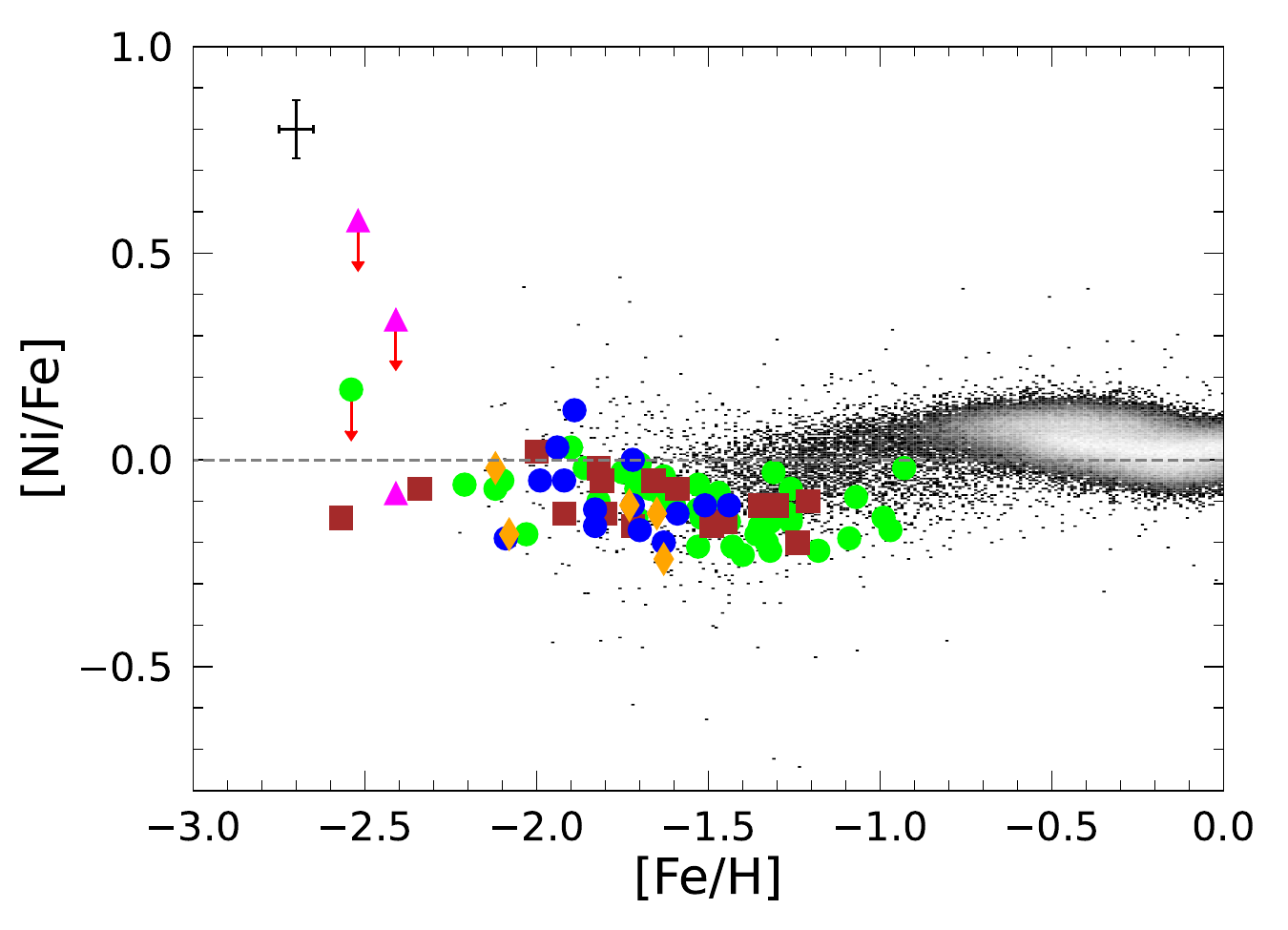}
   \end{minipage}\hfill
   \begin{minipage}{0.49\textwidth}
        \centering
        \includegraphics[width=1.0\textwidth]{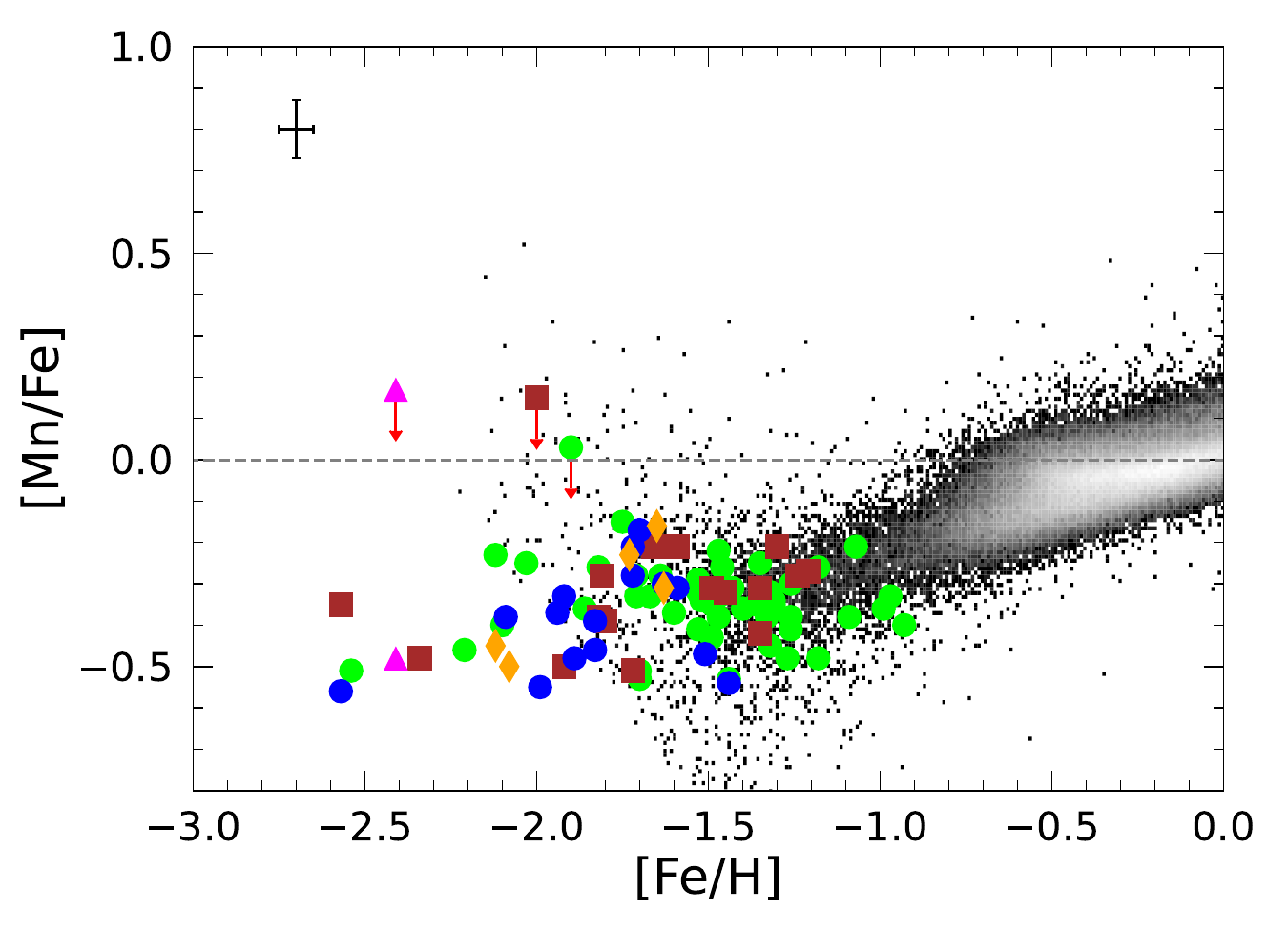} 
        \includegraphics[width=1.0\textwidth]{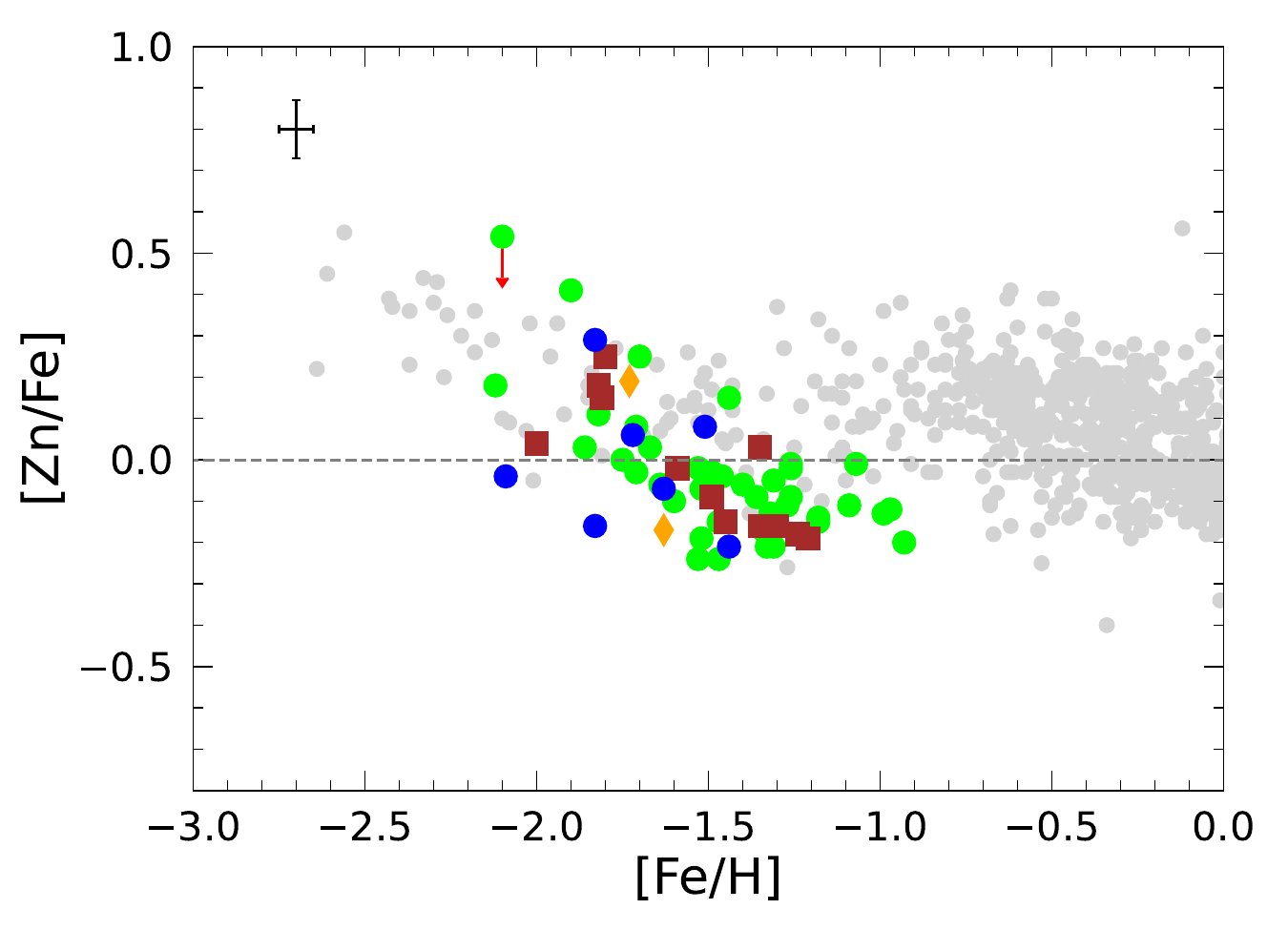}
   \end{minipage}
   \caption{Same illustration as Fig. \ref{FigOddZ} for the iron-peak elements [Sc/Fe], [Mn/Fe], [Ni/Fe] and [Zn/Fe] abundance ratios. The MW literature data for Sc and Zn abundances are taken from \citet{stephens2002,gratton03,reddy2003,reddy06,bensby05,bensby14,roederer2014} and \citet{reggiani2017}.}
              \label{FigIronPeak}%
    \end{figure*}
%-------------------------- Figure -----------------------------
   \begin{figure}
   \centering
   \includegraphics[width=\hsize]{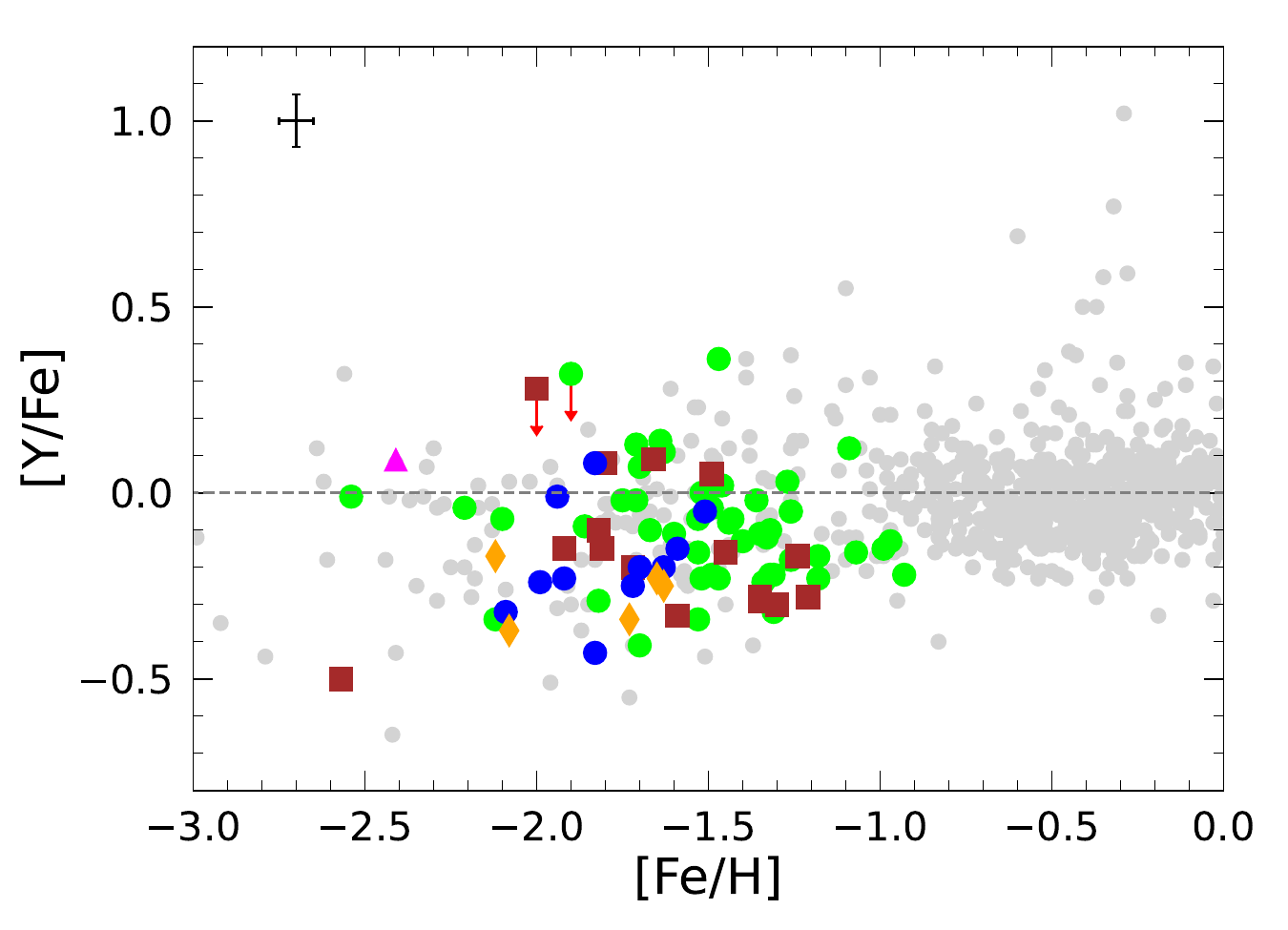}
   \caption{Same illustration as Fig. \ref{FigOddZ} for the neutron capture element [Y/Fe] abundance ratios. The MW literature data for Y abundances are taken from \citet{edvardsson93,fulbright2000,stephens2002,reddy2003,reddy06,bensby05} and \citet{reggiani2017}.}
              \label{FigNeutronCapture}%
    \end{figure}
%-------------------------- Figure -----------------------------
   \begin{figure*}
   \centering
   \includegraphics[width=1.0\textwidth]{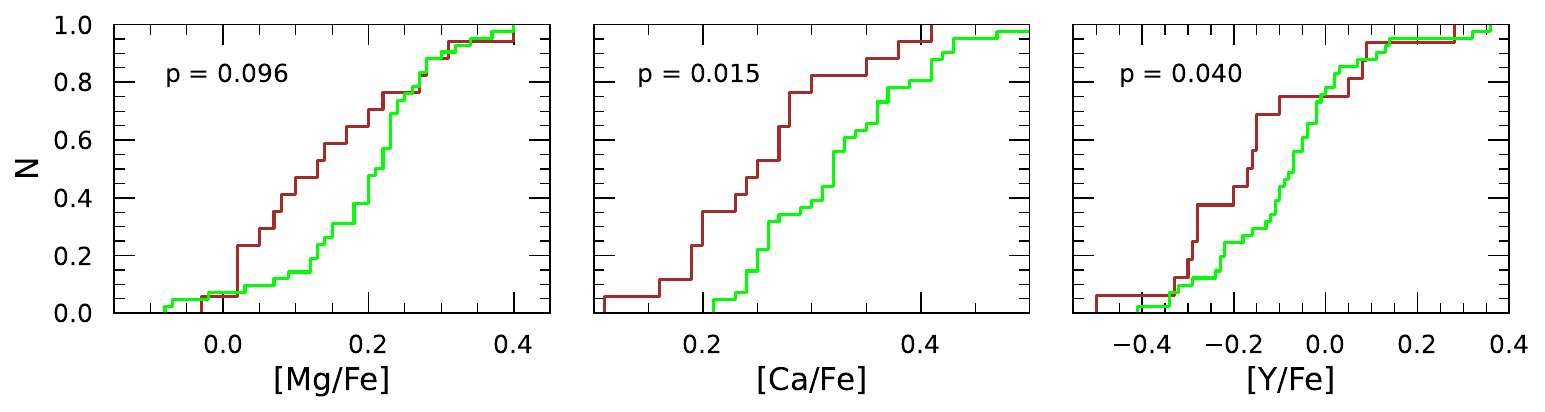}
   \includegraphics[width=1.0\textwidth]{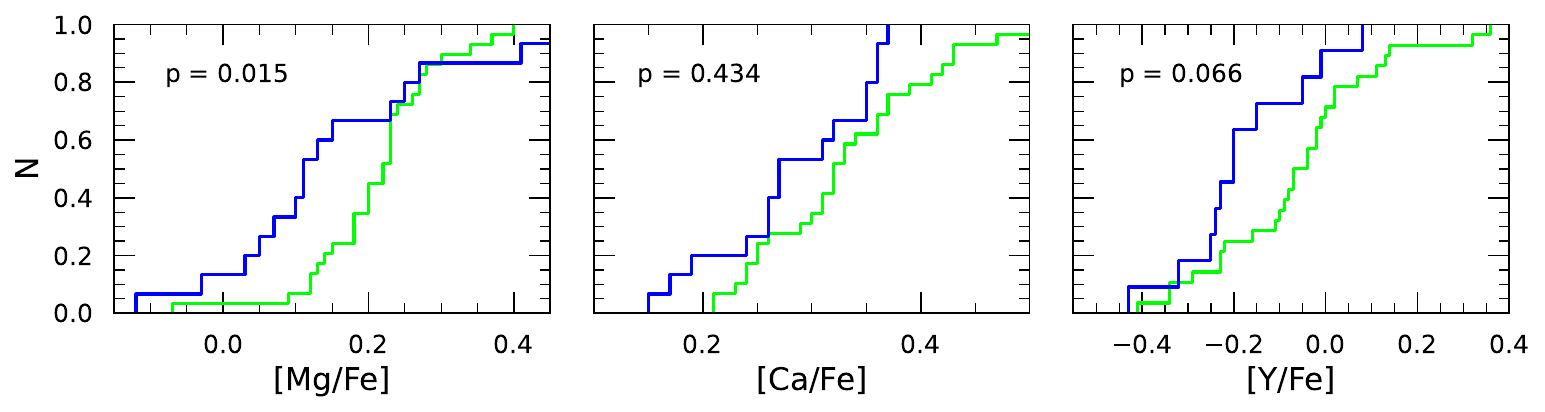}
   \caption{Top panel: comparison of the cumulative distribution functions of [Mg/Fe], [Ca/Fe] and [Y/Fe] for Sequoia and GES stars (brown and green lines, respectively) in the common metallicity range (-2.60 $<\mathrm{[Fe/H]}<$ -1.20 dex). We report the probability that the two distributions are drawn from the same parent sample according to a Kolmogorov-Smirnov test. Bottom panel: same as the top panel but for for Antaeus and GES stars (blue and green lines, respectively) in the common metallicity range (-2.60 $<\mathrm{[Fe/H]}<$ -1.45 dex).}
              \label{FigECDF}%
    \end{figure*}
%------------------------ Figure ---------------------------
   \begin{figure*}
   \centering
   \begin{minipage}{0.33\textwidth}
        \centering
        \includegraphics[width=1.00\textwidth]{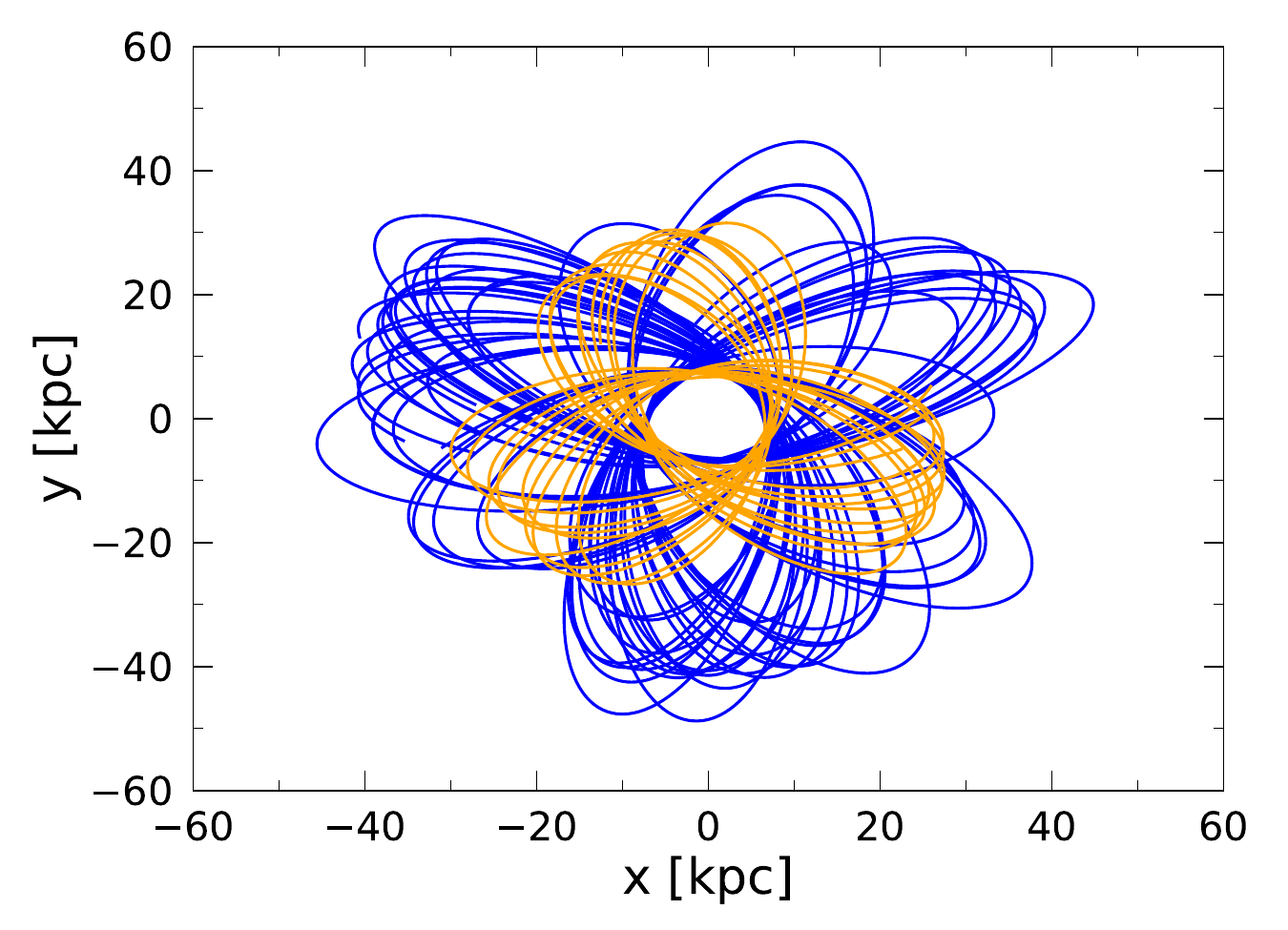} 
   \end{minipage}
   \begin{minipage}{0.33\textwidth}
        \centering
        \includegraphics[width=1.00\textwidth]{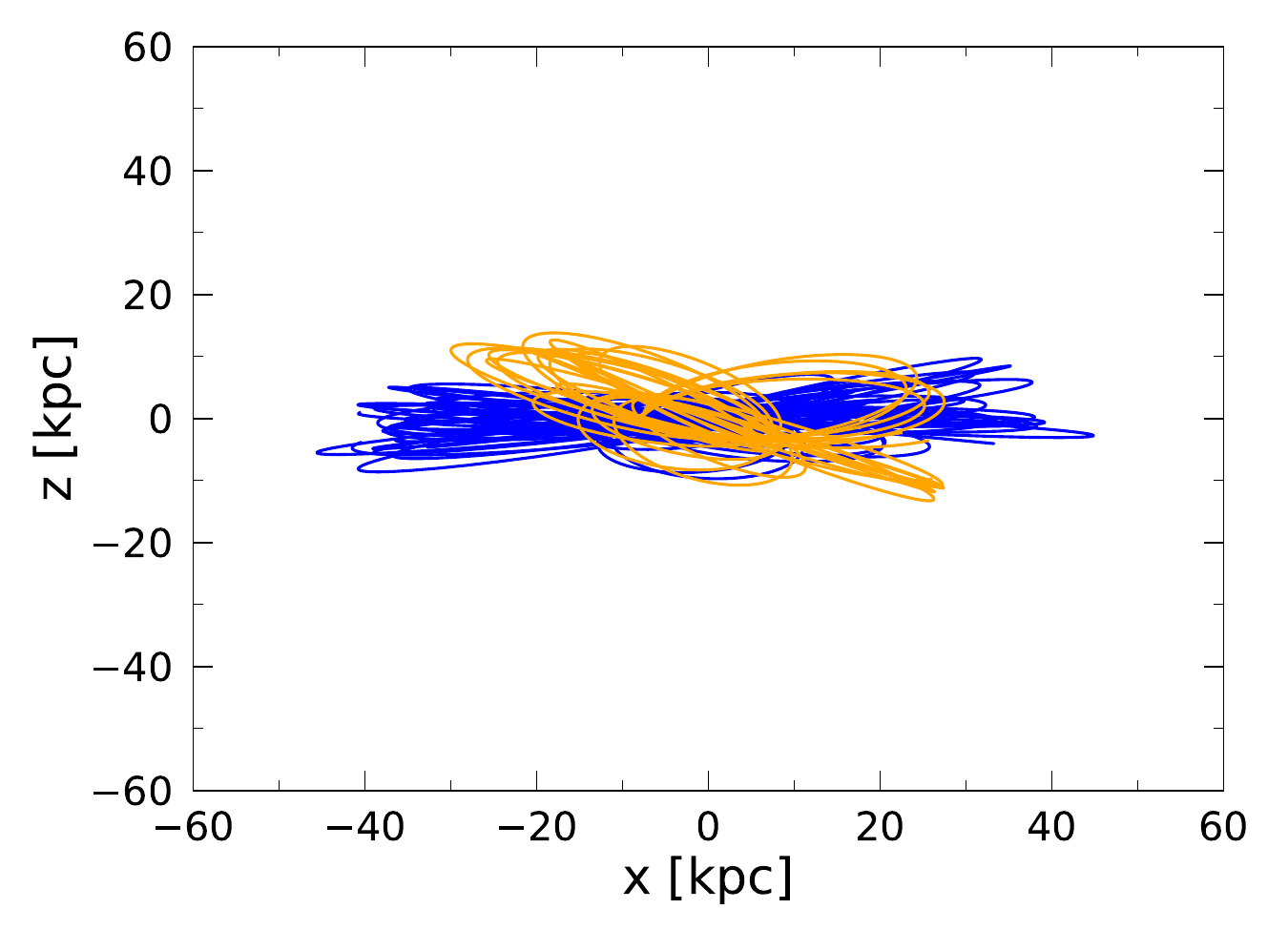}
   \end{minipage}
   \begin{minipage}{0.33\textwidth}
        \centering
        \includegraphics[width=1.00\textwidth]{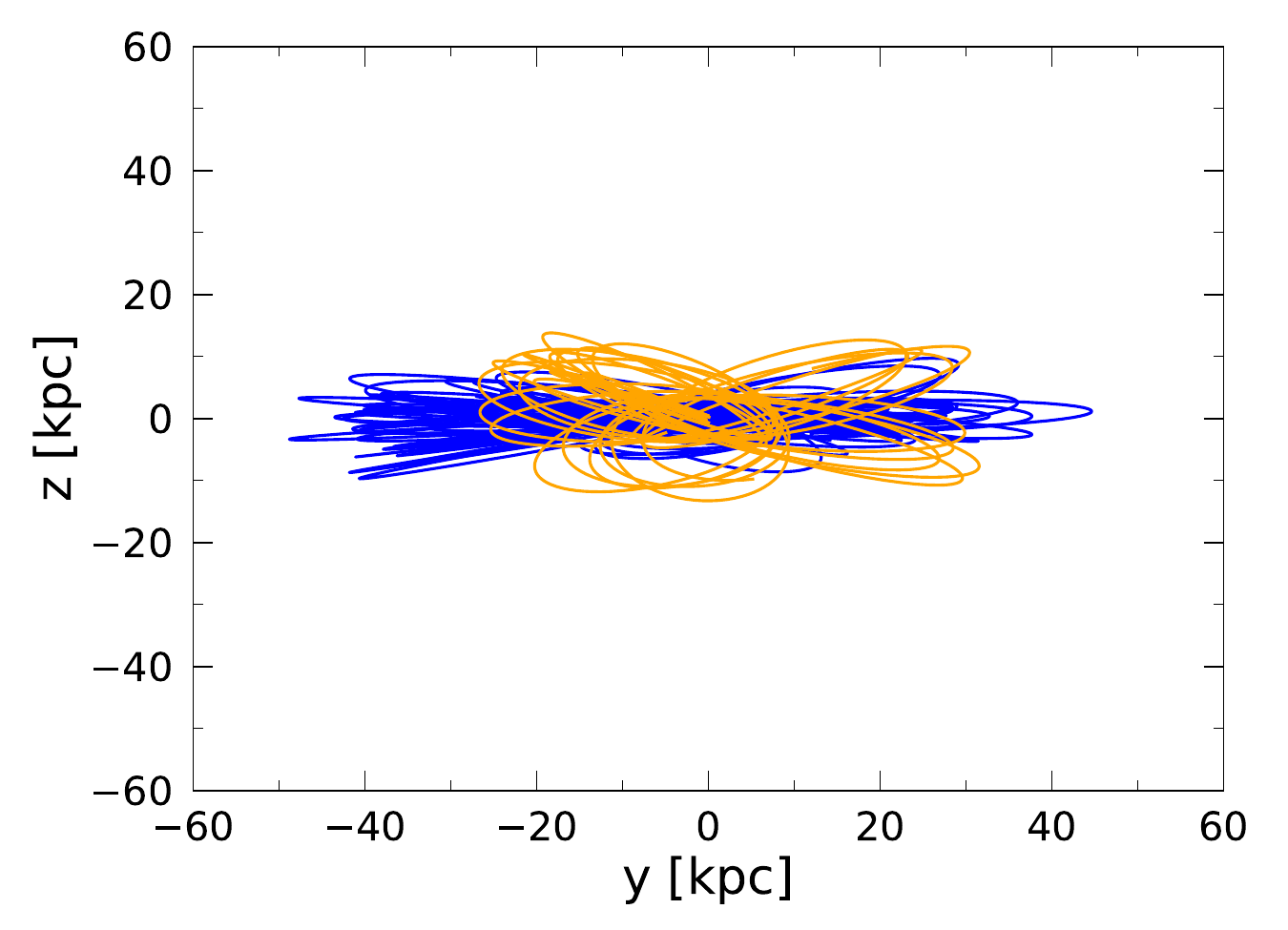}
   \end{minipage}
   \caption{Orbits of Antaeus (blue lines) and ED-3 (orange lines) stars in the Galactocentric reference frame integrated backwards in time for 2.5 Gyr in a \citet{mcmillan17} potential for the MW.}
              \label{Fig_orbits_ED3_LRL64}%
    \end{figure*}    
%---------------------------------------------------------------
%-------------------------------------------------------------

   %Chemical elements are produced in the stellar core through various nucleosynthetic pathways and injected into the interstellar medium at different stages of the stellar evolutionary cycle. Therefore, depending on the star formation efficiencies and chemical enrichment histories experienced by the progenitor galaxies, substructures display different patterns in the chemical planes. These differences in elemental abundances can be used to trace the formation history of the MW, supplying valuable information into the processes that govern galaxy evolution. In fact, the distinctive chemical differences observed among halo substructures play a crucial role in unraveling their intricate dynamical degeneration within the IoM  space. By examining the elemental abundances, particularly in the context of the RH of the MW, we can discern the unique signatures that distinguish overlapping substructures, enabling a comprehensive understanding of their dynamical properties and facilitating a more precise characterization of their origin. 
   In this section we present the chemical abundances of the substructures in the RH of the MW. Having inspected the [Fe/H] vs. [X/Fe] diagrams for all the elements we measured, we discuss here only those that appear to provide the clearest view of the substructures we are considering. In particular, we will focus on the abundances of Na, Mg, Al, Ca, Sc, TiI, TiII, Mn, Ni, Zn and Y. It is of significance to acknowledge that, due to impact of the combination of the atmospheric parameters onto the shape of absorption lines, our capacity to measure the chemical abundances for every depicted chemical element across all stars is restricted. Consequently, certain stars may not feature in specific figures.

\subsection{Metallicity distribution functions}

   We can now examine the metallicity distribution functions (MDFs) of the six substructures that were built exploiting only the IoM. Fig. \ref{MDF} displays the results, which suggest that the global MDF of our sample is bimodal, with a primary peak at $\mathrm{[Fe/H]} \simeq -1.4$ dex and a secondary peak at $\mathrm{[Fe/H]} \simeq -2.4$ dex. The high-metallicity end of the RH stars distribution is dominated by GES, with a median metallicity of [Fe/H] = $-1.47$ dex and $\sigma_{[Fe/H]}$ = $0.32$. Our result is consistent with \citet{bonifacio21} and 0.2-0.3 dex more metal-poor than typical literature values \citep{vincenzo2019,naidu20,feuillet21,buder22,bellazz23}. Once again, we wish to emphasize the importance of our adopted selection criterion for GES stars. This criterion is structured in such a way that (i) prevent us to include stars from the Solar Neighborhood with [Fe/H] $> -0.80$ dex and (ii) we exclusively observe the higher-energy populations of GES. As shown in \citet{koppelman2020}, the high-energy debris of a merger event typically consists of the first stars to have been lost by a progenitor, meaning those in its peripheries. As commonly observed in dwarf spheroidal galaxies orbiting the Milky Way \citep[see, e.g.,][]{tolstoy2023} it is reasonable to expect a negative metallicity gradient that would explain why our targets are slightly more metal-poor than lower-energy GES samples. Stars assigned to Sequoia are, on average, slightly more metal-poor with respect to GES and display a broad MDF, with a median metallicity of [Fe/H] = $-1.66$ dex and $\sigma_{[Fe/H]}$ = $0.37$, that is consistent with the result by D23. %While the bulk of the Sequoia population has a metallicity within $-2.0$ $<$ [Fe/H] $<$ $-1.1$ dex,
   The two most metal-poor stars in Sequoia have a metallicity that overlaps with the one of ED-2. However, these 2 stars occupy a place in the IoM space that is significantly different with respect to the ED-2 clump. Moreover, they do not exhibit the distinctive high [Ti/Fe] ratios observed in the stars of ED-2 at comparable metallicity (see Fig. \ref{FigAlpha}). As a result, we exclude a possible association of these stars with the ED-2 substructure based on their chemical features. Antaeus displays a coherent MDF with a median metallicity of [Fe/H] = $-1.83$ dex and $\sigma_{[Fe/H]}$ = $0.27$, that is consistent with the value of [Fe/H] = $-1.74$ dex provided by \citet{oria2022}. The MDF of ED-3 resembles the one of Antaeus with very small dispersion, having a median metallicity of [Fe/H] = $-1.73$ dex and $\sigma_{[Fe/H]}$ = $0.21$, thus hinting that the two substructures might be related. ED-2 is the most metal-poor substructure and has an extremely tight MDF with a median metallicity of [Fe/H] = $-2.43$ dex and $\sigma_{[Fe/H]}$ = $0.04$, which is consistent with the metallicity range reported by \citet{balbinot2023} for the stars associated to this substructure. In the end, Clump-6 and Clump-7 show rather flat, unpeaked and broad MDFs, with median metallicity of [Fe/H] = $-1.89$ and $-2.02$ dex and $\sigma_{[Fe/H]}$ = $0.48$ and $0.29$ respectively.
   
%\subsection{Eccentricity}
   To have a clearer view, we explore the space defined by the orbital eccentricity, defined as $ecc = (apo - peri)/(apo + peri)$, where $apo$ and $peri$ are the orbital apocenter and pericenter, respectively, that is a dynamical parameter that can help in efficiently differentiating the various substructures. Each substructure comprises stars that predominantly follow orbits characterized by very similar eccentricity, see Fig. \ref{eccDF}. The exceptions are Clump-6 and Clump-7, whose stars are spread across the range of $ecc \simeq 0.1-0.8$. In the left panel of Fig. \ref{ecc} we plot the behaviour of $ecc$ as a function of the metallicity [Fe/H], finding that, in our sample, only stars with [Fe/H] $< -1.5$ dex are moving on low eccentricity orbits ($ecc < 0.4$). As evident from the right-hand panel of Fig. \ref{ecc}, RH substructures are well separated in the $ecc$ - $Z_{\mathrm{max}}$ space, too, where  $Z_{\mathrm{max}}$ indicates the maximum absolute height over the Galactic plane reached by a star during its orbit. Once again the exceptions are Clump-6 and Clump-7. The wide spread in the eccentricity values for the stars in these substructures, coupled with the shape of their MDFs, compel us to attribute the existence of these cluster to a combination of our selection criterion and the choice of hyperparameters in the clustering algorithm. As a result, we do not consider Clump-6 and Clump-7 to be genuine substructures or independent accretion events. Thus, we will exclude stars associated to these two clusters from the rest of the analysis.
   
   Based on the MDF and the eccentricity parameter, three substructures emerge as notably more coherent. GES distinguishes itself as the most metal-rich component, characterized by a peak in the MDF that closely mirrors the primary peak observed in the overall sample. Furthermore, GES comprises stars that exhibit the highest eccentricities in their orbital motion. Then, Antaeus is identified as another well-defined substructure, with its constituent stars primarily having orbits characterized by a relatively high eccentricity ($\simeq 0.7$). Finally, ED-2 exhibits distinctive features in its MDF and $ecc$, both being characterized by a remarkably narrow distribution. It is interesting to note that the 2 stars that were incorporated into the ED-2 clump by the clustering algorithm have [Fe/H] $\simeq$ $-1.90$ dex and also showcase a behaviour more akin to that of Sequoia in eccentricty ($ecc > 0.6$) and [TiI/Fe]. This values support our choice to incorporate them as part of Sequoia.

\subsection{Odd-Z elements}\label{oddz}

  Odd-Z elements, like Na and Al, are believed to be mainly synthesised in massive stars exploding as core-collapse supernovae (CC-SNe) at low metallicity and in Asymptotic Giant Branch (AGB) stars at intermediate metallicity \citep{nomoto2013}. Since they are formed starting from a neutron excess produced in the CNO cycle, their yield depends on the metallicity of the progenitor \citep{kobayashi2020}.
  
  Recent studies have demonstrated that stars that have formed in-situ are enhanced in odd-Z elements abundances with respect to accreted populations and stars in dwarf satellite galaxies of the MW \citep{tolstoy09,feuillet21,hasselquist21,belokurov22}.

  We show in Figure \ref{FigOddZ} that the stars from our sample display sub-solar values of [Na/Fe], in agreement with the findings by \citet{nissen&schuster2010}. Moreover, it is evident that Antaeus displays a depletion in Na relative to GES. This distinction is discernible through the positioning of its stars on a distinct sequence, consistently registering values $\sim 0.2$ dex lower. The sequence defined by Antaeus closely resembles the pattern found by \citet{matsuno22} for the stars belonging to Sequoia, whereas our stars associated with Sequoia exhibit behaviour more akin to that of GES. However, \citet{matsuno22} employed a relatively wide retrograde region within the IoM space to identify members of Sequoia (with $L_{\mathrm{z}}$ values ranging from -1600 to -3100 kpc km $\mathrm{s^{-1}}$), making very likely that some Antaeus stars inadvertently contaminate their Sequoia sample, thus potentially contributing to the observed similarities. Figure \ref{FigOddZ} also shows that the [Al/Fe] trend of our targets is almost flat at [Al/Fe] $\sim$ -0.3 dex for [Fe/H] $<$ -1.0 dex. The [Al/Fe] abundances observed in the majority of the target stars show a distinctively lower level compared to the typical values observed in the MW disc, thereby indicating a consistency with the position occupied by accreted systems. Interestingly, we find a spread of about 0.5 dex in the values of [Na/Fe] and [Al/Fe] for the associated stars, which might hint at a different fraction of massive stars exploding as hypernovae in the progenitors \citep{smiljanic2016}. %No evident differences can be spotted among different substructures. 

%--------------------------------------- 
\subsection{$\alpha$-elements}\label{alpha}

  With the notable exception of oxygen, the $\alpha$-elements are mainly synthesised in the core of massive stars and diffused into the interstellar medium through CC-SNe, however a small fraction is produced also in SN Ia \citep{kobayashi2020}. Thus, the chemical patterns of these elements can be used as a proxy of the different timescales of star formation. A typical indicator of the efficiency of a galaxy's star formation is the metallicity of the knee in the [$\alpha$/Fe] vs. [Fe/H] plane, which denotes the beginning of a sizable contribution to the chemical enrichment by SNe Ia \citep{tinsley79,matteucci&greggio86}.
  
  Mg is synthesised through the hydrostatic burning of He, C and Ne in massive stars (30 - 35 $\mathrm{M_{\odot}}$) with a negligible contribution by SN Ia \citep{woosley&weaver1995,arnett1996}. The explosive $\alpha$-elements (Ca and Ti) are mainly produced in less massive stars (15 - 25 $\mathrm{M_{\odot}}$), with a considerable fraction formed also in SN Ia \citep{kobayashi2020}.
  
  As shown in Fig. \ref{FigAlpha}, the interpretation of our findings is rather challenging. The spread of the measured [$\alpha$/Fe] is significantly larger than our uncertainties. The fact that our sample is systematically more metal-poor than typical disc stars makes the comparison with the APOGEE control sample not trivial (see Mg and Ca in Fig. \ref{FigAlpha}). Indeed, the [Mg/Fe] of our sample seems systematically lower at [Fe/H] $>$ -1.5 dex, when the patterns expected for a dwarf galaxy start to part ways from that expected for the MW \citep{tolstoy09,nissen&schuster2010,hayes2018,horta23}. In the most metal-poor regime the two populations mix, but this is not surprising given that SN Ia still have to start exploding. Conversely, for elements like Ti, where the reliability of APOGEE derived abundances is uncertain, we have opted to juxtapose our findings with established high-resolution spectroscopy literature  data. As demonstrated in Fig. \ref{FigAlpha}, our results exhibit a strong concordance with those obtained MW halo. Apart from this general behaviour, we now focus in detail on the $\alpha$-elements trends of the six substructures, and discuss the possible origin of the distinct chemical patterns: 
  
  \begin{itemize}
      \item \textit{Gaia-Enceladus}. The Mg abundance ratios of the GES stars overlap the metal-poor tail of the distribution of the low-$\alpha$ MW population, as already observed in previous works \citep{helmi2018,mackereth19}. The reason for the difference in abundance between GES and high-$\alpha$ stars is commonly believed to be the result of chemical enrichment by SN Ia. In fact, in-situ stars have not undergone significant enrichment by SN Ia at [Fe/H] $\simeq$ -1.0 dex, whereas a dwarf galaxy like GES has already experienced such enrichment due to a less efficient star formation history. 
      \item \textit{Sequoia}. The stars associated to Sequoia exhibit flat [$\alpha$/Fe] patterns across the entire metallicity range. In particular, Sequoia occupies the same region of GES at [Fe/H] $> -1.5$ dex, especially in the chemical planes of TiI. At lower metallicity, however, the behaviour exhibited by the Sequoia substructure more closely resembles that of the smaller retrograde substructures. At the same time, Sequoia seems to display [Mg/Fe] and [Ca/Fe] ratios 0.1 - 0.2 dex lower than GES, in agreement with the results by \citet{matsuno22}. Based on these findings, we favor the interpretation of Sequoia as an independent merger event with respect to GES. Lower [$\alpha$/Fe] values cannot be explained by the scenario that considers Sequoia as the outskirt of GES that were first lost during the tidal disruption of the galaxy, as this would mean that the outer regions were polluted by SN Ia ejecta before the inner ones, which is a feature yet not observed in existing dwarf MW satellites.
      \item \textit{Antaeus}. The abundances of the $\alpha$-elements suggest that Antaeus follows a distinctive chemical sequence in the all the [$\alpha$/Fe] planes at lower values at fixed metallicity with respect to the dominant components in the RH (GES and Sequoia). At [Fe/H] $\simeq$ -1.5 dex, the [$\alpha$/Fe] of Antaeus are consistently 0.1 - 0.2 dex lower than those observed in GES. We interpret this feature as evidence for the existence of Antaeus as an independent merging event, characterised by a progenitor with lower SFR compared to GES and Sequoia. 
      \item \textit{ED-3}. This substructure is located in a region of the [$\alpha$/Fe] chemical plane that is consistent with that of Antaeus. Being constituted of only 5 stars, ED-3 overlaps with the Antaeus clump across its entire metallicity range. The two systems display a high degree of proximity in terms of their $E$ and $L_{\mathrm{z}}$, resulting in significant overlap within this parameter space. Based on this new chemo-dynamical evidence, we conclude that Antaeus / L-RL64 and ED-3 are actually part of the same relic. This evidence is confirmed in all of the chemical planes, as we will show in the next paragraphs.      
      \item \textit{ED-2}. The 4 stars associated with ED-2 occupy the very metal-poor end of the chemical planes, being all located at -2.6 $<$ [Fe/H] $<$ -2.4 dex. This system exhibits a markedly narrow metallicity distribution, and its chemical composition is generally consistent with that of the other RH substructures, with the exception of Ti, where it seems to be enhanced. Its position in the [$\alpha$/Fe] plane seems to recall the one of I'itoi \citep{naidu2022}. %Unfortunately, the $\alpha$-element abundances observed in the two stars with ambiguous dynamical associations do not provide a reliable discriminant for determining their membership in either ED-2 or Sequoia.
      Given the small sample, any conclusions inferred about the chemical pattern of this system would be premature and require further investigation on a much extended sample of stars. 
      %\item \textit{Clump-6}. This substructure exhibits an unusually broad metallicty distribution, while also showing chemical patterns that strongly resemble those of GES across all $\alpha$-element chemical planes. Given these chemical trends and its peculiar position in the IoM space (see Fig. \ref{FigIoM}), it could be inferred that Clump-6 may have originated from a dwarf galaxy with a comparable characteristics to GES, albeit accreted at an earlier epoch. However, such a scenario is improbable, leading us to lean towards the interpretation that this system may be an artificial construct, resulting either from a selection bias introduced by our criterion (see Section \ref{selection}) or from a significant contamination by GES stars \citep{belokurov23,davies2023}. As a result, any conclusions drawn regarding its chemical characteristics would be likely unreliable and warrant further investigation.
      %\item \textit{Clump-7}. This clump features a very low number of stars, among which there is a significant spread in the chemical abundances of $\alpha$-elements. This behaviour prevent us to discern a clear chemical pattern. As for Clump-6, these findings suggest that the nature of this substructure may be artificial.
  \end{itemize}

\subsection{Iron-peak elements}\label{ironpeak}

  The production of iron-peak elements is sourced from a variety of nucleosynthesis formation pathways, with SN Ia, CC-SNe and hypernovae all making a sizeable contribution to the overall production of these chemical elements \citep{romano2010,leung2018,kobayashi2020,lach2020}.
  
  Sc is mainly produced in CC-SNe in the burning of C and Ne \citep{woosley2002}, with a negligible contribution by SN Ia \citep{kobayashi2020}. Mn is almost entirely produced by SN Ia, with a relative contribution by CC-SNe and hypernovae even lower than Fe \citep{kobayashi2009,kobayashi2020}. SN Ia are also the predominant channel of formation of Ni, although it can be produced in CC-SNe, as for Fe. In the end, Zn is synthesised in the highest energy hypernovae.
  
  %In the following paragraph, we aim to describe the behaviour of iron-peak elements, with particular focus on Ni and Mn.
  We note that the all the RH substructures exhibits abundances that are remarkably similar to those observed in accreted populations and low-mass satellite galaxies of the MW \citep{NS97,tolstoy09}. %Specifically, we find that the abundances of Ni and Mn in these system are lower than those observed both of the low-$\alpha$ and the high-$\alpha$ sequences observed in the MW. %In particular, we found that GES dominates the sample at [Fe/H] $>$ -1.5 and displays a [Mn/Fe] $\sim$ -0.3 which is slightly higher than the one of Antaeus ([Mn/Fe] $\sim$ -0.5).
  Interpretation of these results is complex, as the small differences between the various substructures can be explained by both the diverse production sites of these elements and the metallicity dependencies of their yields at low metallicities \citep{weinberg2022}. The chemical patterns of the iron-peak elements for the different substructures seem to reveal that GES and Sequoia display a similar distribution. Moreover, Antaeus and ED-3 exhibit once again an overlap in their respective loci within these chemical plane, providing additional support for the hypothesis that they are associated with the same accretion event.

\subsection{Neutron capture elements}\label{neutronecapture}

  Of all neutron capture elements, here we discuss only yttrium (Y). This element is predominantly formed by the slow neutron capture process (\textit{s}-process, i.e., neutron captures slower than the $\beta$-decay).
  
  Y is predominantly formed by \textit{s}-processes in the envelope of low-mass (1 - 4 $\mathrm{M_{\odot}}$) AGB stars and, to a lesser extent, in massive stars \citep{busso1999,pignatari2010,cristallo2015}. At low metallicities, magneto-rotational SNe and mergers of neutron stars contribute significantly to the production of these neutron-capture elements via \textit{r}-processes, thus explaining the observed scatter \citep{kasen2017, molero2023}.
  
  Despite the huge scatter between abundances of stars born in the same progenitor, the results in Fig. \ref{FigNeutronCapture} show that the average abundances for the Sequoia substructure is comparable to the one of GES, as previously hinted also by \citet{aguado21}. Within the common metallicity range, Antaeus and ED-3 exhibit a pronounced degree of overlapping, consistently displaying average abundances that are lower ($\sim$ 0.1 - 0.2 dex) than those observed in GES.

\section{Discussion and conclusions}\label{conclusion}

    %Establishing an unambiguous link between halo substructures identified solely based on orbital information and accretion events poses an extremely challenging task. As extensively shown by numerical simulations \citep{koppelman2020,naidu2021,belokurov23}, substructures can potentially suffer from contamination by stars originating in other progenitors, or stars born from the same progenitor can originate separated pile-ups in the dynamical space. When trying to solve this dynamical degeneration, the chemical composition of accreted stars comes to help, since it is imprinted in the stellar atmosphere and reflects the chemical enrichment history of the progenitor galaxy.
    In this first paper of a series, we have introduced the WRS project, describing the methodology employed and highlighting the scientific goals it aims to pursue. To demonstrate the potential of this project, we narrowed our focus to the subset composed of 90 stars that were associated by the \texttt{DBSCAN} clustering algorithm with various RH substructures within the IoM space \citep[see][]{dodd23}, analyzing their chemical properties. The interpretation of the chemical patterns of unassociated stars will be left to future contributions to the series. Hereafter, we present a concise summary of our principal findings:
    
    \begin{enumerate}
      \item A selection criterion based on tangential velocity only (thus not requiring the $V_{\mathrm{los}}$ information) is able to select stars that move on retrograde motions with a 91.4$\%$ efficiency (by requiring $V_{\mathrm{T}} > 400$ km $\mathrm{s^{-1}}$). All of the stars included in our sample consistently exhibit chemical pattern across all chemical element planes that aligns with the characteristic abundance patterns displayed by accreted systems and dwarf galaxies.
      \item In spite of selection criteria leaving out of our sample the bulk of GES stars at higher $L_{\mathrm{z}}$ and lower energy, GES stands out as the largest and most metal-rich component within the RH, which supports the hypothesis that it represents the remnant of the most massive galaxy that was accreted by the MW. 
      \item The chemical patterns of GES and Sequoia display similar trends at [Fe/H] $> -1.5$ dex. However, significant distinctions become evident at lower metallicity, specifically characterized by a discernible difference in the  $\alpha$-elements. In particular, as shown in the top panel of Fig. \ref{FigECDF}, Sequoia exhibit a pronounced excess of stars with lower [Mg/Fe] and [Ca/Fe] ratios with respect to GES. Consequently, this discrepancy leads us to interpret Sequoia as an independent accretion event distinct from GES. 
      \item For the first time, we have achieved a comprehensive chemical characterization of Antaeus, encompassing detailed chemical abundances for numerous elements. Our analysis reveals distinct behaviour in the chemical space compared to GES within the same metallicity range (see bottom panel of Fig. \ref{FigECDF}), reinforcing the assertion that it constitutes the remnant of an accretion event independent of GES.
      \item Antaeus and ED-3 exhibit analogous positions in the IoM spaces, albeit with some distinctions in $L_{\perp}$. Additionally, they share similarities in terms of their metallicity distribution function and orbital eccentricity. Furthermore, the chemical patterns observed in Antaeus and ED-3 are consistent across all chemical elements. In the end, we integrated backwards in time for 2.5 Gyr the orbits of the stars associated to these two substructures in a \citet{mcmillan17} potential for the MW. As depicted in Fig. \ref{Fig_orbits_ED3_LRL64}, stars from the two substructures appear to move on very similar orbits. Collectively, these lines of evidence compel us to interpret them as components of the same merging event. Henceforth, we will refer to this merging event as Antaeus.
      %Additionally, the chemical composition of Antaeus distinctly deviates from that of GES. This divergence prompts us to characterize it as the remnant of a independent progenitor.
      \item ED-2 is the most metal-poor component, with a remarkably narrow MDF peaked at [Fe/H] = $-2.43$ dex, in agreement with the findings by \citet{balbinot2023}. The limited number of stars within the ED-2 substructure, coupled with the specific combination of their stellar parameters ($T_{\mathrm{eff}} >$ 6000 K and [Fe/H] < -2.4 dex) that lead to the manifestation of weak spectral lines, restricts our capacity to attain detailed chemical abundances across all elements. As a consequence, we are unable to definitively assert whether ED-2 constitutes the remnant of a globular cluster or a dwarf galaxy. In this regard, a search for the characteristic anti-correlations commonly associated with globular clusters will serve to dispel this doubt.     
      %Our attempt to increase the number of stars in this substructure and assume a broader MDF holds merit, even if the chemical signatures do not permit a definitive association with certainty.
      %Despite its narrow MDF, we concur with \citet{balbinot2023} suggestion that ED-2 represents the remnant of a dwarf galaxy. 
    \end{enumerate}
    
    In light of the aforementioned complexities in identifying and interpreting MW halo substructures based on orbital information alone, the significance of our results lies in the fact that they are based on a pre-established classification, adding clarity amidst the ongoing confusion surrounding the newly identified substructures. While our study represents a crucial first step towards unraveling their true nature, it is crucial to acknowledge the need for more extensive investigations. Indeed, detailed chemical abundance measurements for a broader range of stars in the Galactic halo remain lacking. Future investigations by WRS will target RH stars that have already been associated to different disrupted progenitors, in order to disentangle ambiguous associations and shed lights on the chemical features of the retrograde substructures. As we continue to expand our knowledge, the pursuit of larger datasets of high-resolution spectroscopy, alongside future large spectroscopic surveys of MW halo stars and dwarf galaxies, e.g. WEAVE \citep{dalton2020} and 4MOST \citep{deJong2019}, will undoubtedly propel us towards a more comprehensive understanding of the Galactic halo's diverse substructures, thus better constraining the evolutionary history of the Milky Way.

\begin{acknowledgements}

    Based on data acquired using the Large Binocular Telescope (LBT) through the programme IT-2021B-006 (P.I. M. Bellazzini) and the strategic programme IT-2022B-004 (P.I. M. Bellazzini). The LBT is an international collaboration among institutions in the United States, Italy, and Germany. LBT Corporation partners are The University of Arizona on behalf of the Arizona university system; Istituto Nazionale di Astrofisica, Italy; LBT Beteiligungsgesellschaft, Germany, representing the Max-Planck Society, the Astrophysical Institute Potsdam, and Heidelberg University; The Ohio State University; and The Research Corporation, on behalf of The University of Notre Dame, University of Minnesota, and University of Virginia.

    Based on observations collected at the ESO-VLT under the programme 0109.B-0522 (P.I. A. Mucciarelli).

    MB, EC, DM and AM acknowledge the support to this study by the PRIN INAF 2023 grant ObFu \textit{CHAM - Chemodynamics of the Accreted Halo of the Milky Way} (P.I. M. Bellazzini).
    
    This research is funded by the project \textit{LEGO – Reconstructing the building blocks of the Galaxy by chemical tagging} (P.I. A. Mucciarelli), granted by the Italian MUR through contract PRIN 2022LLP8TK\_001.

    MB and DM acknowledge the support to activities related to the ESA/Gaia mission by the Italian Space Agency (ASI) through contract 2018-24-HH.0 and its addendum 2018-24-HH.1-2022 to the National Institute for Astrophysics (INAF). 

    This work is a part of a project that has received funding from the European Research Council (ERC) under the European Union’s Horizon 2020 research and innovation programme (CartographY; grant agreement ID 804752).

    AS acknowledge the support of the Science and Technology Facilities Council.

    AS acknowledges support from the European Research Council Consolidator Grant funding scheme (project ASTEROCHRONOMETRY, G.A. n. 772293, \url{http://www.asterochronometry.eu}). Funding for the Stellar Astrophysics Centre is provided by The Danish National Research Foundation (Grant agreement No.~DNRF106).

    EC and AM are grateful to R.Lallement for her help in deriving the color excess for the target stars.

    We thank the anonymous referee for constructive comments that helped us to improve the quality of the manuscript.

    This work has made use of data from the European Space Agency (ESA) mission Gaia \url{https://www.cosmos.esa.int/gaia}), processed by the Gaia Data Processing and Analysis Consortium (DPAC, \url{https://www.cosmos.esa.int/web/gaia/dpac/consortium}). Funding for the DPAC has been provided by national institutions, in particular the institutions participating in the Gaia Multilateral Agreement. 
    
    This work made use of SDSS-IV data. Funding for the Sloan Digital Sky Survey IV has been provided by the Alfred P. Sloan Foundation, the U.S. Department of Energy Office of Science, and the Participating Institutions. SDSS-IV acknowledges support and resources from the Center for High Performance Computing  at the University of Utah. The SDSS website is \url{www.sdss4.org}. SDSS-IV is managed by the Astrophysical Research Consortium for the Participating Institutions of the SDSS Collaboration including the Brazilian Participation Group, the Carnegie Institution for Science, Carnegie Mellon University, Center for Astrophysics | Harvard \& Smithsonian, the Chilean Participation Group, the French Participation Group, Instituto de Astrof\'isica de Canarias, The Johns Hopkins University, Kavli Institute for the Physics and Mathematics of the Universe (IPMU) / University of Tokyo, the Korean Participation Group, Lawrence Berkeley National Laboratory, Leibniz Institut f\"ur Astrophysik Potsdam (AIP),  Max-Planck-Institut f\"ur Astronomie (MPIA Heidelberg), Max-Planck-Institut f\"ur Astrophysik (MPA Garching), Max-Planck-Institut f\"ur Extraterrestrische Physik (MPE), National Astronomical Observatories of China, New Mexico State University, New York University, University of Notre Dame, Observat\'ario Nacional / MCTI, The Ohio State University, Pennsylvania State University, Shanghai Astronomical Observatory, United Kingdom Participation Group, Universidad Nacional Aut\'onoma de M\'exico, University of Arizona, University of Colorado Boulder, University of Oxford, University of Portsmouth, University of Utah, University of Virginia, University of Washington, University of Wisconsin, Vanderbilt University, and Yale University.

\end{acknowledgements}

\bibliographystyle{aa}
\bibliography{aanda_retrograde.bib}

\begin{thebibliography}{140}
\expandafter\ifx\csname natexlab\endcsname\relax\def\natexlab#1{#1}\fi

\bibitem[{{Abdurro'uf} {et~al.}(2022){Abdurro'uf}, {Accetta}, {Aerts}, {Silva Aguirre}, {Ahumada}, {Ajgaonkar}, {Filiz Ak}, {Alam}, {Allende Prieto}, {Almeida}, {Anders}, {Anderson}, {Andrews}, {Anguiano}, {Aquino-Ort{\'\i}z}, {Arag{\'o}n-Salamanca}, {Argudo-Fern{\'a}ndez}, {Ata}, {Aubert}, {Avila-Reese}, {Badenes}, {Barb{\'a}}, {Barger}, {Barrera-Ballesteros}, {Beaton}, {Beers}, {Belfiore}, {Bender}, {Bernardi}, {Bershady}, {Beutler}, {Bidin}, {Bird}, {Bizyaev}, {Blanc}, {Blanton}, {Boardman}, {Bolton}, {Boquien}, {Borissova}, {Bovy}, {Brandt}, {Brown}, {Brownstein}, {Brusa}, {Buchner}, {Bundy}, {Burchett}, {Bureau}, {Burgasser}, {Cabang}, {Campbell}, {Cappellari}, {Carlberg}, {Wanderley}, {Carrera}, {Cash}, {Chen}, {Chen}, {Cherinka}, {Chiappini}, {Choi}, {Chojnowski}, {Chung}, {Clerc}, {Cohen}, {Comerford}, {Comparat}, {da Costa}, {Covey}, {Crane}, {Cruz-Gonzalez}, {Culhane}, {Cunha}, {Dai}, {Damke}, {Darling}, {Davidson}, {Davies}, {Dawson}, {De Lee}, {Diamond-Stanic}, {Cano-D{\'\i}az}, {S{\'a}nchez},
  {Donor}, {Duckworth}, {Dwelly}, {Eisenstein}, {Elsworth}, {Emsellem}, {Eracleous}, {Escoffier}, {Fan}, {Farr}, {Feng}, {Fern{\'a}ndez-Trincado}, {Feuillet}, {Filipp}, {Fillingham}, {Frinchaboy}, {Fromenteau}, {Galbany}, {Garc{\'\i}a}, {Garc{\'\i}a-Hern{\'a}ndez}, {Ge}, {Geisler}, {Gelfand}, {G{\'e}ron}, {Gibson}, {Goddy}, {Godoy-Rivera}, {Grabowski}, {Green}, {Greener}, {Grier}, {Griffith}, {Guo}, {Guy}, {Hadjara}, {Harding}, {Hasselquist}, {Hayes}, {Hearty}, {Hern{\'a}ndez}, {Hill}, {Hogg}, {Holtzman}, {Horta}, {Hsieh}, {Hsu}, {Hsu}, {Huber}, {Huertas-Company}, {Hutchinson}, {Hwang}, {Ibarra-Medel}, {Chitham}, {Ilha}, {Imig}, {Jaekle}, {Jayasinghe}, {Ji}, {Johnson}, {Jones}, {J{\"o}nsson}, {Katkov}, {Khalatyan}, {Kinemuchi}, {Kisku}, {Knapen}, {Kneib}, {Kollmeier}, {Kong}, {Kounkel}, {Kreckel}, {Krishnarao}, {Lacerna}, {Lane}, {Langgin}, {Lavender}, {Law}, {Lazarz}, {Leung}, {Leung}, {Lewis}, {Li}, {Li}, {Lian}, {Liang}, {Lin}, {Lin}, {Lin}, {Lintott}, {Long}, {Longa-Pe{\~n}a}, {L{\'o}pez-Cob{\'a}}, {Lu},
  {Lundgren}, {Luo}, {Mackereth}, {de la Macorra}, {Mahadevan}, {Majewski}, {Manchado}, {Mandeville}, {Maraston}, {Margalef-Bentabol}, {Masseron}, {Masters}, {Mathur}, {McDermid}, {Mckay}, {Merloni}, {Merrifield}, {Meszaros}, {Miglio}, {Di Mille}, {Minniti}, {Minsley}, {Monachesi}, {Moon}, {Mosser}, {Mulchaey}, {Muna}, {Mu{\~n}oz}, {Myers}, {Myers}, {Nadathur}, {Nair}, {Nandra}, {Neumann}, {Newman}, {Nidever}, {Nikakhtar}, {Nitschelm}, {O'Connell}, {Garma-Oehmichen}, {Luan Souza de Oliveira}, {Olney}, {Oravetz}, {Ortigoza-Urdaneta}, {Osorio}, {Otter}, {Pace}, {Padilla}, {Pan}, {Pan}, {Parikh}, {Parker}, {Peirani}, {Pe{\~n}a Ram{\'\i}rez}, {Penny}, {Percival}, {Perez-Fournon}, {Pinsonneault}, {Poidevin}, {Poovelil}, {Price-Whelan}, {B{\'a}rbara de Andrade Queiroz}, {Raddick}, {Ray}, {Rembold}, {Riddle}, {Riffel}, {Riffel}, {Rix}, {Robin}, {Rodr{\'\i}guez-Puebla}, {Roman-Lopes}, {Rom{\'a}n-Z{\'u}{\~n}iga}, {Rose}, {Ross}, {Rossi}, {Rubin}, {Salvato}, {S{\'a}nchez}, {S{\'a}nchez-Gallego}, {Sanderson}, {Santana
  Rojas}, {Sarceno}, {Sarmiento}, {Sayres}, {Sazonova}, {Schaefer}, {Schiavon}, {Schlegel}, {Schneider}, {Schultheis}, {Schwope}, {Serenelli}, {Serna}, {Shao}, {Shapiro}, {Sharma}, {Shen}, {Shetrone}, {Shu}, {Simon}, {Skrutskie}, {Smethurst}, {Smith}, {Sobeck}, {Spoo}, {Sprague}, {Stark}, {Stassun}, {Steinmetz}, {Stello}, {Stone-Martinez}, {Storchi-Bergmann}, {Stringfellow}, {Stutz}, {Su}, {Taghizadeh-Popp}, {Talbot}, {Tayar}, {Telles}, {Teske}, {Thakar}, {Theissen}, {Tkachenko}, {Thomas}, {Tojeiro}, {Hernandez Toledo}, {Troup}, {Trump}, {Trussler}, {Turner}, {Tuttle}, {Unda-Sanzana}, {V{\'a}zquez-Mata}, {Valentini}, {Valenzuela}, {Vargas-Gonz{\'a}lez}, {Vargas-Maga{\~n}a}, {Alfaro}, {Villanova}, {Vincenzo}, {Wake}, {Warfield}, {Washington}, {Weaver}, {Weijmans}, {Weinberg}, {Weiss}, {Westfall}, {Wild}, {Wilde}, {Wilson}, {Wilson}, {Wilson}, {Wolf}, {Wood-Vasey}, {Yan}, {Zamora}, {Zasowski}, {Zhang}, {Zhao}, {Zheng}, {Zheng}, \& {Zhu}}]{apogee22}
{Abdurro'uf}, {Accetta}, K., {Aerts}, C., {et~al.} 2022, \apjs, 259, 35

\bibitem[{{Aguado} {et~al.}(2021){Aguado}, {Belokurov}, {Myeong}, {Evans}, {Kobayashi}, {Sbordone}, {Chanam{\'e}}, {Navarrete}, \& {Koposov}}]{aguado21}
{Aguado}, D.~S., {Belokurov}, V., {Myeong}, G.~C., {et~al.} 2021, \apjl, 908, L8

\bibitem[{{Amarante} {et~al.}(2022){Amarante}, {Debattista}, {Beraldo e Silva}, {Laporte}, \& {Deg}}]{amarante2022}
{Amarante}, J. A.~S., {Debattista}, V.~P., {Beraldo e Silva}, L., {Laporte}, C. F.~P., \& {Deg}, N. 2022, \apj, 937, 12

\bibitem[{{Andrae} {et~al.}(2018){Andrae}, {Fouesneau}, {Creevey}, {Ordenovic}, {Mary}, {Burlacu}, {Chaoul}, {Jean-Antoine-Piccolo}, {Kordopatis}, {Korn}, {Lebreton}, {Panem}, {Pichon}, {Th{\'e}venin}, {Walmsley}, \& {Bailer-Jones}}]{andrae18}
{Andrae}, R., {Fouesneau}, M., {Creevey}, O., {et~al.} 2018, \aap, 616, A8

\bibitem[{{Andrew} {et~al.}(2022){Andrew}, {Penoyre}, {Belokurov}, {Evans}, \& {Oh}}]{shion22}
{Andrew}, S., {Penoyre}, Z., {Belokurov}, V., {Evans}, N.~W., \& {Oh}, S. 2022, \mnras, 516, 3661

\bibitem[{{Arnett}(1996)}]{arnett1996}
{Arnett}, D. 1996, {Supernovae and Nucleosynthesis: An Investigation of the History of Matter from the Big Bang to the Present}

\bibitem[{{Balbinot} {et~al.}(2023){Balbinot}, {Helmi}, {Callingham}, {Matsuno}, {Dodd}, \& {Ruiz-Lara}}]{balbinot2023}
{Balbinot}, E., {Helmi}, A., {Callingham}, T., {et~al.} 2023, arXiv e-prints, arXiv:2306.02756

\bibitem[{{Barklem} {et~al.}(2005){Barklem}, {Christlieb}, {Beers}, {Hill}, {Bessell}, {Holmberg}, {Marsteller}, {Rossi}, {Zickgraf}, \& {Reimers}}]{barklem05}
{Barklem}, P.~S., {Christlieb}, N., {Beers}, T.~C., {et~al.} 2005, \aap, 439, 129

\bibitem[{{Bellazzini} {et~al.}(2023){Bellazzini}, {Massari}, {De Angeli}, {Mucciarelli}, {Bragaglia}, {Riello}, \& {Montegriffo}}]{bellazz23}
{Bellazzini}, M., {Massari}, D., {De Angeli}, F., {et~al.} 2023, \aap, 674, A194

\bibitem[{{Belokurov} {et~al.}(2018){Belokurov}, {Erkal}, {Evans}, {Koposov}, \& {Deason}}]{belokurov2018}
{Belokurov}, V., {Erkal}, D., {Evans}, N.~W., {Koposov}, S.~E., \& {Deason}, A.~J. 2018, \mnras, 478, 611

\bibitem[{{Belokurov} \& {Kravtsov}(2022)}]{belokurov22}
{Belokurov}, V. \& {Kravtsov}, A. 2022, \mnras, 514, 689

\bibitem[{{Belokurov} {et~al.}(2020){Belokurov}, {Sanders}, {Fattahi}, {Smith}, {Deason}, {Evans}, \& {Grand}}]{belokurov2020}
{Belokurov}, V., {Sanders}, J.~L., {Fattahi}, A., {et~al.} 2020, \mnras, 494, 3880

\bibitem[{{Belokurov} {et~al.}(2023){Belokurov}, {Vasiliev}, {Deason}, {Koposov}, {Fattahi}, {Dillamore}, {Davies}, \& {Grand}}]{belokurov23}
{Belokurov}, V., {Vasiliev}, E., {Deason}, A.~J., {et~al.} 2023, \mnras, 518, 6200

\bibitem[{{Bennett} \& {Bovy}(2019)}]{bennetandbovy19}
{Bennett}, M. \& {Bovy}, J. 2019, \mnras, 482, 1417

\bibitem[{{Bensby} {et~al.}(2005){Bensby}, {Feltzing}, {Lundstr{\"o}m}, \& {Ilyin}}]{bensby05}
{Bensby}, T., {Feltzing}, S., {Lundstr{\"o}m}, I., \& {Ilyin}, I. 2005, \aap, 433, 185

\bibitem[{{Bensby} {et~al.}(2014){Bensby}, {Feltzing}, \& {Oey}}]{bensby14}
{Bensby}, T., {Feltzing}, S., \& {Oey}, M.~S. 2014, \aap, 562, A71

\bibitem[{{Bertaux} {et~al.}(2014){Bertaux}, {Lallement}, {Ferron}, {Boonne}, \& {Bodichon}}]{bertaux2014}
{Bertaux}, J.~L., {Lallement}, R., {Ferron}, S., {Boonne}, C., \& {Bodichon}, R. 2014, \aap, 564, A46

\bibitem[{{Bonaca} {et~al.}(2017){Bonaca}, {Conroy}, {Wetzel}, {Hopkins}, \& {Kere{\v{s}}}}]{bonaca17}
{Bonaca}, A., {Conroy}, C., {Wetzel}, A., {Hopkins}, P.~F., \& {Kere{\v{s}}}, D. 2017, \apj, 845, 101

\bibitem[{{Bonaca} {et~al.}(2012){Bonaca}, {Geha}, \& {Kallivayalil}}]{bonaca12}
{Bonaca}, A., {Geha}, M., \& {Kallivayalil}, N. 2012, \apjl, 760, L6

\bibitem[{{Bonaca} {et~al.}(2021){Bonaca}, {Naidu}, {Conroy}, {Caldwell}, {Cargile}, {Han}, {Johnson}, {Kruijssen}, {Myeong}, {Speagle}, {Ting}, \& {Zaritsky}}]{bonaca2021}
{Bonaca}, A., {Naidu}, R.~P., {Conroy}, C., {et~al.} 2021, \apjl, 909, L26

\bibitem[{{Bonifacio} {et~al.}(2021){Bonifacio}, {Monaco}, {Salvadori}, {Caffau}, {Spite}, {Sbordone}, {Spite}, {Ludwig}, {Di Matteo}, {Haywood}, {Fran{\c{c}}ois}, {Koch-Hansen}, {Christlieb}, \& {Zaggia}}]{bonifacio21}
{Bonifacio}, P., {Monaco}, L., {Salvadori}, S., {et~al.} 2021, \aap, 651, A79

\bibitem[{{Buder} {et~al.}(2022){Buder}, {Lind}, {Ness}, {Feuillet}, {Horta}, {Monty}, {Buck}, {Nordlander}, {Bland-Hawthorn}, {Casey}, {de Silva}, {D'Orazi}, {Freeman}, {Hayden}, {Kos}, {Martell}, {Lewis}, {Lin}, {Schlesinger}, {Sharma}, {Simpson}, {Stello}, {Zucker}, {Zwitter}, {Ciuc{\u{a}}}, {Horner}, {Kobayashi}, {Ting}, {Wyse}, \& {Wyse}}]{buder22}
{Buder}, S., {Lind}, K., {Ness}, M.~K., {et~al.} 2022, \mnras, 510, 2407

\bibitem[{{Buder} {et~al.}(2021){Buder}, {Sharma}, {Kos}, {Amarsi}, {Nordlander}, {Lind}, {Martell}, {Asplund}, {Bland-Hawthorn}, {Casey}, {de Silva}, {D'Orazi}, {Freeman}, {Hayden}, {Lewis}, {Lin}, {Schlesinger}, {Simpson}, {Stello}, {Zucker}, {Zwitter}, {Beeson}, {Buck}, {Casagrande}, {Clark}, {{\v{C}}otar}, {da Costa}, {de Grijs}, {Feuillet}, {Horner}, {Kafle}, {Khanna}, {Kobayashi}, {Liu}, {Montet}, {Nandakumar}, {Nataf}, {Ness}, {Spina}, {Tepper-Garc{\'\i}a}, {Ting}, {Traven}, {Vogrin{\v{c}}i{\v{c}}}, {Wittenmyer}, {Wyse}, {{\v{Z}}erjal}, \& {Galah Collaboration}}]{galah2021}
{Buder}, S., {Sharma}, S., {Kos}, J., {et~al.} 2021, \mnras, 506, 150

\bibitem[{{Busso} {et~al.}(1999){Busso}, {Gallino}, \& {Wasserburg}}]{busso1999}
{Busso}, M., {Gallino}, R., \& {Wasserburg}, G.~J. 1999, \araa, 37, 239

\bibitem[{{Castelli} \& {Kurucz}(2003)}]{castelli&kurucz2003}
{Castelli}, F. \& {Kurucz}, R.~L. 2003, in Modelling of Stellar Atmospheres, ed. N.~{Piskunov}, W.~W. {Weiss}, \& D.~F. {Gray}, Vol. 210, A20

\bibitem[{{Conroy} {et~al.}(2019){Conroy}, {Bonaca}, {Cargile}, {Johnson}, {Caldwell}, {Naidu}, {Zaritsky}, {Fabricant}, {Moran}, {Rhee}, {Szentgyorgyi}, {Berlind}, {Calkins}, {Kattner}, \& {Ly}}]{conroy19}
{Conroy}, C., {Bonaca}, A., {Cargile}, P., {et~al.} 2019, \apj, 883, 107

\bibitem[{{Cristallo} {et~al.}(2015){Cristallo}, {Straniero}, {Piersanti}, \& {Gobrecht}}]{cristallo2015}
{Cristallo}, S., {Straniero}, O., {Piersanti}, L., \& {Gobrecht}, D. 2015, \apjs, 219, 40

\bibitem[{{Dalton} {et~al.}(2020){Dalton}, {Trager}, {Abrams}, {Bonifacio}, {Aguerri}, {Vallenari}, {Bishop}, {Middleton}, {Benn}, {Dee}, {Mignot}, {Lewis}, {Pragt}, {Pico}, {Walton}, {Rey}, {Allende Prieto}, {Lhom{\'e}}, {Balcells}, {Terrett}, {Brock}, {Ridings}, {Skvar{\v{c}}}, {Verheijen}, {Steele}, {Stuik}, {Kroes}, {Tromp}, {Kragt}, {Lesman}, {Mottram}, {Bates}, {Gribbin}, {Burgal}, {Herreros}, {Delgado}, {Martin}, {Cano}, {Navarro}, {Irwin}, {Peralta de Arriba}, {O'Mahoney}, {Bianco}, {Moleinezhad}, {ter Horst}, {Molinari}, {Lodi}, {Guerra}, {Baruffalo}, {Carrasco}, {Farcas}, {Schallig}, {Hughes}, {Hill}, {Smith}, {Drew}, {Poggianti}, {Iovino}, {Pieri}, {Jin}, {Dominguez Palmero}, {Fari{\~n}a}, {Mart{\'\i}n}, {Worley}, {Murphy}, {Guest}, {Morris}, {Elswijk}, {de Haan}, {Hanenburg}, {Salasnich}, {Mayya}, {Izazaga-P{\'e}rez}, {Gafton}, {Caffau}, {Horville}, {Paz Chinch{\'o}n}, {Falcon-Barosso}, {G{\"a}nsicke}, {San Juan}, \& {Hernandez}}]{dalton2020}
{Dalton}, G., {Trager}, S., {Abrams}, D.~C., {et~al.} 2020, in Society of Photo-Optical Instrumentation Engineers (SPIE) Conference Series, Vol. 11447, Society of Photo-Optical Instrumentation Engineers (SPIE) Conference Series, 1144714

\bibitem[{{Das} {et~al.}(2020){Das}, {Hawkins}, \& {Jofr{\'e}}}]{das2020}
{Das}, P., {Hawkins}, K., \& {Jofr{\'e}}, P. 2020, \mnras, 493, 5195

\bibitem[{{Davies} {et~al.}(2023){Davies}, {Vasiliev}, {Belokurov}, {Evans}, \& {Dillamore}}]{davies2023}
{Davies}, E.~Y., {Vasiliev}, E., {Belokurov}, V., {Evans}, N.~W., \& {Dillamore}, A.~M. 2023, \mnras, 519, 530

\bibitem[{{de Jong} {et~al.}(2019){de Jong}, {Agertz}, {Berbel}, {Aird}, {Alexander}, {Amarsi}, {Anders}, {Andrae}, {Ansarinejad}, {Ansorge}, {Antilogus}, {Anwand-Heerwart}, {Arentsen}, {Arnadottir}, {Asplund}, {Auger}, {Azais}, {Baade}, {Baker}, {Baker}, {Balbinot}, {Baldry}, {Banerji}, {Barden}, {Barklem}, {Barth{\'e}l{\'e}my-Mazot}, {Battistini}, {Bauer}, {Bell}, {Bellido-Tirado}, {Bellstedt}, {Belokurov}, {Bensby}, {Bergemann}, {Bestenlehner}, {Bielby}, {Bilicki}, {Blake}, {Bland-Hawthorn}, {Boeche}, {Boland}, {Boller}, {Bongard}, {Bongiorno}, {Bonifacio}, {Boudon}, {Brooks}, {Brown}, {Brown}, {Br{\"u}ggen}, {Brynnel}, {Brzeski}, {Buchert}, {Buschkamp}, {Caffau}, {Caillier}, {Carrick}, {Casagrande}, {Case}, {Casey}, {Cesarini}, {Cescutti}, {Chapuis}, {Chiappini}, {Childress}, {Christlieb}, {Church}, {Cioni}, {Cluver}, {Colless}, {Collett}, {Comparat}, {Cooper}, {Couch}, {Courbin}, {Croom}, {Croton}, {Daguis{\'e}}, {Dalton}, {Davies}, {Davis}, {de Laverny}, {Deason}, {Dionies}, {Disseau}, {Doel},
  {D{\"o}scher}, {Driver}, {Dwelly}, {Eckert}, {Edge}, {Edvardsson}, {Youssoufi}, {Elhaddad}, {Enke}, {Erfanianfar}, {Farrell}, {Fechner}, {Feiz}, {Feltzing}, {Ferreras}, {Feuerstein}, {Feuillet}, {Finoguenov}, {Ford}, {Fotopoulou}, {Fouesneau}, {Frenk}, {Frey}, {Gaessler}, {Geier}, {Gentile Fusillo}, {Gerhard}, {Giannantonio}, {Giannone}, {Gibson}, {Gillingham}, {Gonz{\'a}lez-Fern{\'a}ndez}, {Gonzalez-Solares}, {Gottloeber}, {Gould}, {Grebel}, {Gueguen}, {Guiglion}, {Haehnelt}, {Hahn}, {Hansen}, {Hartman}, {Hauptner}, {Hawkins}, {Haynes}, {Haynes}, {Heiter}, {Helmi}, {Aguayo}, {Hewett}, {Hinton}, {Hobbs}, {Hoenig}, {Hofman}, {Hook}, {Hopgood}, {Hopkins}, {Hourihane}, {Howes}, {Howlett}, {Huet}, {Irwin}, {Iwert}, {Jablonka}, {Jahn}, {Jahnke}, {Jarno}, {Jin}, {Jofre}, {Johl}, {Jones}, {J{\"o}nsson}, {Jordan}, {Karovicova}, {Khalatyan}, {Kelz}, {Kennicutt}, {King}, {Kitaura}, {Klar}, {Klauser}, {Kneib}, {Koch}, {Koposov}, {Kordopatis}, {Korn}, {Kosmalski}, {Kotak}, {Kovalev}, {Kreckel}, {Kripak}, {Krumpe},
  {Kuijken}, {Kunder}, {Kushniruk}, {Lam}, {Lamer}, {Laurent}, {Lawrence}, {Lehmitz}, {Lemasle}, {Lewis}, {Li}, {Lidman}, {Lind}, {Liske}, {Lizon}, {Loveday}, {Ludwig}, {McDermid}, {Maguire}, {Mainieri}, {Mali}, {Mandel}, {Mandel}, {Mannering}, {Martell}, {Martinez Delgado}, {Matijevic}, {McGregor}, {McMahon}, {McMillan}, {Mena}, {Merloni}, {Meyer}, {Michel}, {Micheva}, {Migniau}, {Minchev}, {Monari}, {Muller}, {Murphy}, {Muthukrishna}, {Nandra}, {Navarro}, {Ness}, {Nichani}, {Nichol}, {Nicklas}, {Niederhofer}, {Norberg}, {Obreschkow}, {Oliver}, {Owers}, {Pai}, {Pankratow}, {Parkinson}, {Paschke}, {Paterson}, {Pecontal}, {Parry}, {Phillips}, {Pillepich}, {Pinard}, {Pirard}, {Piskunov}, {Plank}, {Pl{\"u}schke}, {Pons}, {Popesso}, {Power}, {Pragt}, {Pramskiy}, {Pryer}, {Quattri}, {Queiroz}, {Quirrenbach}, {Rahurkar}, {Raichoor}, {Ramstedt}, {Rau}, {Recio-Blanco}, {Reiss}, {Renaud}, {Revaz}, {Rhode}, {Richard}, {Richter}, {Rix}, {Robotham}, {Roelfsema}, {Romaniello}, {Rosario}, {Rothmaier}, {Roukema}, {Ruchti},
  {Rupprecht}, {Rybizki}, {Ryde}, {Saar}, {Sadler}, {Sahl{\'e}n}, {Salvato}, {Sassolas}, {Saunders}, {Saviauk}, {Sbordone}, {Schmidt}, {Schnurr}, {Scholz}, {Schwope}, {Seifert}, {Shanks}, {Sheinis}, {Sivov}, {Sk{\'u}lad{\'o}ttir}, {Smartt}, {Smedley}, {Smith}, {Smith}, {Sorce}, {Spitler}, {Starkenburg}, {Steinmetz}, {Stilz}, {Storm}, {Sullivan}, {Sutherland}, {Swann}, {Tamone}, {Taylor}, {Teillon}, {Tempel}, {ter Horst}, {Thi}, {Tolstoy}, {Trager}, {Traven}, {Tremblay}, {Tresse}, {Valentini}, {van de Weygaert}, {van den Ancker}, {Veljanoski}, {Venkatesan}, {Wagner}, {Wagner}, {Walcher}, {Waller}, {Walton}, {Wang}, {Winkler}, {Wisotzki}, {Worley}, {Worseck}, {Xiang}, {Xu}, {Yong}, {Zhao}, {Zheng}, {Zscheyge}, \& {Zucker}}]{deJong2019}
{de Jong}, R.~S., {Agertz}, O., {Berbel}, A.~A., {et~al.} 2019, The Messenger, 175, 3

\bibitem[{{Dekker} {et~al.}(2000){Dekker}, {D'Odorico}, {Kaufer}, {Delabre}, \& {Kotzlowski}}]{dekker2000}
{Dekker}, H., {D'Odorico}, S., {Kaufer}, A., {Delabre}, B., \& {Kotzlowski}, H. 2000, in Society of Photo-Optical Instrumentation Engineers (SPIE) Conference Series, Vol. 4008, Optical and IR Telescope Instrumentation and Detectors, ed. M.~{Iye} \& A.~F. {Moorwood}, 534--545

\bibitem[{{Dodd} {et~al.}(2023){Dodd}, {Callingham}, {Helmi}, {Matsuno}, {Ruiz-Lara}, {Balbinot}, \& {L{\"o}vdal}}]{dodd23}
{Dodd}, E., {Callingham}, T.~M., {Helmi}, A., {et~al.} 2023, \aap, 670, L2

\bibitem[{{Drimmel} \& {Poggio}(2018)}]{drimmelandpoggio18}
{Drimmel}, R. \& {Poggio}, E. 2018, Research Notes of the American Astronomical Society, 2, 210

\bibitem[{{Edvardsson} {et~al.}(1993){Edvardsson}, {Andersen}, {Gustafsson}, {Lambert}, {Nissen}, \& {Tomkin}}]{edvardsson93}
{Edvardsson}, B., {Andersen}, J., {Gustafsson}, B., {et~al.} 1993, \aaps, 102, 603

\bibitem[{{Ester} {et~al.}(1996){Ester}, {Kriegel}, {Sander}, \& {Xu}}]{ester1996}
{Ester}, M., {Kriegel}, H.-P., {Sander}, J., \& {Xu}, X. 1996, in Second International Conference on Knowledge Discovery and Data Mining (KDD'96). Proceedings of a conference held August 2-4, 226--331

\bibitem[{{Fern{\'a}ndez-Alvar} {et~al.}(2018){Fern{\'a}ndez-Alvar}, {Carigi}, {Schuster}, {Hayes}, {{\'A}vila-Vergara}, {Majewski}, {Allende Prieto}, {Beers}, {S{\'a}nchez}, {Zamora}, {Garc{\'\i}a-Hern{\'a}ndez}, {Tang}, {Fern{\'a}ndez-Trincado}, {Tissera}, {Geisler}, \& {Villanova}}]{fernandezalvar2018}
{Fern{\'a}ndez-Alvar}, E., {Carigi}, L., {Schuster}, W.~J., {et~al.} 2018, \apj, 852, 50

\bibitem[{{Feuillet} {et~al.}(2020){Feuillet}, {Feltzing}, {Sahlholdt}, \& {Casagrande}}]{feuillet2020}
{Feuillet}, D.~K., {Feltzing}, S., {Sahlholdt}, C.~L., \& {Casagrande}, L. 2020, \mnras, 497, 109

\bibitem[{{Feuillet} {et~al.}(2021){Feuillet}, {Sahlholdt}, {Feltzing}, \& {Casagrande}}]{feuillet21}
{Feuillet}, D.~K., {Sahlholdt}, C.~L., {Feltzing}, S., \& {Casagrande}, L. 2021, \mnras, 508, 1489

\bibitem[{{Font} {et~al.}(2020){Font}, {McCarthy}, {Poole-Mckenzie}, {Stafford}, {Brown}, {Schaye}, {Crain}, {Theuns}, \& {Schaller}}]{font2020}
{Font}, A.~S., {McCarthy}, I.~G., {Poole-Mckenzie}, R., {et~al.} 2020, \mnras, 498, 1765

\bibitem[{{Freeman} \& {Bland-Hawthorn}(2002)}]{freeman02}
{Freeman}, K. \& {Bland-Hawthorn}, J. 2002, \araa, 40, 487

\bibitem[{{Fulbright}(2000)}]{fulbright2000}
{Fulbright}, J.~P. 2000, \aj, 120, 1841

\bibitem[{{Gaia Collaboration} {et~al.}(2018{\natexlab{a}}){Gaia Collaboration}, {Babusiaux}, {van Leeuwen}, {Barstow}, {Jordi}, {Vallenari}, {Bossini}, {Bressan}, {Cantat-Gaudin}, {van Leeuwen}, {Brown}, {Prusti}, {de Bruijne}, {Bailer-Jones}, {Biermann}, {Evans}, {Eyer}, {Jansen}, {Klioner}, {Lammers}, {Lindegren}, {Luri}, {Mignard}, {Panem}, {Pourbaix}, {Randich}, {Sartoretti}, {Siddiqui}, {Soubiran}, {Walton}, {Arenou}, {Bastian}, {Cropper}, {Drimmel}, {Katz}, {Lattanzi}, {Bakker}, {Cacciari}, {Casta{\~n}eda}, {Chaoul}, {Cheek}, {De Angeli}, {Fabricius}, {Guerra}, {Holl}, {Masana}, {Messineo}, {Mowlavi}, {Nienartowicz}, {Panuzzo}, {Portell}, {Riello}, {Seabroke}, {Tanga}, {Th{\'e}venin}, {Gracia-Abril}, {Comoretto}, {Garcia-Reinaldos}, {Teyssier}, {Altmann}, {Andrae}, {Audard}, {Bellas-Velidis}, {Benson}, {Berthier}, {Blomme}, {Burgess}, {Busso}, {Carry}, {Cellino}, {Clementini}, {Clotet}, {Creevey}, {Davidson}, {De Ridder}, {Delchambre}, {Dell'Oro}, {Ducourant}, {Fern{\'a}ndez-Hern{\'a}ndez},
  {Fouesneau}, {Fr{\'e}mat}, {Galluccio}, {Garc{\'\i}a-Torres}, {Gonz{\'a}lez-N{\'u}{\~n}ez}, {Gonz{\'a}lez-Vidal}, {Gosset}, {Guy}, {Halbwachs}, {Hambly}, {Harrison}, {Hern{\'a}ndez}, {Hestroffer}, {Hodgkin}, {Hutton}, {Jasniewicz}, {Jean-Antoine-Piccolo}, {Jordan}, {Korn}, {Krone-Martins}, {Lanzafame}, {Lebzelter}, {L{\"o}ffler}, {Manteiga}, {Marrese}, {Mart{\'\i}n-Fleitas}, {Moitinho}, {Mora}, {Muinonen}, {Osinde}, {Pancino}, {Pauwels}, {Petit}, {Recio-Blanco}, {Richards}, {Rimoldini}, {Robin}, {Sarro}, {Siopis}, {Smith}, {Sozzetti}, {S{\"u}veges}, {Torra}, {van Reeven}, {Abbas}, {Abreu Aramburu}, {Accart}, {Aerts}, {Altavilla}, {{\'A}lvarez}, {Alvarez}, {Alves}, {Anderson}, {Andrei}, {Anglada Varela}, {Antiche}, {Antoja}, {Arcay}, {Astraatmadja}, {Bach}, {Baker}, {Balaguer-N{\'u}{\~n}ez}, {Balm}, {Barache}, {Barata}, {Barbato}, {Barblan}, {Barklem}, {Barrado}, {Barros}, {Bartholom{\'e} Mu{\~n}oz}, {Bassilana}, {Becciani}, {Bellazzini}, {Berihuete}, {Bertone}, {Bianchi}, {Bienaym{\'e}}, {Blanco-Cuaresma},
  {Boch}, {Boeche}, {Bombrun}, {Borrachero}, {Bouquillon}, {Bourda}, {Bragaglia}, {Bramante}, {Breddels}, {Brouillet}, {Br{\"u}semeister}, {Brugaletta}, {Bucciarelli}, {Burlacu}, {Busonero}, {Butkevich}, {Buzzi}, {Caffau}, {Cancelliere}, {Cannizzaro}, {Carballo}, {Carlucci}, {Carrasco}, {Casamiquela}, {Castellani}, {Castro-Ginard}, {Charlot}, {Chemin}, {Chiavassa}, {Cocozza}, {Costigan}, {Cowell}, {Crifo}, {Crosta}, {Crowley}, {Cuypers}, {Dafonte}, {Damerdji}, {Dapergolas}, {David}, {David}, {de Laverny}, {De Luise}, {De March}, {de Martino}, {de Souza}, {de Torres}, {Debosscher}, {del Pozo}, {Delbo}, {Delgado}, {Delgado}, {Diakite}, {Diener}, {Distefano}, {Dolding}, {Drazinos}, {Dur{\'a}n}, {Edvardsson}, {Enke}, {Eriksson}, {Esquej}, {Eynard Bontemps}, {Fabre}, {Fabrizio}, {Faigler}, {Falc{\~a}o}, {Farr{\`a}s Casas}, {Federici}, {Fedorets}, {Fernique}, {Figueras}, {Filippi}, {Findeisen}, {Fonti}, {Fraile}, {Fraser}, {Fr{\'e}zouls}, {Gai}, {Galleti}, {Garabato}, {Garc{\'\i}a-Sedano}, {Garofalo}, {Garralda},
  {Gavel}, {Gavras}, {Gerssen}, {Geyer}, {Giacobbe}, {Gilmore}, {Girona}, {Giuffrida}, {Glass}, {Gomes}, {Granvik}, {Gueguen}, {Guerrier}, {Guiraud}, {Guti{\'e}}, {Haigron}, {Hatzidimitriou}, {Hauser}, {Haywood}, {Heiter}, {Helmi}, {Heu}, {Hilger}, {Hobbs}, {Hofmann}, {Holland}, {Huckle}, {Hypki}, {Icardi}, {Jan{\ss}en}, {Jevardat de Fombelle}, {Jonker}, {Juh{\'a}sz}, {Julbe}, {Karampelas}, {Kewley}, {Klar}, {Kochoska}, {Kohley}, {Kolenberg}, {Kontizas}, {Kontizas}, {Koposov}, {Kordopatis}, {Kostrzewa-Rutkowska}, {Koubsky}, {Lambert}, {Lanza}, {Lasne}, {Lavigne}, {Le Fustec}, {Le Poncin-Lafitte}, {Lebreton}, {Leccia}, {Leclerc}, {Lecoeur-Taibi}, {Lenhardt}, {Leroux}, {Liao}, {Licata}, {Lindstr{\o}m}, {Lister}, {Livanou}, {Lobel}, {L{\'o}pez}, {Managau}, {Mann}, {Mantelet}, {Marchal}, {Marchant}, {Marconi}, {Marinoni}, {Marschalk{\'o}}, {Marshall}, {Martino}, {Marton}, {Mary}, {Massari}, {Matijevi{\v{c}}}, {Mazeh}, {McMillan}, {Messina}, {Michalik}, {Millar}, {Molina}, {Molinaro}, {Moln{\'a}r}, {Montegriffo},
  {Mor}, {Morbidelli}, {Morel}, {Morris}, {Mulone}, {Muraveva}, {Musella}, {Nelemans}, {Nicastro}, {Noval}, {O'Mullane}, {Ord{\'e}novic}, {Ord{\'o}{\~n}ez-Blanco}, {Osborne}, {Pagani}, {Pagano}, {Pailler}, {Palacin}, {Palaversa}, {Panahi}, {Pawlak}, {Piersimoni}, {Pineau}, {Plachy}, {Plum}, {Poggio}, {Poujoulet}, {Pr{\v{s}}a}, {Pulone}, {Racero}, {Ragaini}, {Rambaux}, {Ramos-Lerate}, {Regibo}, {Reyl{\'e}}, {Riclet}, {Ripepi}, {Riva}, {Rivard}, {Rixon}, {Roegiers}, {Roelens}, {Romero-G{\'o}mez}, {Rowell}, {Royer}, {Ruiz-Dern}, {Sadowski}, {Sagrist{\`a} Sell{\'e}s}, {Sahlmann}, {Salgado}, {Salguero}, {Sanna}, {Santana-Ros}, {Sarasso}, {Savietto}, {Schultheis}, {Sciacca}, {Segol}, {Segovia}, {S{\'e}gransan}, {Shih}, {Siltala}, {Silva}, {Smart}, {Smith}, {Solano}, {Solitro}, {Sordo}, {Soria Nieto}, {Souchay}, {Spagna}, {Spoto}, {Stampa}, {Steele}, {Steidelm{\"u}ller}, {Stephenson}, {Stoev}, {Suess}, {Surdej}, {Szabados}, {Szegedi-Elek}, {Tapiador}, {Taris}, {Tauran}, {Taylor}, {Teixeira}, {Terrett},
  {Teyssandier}, {Thuillot}, {Titarenko}, {Torra Clotet}, {Turon}, {Ulla}, {Utrilla}, {Uzzi}, {Vaillant}, {Valentini}, {Valette}, {van Elteren}, {Van Hemelryck}, {Vaschetto}, {Vecchiato}, {Veljanoski}, {Viala}, {Vicente}, {Vogt}, {von Essen}, {Voss}, {Votruba}, {Voutsinas}, {Walmsley}, {Weiler}, {Wertz}, {Wevers}, {Wyrzykowski}, {Yoldas}, {{\v{Z}}erjal}, {Ziaeepour}, {Zorec}, {Zschocke}, {Zucker}, {Zurbach}, \& {Zwitter}}]{GC18_extinction}
{Gaia Collaboration}, {Babusiaux}, C., {van Leeuwen}, F., {et~al.} 2018{\natexlab{a}}, \aap, 616, A10

\bibitem[{{Gaia Collaboration} {et~al.}(2018{\natexlab{b}}){Gaia Collaboration}, {Brown}, {Vallenari}, {Prusti}, {de Bruijne}, {Babusiaux}, {Bailer-Jones}, {Biermann}, {Evans}, {Eyer}, {Jansen}, {Jordi}, {Klioner}, {Lammers}, {Lindegren}, {Luri}, {Mignard}, {Panem}, {Pourbaix}, {Randich}, {Sartoretti}, {Siddiqui}, {Soubiran}, {van Leeuwen}, {Walton}, {Arenou}, {Bastian}, {Cropper}, {Drimmel}, {Katz}, {Lattanzi}, {Bakker}, {Cacciari}, {Casta{\~n}eda}, {Chaoul}, {Cheek}, {De Angeli}, {Fabricius}, {Guerra}, {Holl}, {Masana}, {Messineo}, {Mowlavi}, {Nienartowicz}, {Panuzzo}, {Portell}, {Riello}, {Seabroke}, {Tanga}, {Th{\'e}venin}, {Gracia-Abril}, {Comoretto}, {Garcia-Reinaldos}, {Teyssier}, {Altmann}, {Andrae}, {Audard}, {Bellas-Velidis}, {Benson}, {Berthier}, {Blomme}, {Burgess}, {Busso}, {Carry}, {Cellino}, {Clementini}, {Clotet}, {Creevey}, {Davidson}, {De Ridder}, {Delchambre}, {Dell'Oro}, {Ducourant}, {Fern{\'a}ndez-Hern{\'a}ndez}, {Fouesneau}, {Fr{\'e}mat}, {Galluccio}, {Garc{\'\i}a-Torres},
  {Gonz{\'a}lez-N{\'u}{\~n}ez}, {Gonz{\'a}lez-Vidal}, {Gosset}, {Guy}, {Halbwachs}, {Hambly}, {Harrison}, {Hern{\'a}ndez}, {Hestroffer}, {Hodgkin}, {Hutton}, {Jasniewicz}, {Jean-Antoine-Piccolo}, {Jordan}, {Korn}, {Krone-Martins}, {Lanzafame}, {Lebzelter}, {L{\"o}ffler}, {Manteiga}, {Marrese}, {Mart{\'\i}n-Fleitas}, {Moitinho}, {Mora}, {Muinonen}, {Osinde}, {Pancino}, {Pauwels}, {Petit}, {Recio-Blanco}, {Richards}, {Rimoldini}, {Robin}, {Sarro}, {Siopis}, {Smith}, {Sozzetti}, {S{\"u}veges}, {Torra}, {van Reeven}, {Abbas}, {Abreu Aramburu}, {Accart}, {Aerts}, {Altavilla}, {{\'A}lvarez}, {Alvarez}, {Alves}, {Anderson}, {Andrei}, {Anglada Varela}, {Antiche}, {Antoja}, {Arcay}, {Astraatmadja}, {Bach}, {Baker}, {Balaguer-N{\'u}{\~n}ez}, {Balm}, {Barache}, {Barata}, {Barbato}, {Barblan}, {Barklem}, {Barrado}, {Barros}, {Barstow}, {Bartholom{\'e} Mu{\~n}oz}, {Bassilana}, {Becciani}, {Bellazzini}, {Berihuete}, {Bertone}, {Bianchi}, {Bienaym{\'e}}, {Blanco-Cuaresma}, {Boch}, {Boeche}, {Bombrun}, {Borrachero},
  {Bossini}, {Bouquillon}, {Bourda}, {Bragaglia}, {Bramante}, {Breddels}, {Bressan}, {Brouillet}, {Br{\"u}semeister}, {Brugaletta}, {Bucciarelli}, {Burlacu}, {Busonero}, {Butkevich}, {Buzzi}, {Caffau}, {Cancelliere}, {Cannizzaro}, {Cantat-Gaudin}, {Carballo}, {Carlucci}, {Carrasco}, {Casamiquela}, {Castellani}, {Castro-Ginard}, {Charlot}, {Chemin}, {Chiavassa}, {Cocozza}, {Costigan}, {Cowell}, {Crifo}, {Crosta}, {Crowley}, {Cuypers}, {Dafonte}, {Damerdji}, {Dapergolas}, {David}, {David}, {de Laverny}, {De Luise}, {De March}, {de Martino}, {de Souza}, {de Torres}, {Debosscher}, {del Pozo}, {Delbo}, {Delgado}, {Delgado}, {Di Matteo}, {Diakite}, {Diener}, {Distefano}, {Dolding}, {Drazinos}, {Dur{\'a}n}, {Edvardsson}, {Enke}, {Eriksson}, {Esquej}, {Eynard Bontemps}, {Fabre}, {Fabrizio}, {Faigler}, {Falc{\~a}o}, {Farr{\`a}s Casas}, {Federici}, {Fedorets}, {Fernique}, {Figueras}, {Filippi}, {Findeisen}, {Fonti}, {Fraile}, {Fraser}, {Fr{\'e}zouls}, {Gai}, {Galleti}, {Garabato}, {Garc{\'\i}a-Sedano}, {Garofalo},
  {Garralda}, {Gavel}, {Gavras}, {Gerssen}, {Geyer}, {Giacobbe}, {Gilmore}, {Girona}, {Giuffrida}, {Glass}, {Gomes}, {Granvik}, {Gueguen}, {Guerrier}, {Guiraud}, {Guti{\'e}rrez-S{\'a}nchez}, {Haigron}, {Hatzidimitriou}, {Hauser}, {Haywood}, {Heiter}, {Helmi}, {Heu}, {Hilger}, {Hobbs}, {Hofmann}, {Holland}, {Huckle}, {Hypki}, {Icardi}, {Jan{\ss}en}, {Jevardat de Fombelle}, {Jonker}, {Juh{\'a}sz}, {Julbe}, {Karampelas}, {Kewley}, {Klar}, {Kochoska}, {Kohley}, {Kolenberg}, {Kontizas}, {Kontizas}, {Koposov}, {Kordopatis}, {Kostrzewa-Rutkowska}, {Koubsky}, {Lambert}, {Lanza}, {Lasne}, {Lavigne}, {Le Fustec}, {Le Poncin-Lafitte}, {Lebreton}, {Leccia}, {Leclerc}, {Lecoeur-Taibi}, {Lenhardt}, {Leroux}, {Liao}, {Licata}, {Lindstr{\o}m}, {Lister}, {Livanou}, {Lobel}, {L{\'o}pez}, {Managau}, {Mann}, {Mantelet}, {Marchal}, {Marchant}, {Marconi}, {Marinoni}, {Marschalk{\'o}}, {Marshall}, {Martino}, {Marton}, {Mary}, {Massari}, {Matijevi{\v{c}}}, {Mazeh}, {McMillan}, {Messina}, {Michalik}, {Millar}, {Molina}, {Molinaro},
  {Moln{\'a}r}, {Montegriffo}, {Mor}, {Morbidelli}, {Morel}, {Morris}, {Mulone}, {Muraveva}, {Musella}, {Nelemans}, {Nicastro}, {Noval}, {O'Mullane}, {Ord{\'e}novic}, {Ord{\'o}{\~n}ez-Blanco}, {Osborne}, {Pagani}, {Pagano}, {Pailler}, {Palacin}, {Palaversa}, {Panahi}, {Pawlak}, {Piersimoni}, {Pineau}, {Plachy}, {Plum}, {Poggio}, {Poujoulet}, {Pr{\v{s}}a}, {Pulone}, {Racero}, {Ragaini}, {Rambaux}, {Ramos-Lerate}, {Regibo}, {Reyl{\'e}}, {Riclet}, {Ripepi}, {Riva}, {Rivard}, {Rixon}, {Roegiers}, {Roelens}, {Romero-G{\'o}mez}, {Rowell}, {Royer}, {Ruiz-Dern}, {Sadowski}, {Sagrist{\`a} Sell{\'e}s}, {Sahlmann}, {Salgado}, {Salguero}, {Sanna}, {Santana-Ros}, {Sarasso}, {Savietto}, {Schultheis}, {Sciacca}, {Segol}, {Segovia}, {S{\'e}gransan}, {Shih}, {Siltala}, {Silva}, {Smart}, {Smith}, {Solano}, {Solitro}, {Sordo}, {Soria Nieto}, {Souchay}, {Spagna}, {Spoto}, {Stampa}, {Steele}, {Steidelm{\"u}ller}, {Stephenson}, {Stoev}, {Suess}, {Surdej}, {Szabados}, {Szegedi-Elek}, {Tapiador}, {Taris}, {Tauran}, {Taylor},
  {Teixeira}, {Terrett}, {Teyssandier}, {Thuillot}, {Titarenko}, {Torra Clotet}, {Turon}, {Ulla}, {Utrilla}, {Uzzi}, {Vaillant}, {Valentini}, {Valette}, {van Elteren}, {Van Hemelryck}, {van Leeuwen}, {Vaschetto}, {Vecchiato}, {Veljanoski}, {Viala}, {Vicente}, {Vogt}, {von Essen}, {Voss}, {Votruba}, {Voutsinas}, {Walmsley}, {Weiler}, {Wertz}, {Wevers}, {Wyrzykowski}, {Yoldas}, {{\v{Z}}erjal}, {Ziaeepour}, {Zorec}, {Zschocke}, {Zucker}, {Zurbach}, \& {Zwitter}}]{GaiaDR2}
{Gaia Collaboration}, {Brown}, A.~G.~A., {Vallenari}, A., {et~al.} 2018{\natexlab{b}}, \aap, 616, A1

\bibitem[{{Gaia Collaboration} {et~al.}(2021){Gaia Collaboration}, {Brown}, {Vallenari}, {Prusti}, {de Bruijne}, {Babusiaux}, {Biermann}, {Creevey}, {Evans}, {Eyer}, {Hutton}, {Jansen}, {Jordi}, {Klioner}, {Lammers}, {Lindegren}, {Luri}, {Mignard}, {Panem}, {Pourbaix}, {Randich}, {Sartoretti}, {Soubiran}, {Walton}, {Arenou}, {Bailer-Jones}, {Bastian}, {Cropper}, {Drimmel}, {Katz}, {Lattanzi}, {van Leeuwen}, {Bakker}, {Cacciari}, {Casta{\~n}eda}, {De Angeli}, {Ducourant}, {Fabricius}, {Fouesneau}, {Fr{\'e}mat}, {Guerra}, {Guerrier}, {Guiraud}, {Jean-Antoine Piccolo}, {Masana}, {Messineo}, {Mowlavi}, {Nicolas}, {Nienartowicz}, {Pailler}, {Panuzzo}, {Riclet}, {Roux}, {Seabroke}, {Sordo}, {Tanga}, {Th{\'e}venin}, {Gracia-Abril}, {Portell}, {Teyssier}, {Altmann}, {Andrae}, {Bellas-Velidis}, {Benson}, {Berthier}, {Blomme}, {Brugaletta}, {Burgess}, {Busso}, {Carry}, {Cellino}, {Cheek}, {Clementini}, {Damerdji}, {Davidson}, {Delchambre}, {Dell'Oro}, {Fern{\'a}ndez-Hern{\'a}ndez}, {Galluccio}, {Garc{\'\i}a-Lario},
  {Garcia-Reinaldos}, {Gonz{\'a}lez-N{\'u}{\~n}ez}, {Gosset}, {Haigron}, {Halbwachs}, {Hambly}, {Harrison}, {Hatzidimitriou}, {Heiter}, {Hern{\'a}ndez}, {Hestroffer}, {Hodgkin}, {Holl}, {Jan{\ss}en}, {Jevardat de Fombelle}, {Jordan}, {Krone-Martins}, {Lanzafame}, {L{\"o}ffler}, {Lorca}, {Manteiga}, {Marchal}, {Marrese}, {Moitinho}, {Mora}, {Muinonen}, {Osborne}, {Pancino}, {Pauwels}, {Petit}, {Recio-Blanco}, {Richards}, {Riello}, {Rimoldini}, {Robin}, {Roegiers}, {Rybizki}, {Sarro}, {Siopis}, {Smith}, {Sozzetti}, {Ulla}, {Utrilla}, {van Leeuwen}, {van Reeven}, {Abbas}, {Abreu Aramburu}, {Accart}, {Aerts}, {Aguado}, {Ajaj}, {Altavilla}, {{\'A}lvarez}, {{\'A}lvarez Cid-Fuentes}, {Alves}, {Anderson}, {Anglada Varela}, {Antoja}, {Audard}, {Baines}, {Baker}, {Balaguer-N{\'u}{\~n}ez}, {Balbinot}, {Balog}, {Barache}, {Barbato}, {Barros}, {Barstow}, {Bartolom{\'e}}, {Bassilana}, {Bauchet}, {Baudesson-Stella}, {Becciani}, {Bellazzini}, {Bernet}, {Bertone}, {Bianchi}, {Blanco-Cuaresma}, {Boch}, {Bombrun}, {Bossini},
  {Bouquillon}, {Bragaglia}, {Bramante}, {Breedt}, {Bressan}, {Brouillet}, {Bucciarelli}, {Burlacu}, {Busonero}, {Butkevich}, {Buzzi}, {Caffau}, {Cancelliere}, {C{\'a}novas}, {Cantat-Gaudin}, {Carballo}, {Carlucci}, {Carnerero}, {Carrasco}, {Casamiquela}, {Castellani}, {Castro-Ginard}, {Castro Sampol}, {Chaoul}, {Charlot}, {Chemin}, {Chiavassa}, {Cioni}, {Comoretto}, {Cooper}, {Cornez}, {Cowell}, {Crifo}, {Crosta}, {Crowley}, {Dafonte}, {Dapergolas}, {David}, {David}, {de Laverny}, {De Luise}, {De March}, {De Ridder}, {de Souza}, {de Teodoro}, {de Torres}, {del Peloso}, {del Pozo}, {Delbo}, {Delgado}, {Delgado}, {Delisle}, {Di Matteo}, {Diakite}, {Diener}, {Distefano}, {Dolding}, {Eappachen}, {Edvardsson}, {Enke}, {Esquej}, {Fabre}, {Fabrizio}, {Faigler}, {Fedorets}, {Fernique}, {Fienga}, {Figueras}, {Fouron}, {Fragkoudi}, {Fraile}, {Franke}, {Gai}, {Garabato}, {Garcia-Gutierrez}, {Garc{\'\i}a-Torres}, {Garofalo}, {Gavras}, {Gerlach}, {Geyer}, {Giacobbe}, {Gilmore}, {Girona}, {Giuffrida}, {Gomel}, {Gomez},
  {Gonzalez-Santamaria}, {Gonz{\'a}lez-Vidal}, {Granvik}, {Guti{\'e}rrez-S{\'a}nchez}, {Guy}, {Hauser}, {Haywood}, {Helmi}, {Hidalgo}, {Hilger}, {H{\l}adczuk}, {Hobbs}, {Holland}, {Huckle}, {Jasniewicz}, {Jonker}, {Juaristi Campillo}, {Julbe}, {Karbevska}, {Kervella}, {Khanna}, {Kochoska}, {Kontizas}, {Kordopatis}, {Korn}, {Kostrzewa-Rutkowska}, {Kruszy{\'n}ska}, {Lambert}, {Lanza}, {Lasne}, {Le Campion}, {Le Fustec}, {Lebreton}, {Lebzelter}, {Leccia}, {Leclerc}, {Lecoeur-Taibi}, {Liao}, {Licata}, {Lindstr{\o}m}, {Lister}, {Livanou}, {Lobel}, {Madrero Pardo}, {Managau}, {Mann}, {Marchant}, {Marconi}, {Marcos Santos}, {Marinoni}, {Marocco}, {Marshall}, {Martin Polo}, {Mart{\'\i}n-Fleitas}, {Masip}, {Massari}, {Mastrobuono-Battisti}, {Mazeh}, {McMillan}, {Messina}, {Michalik}, {Millar}, {Mints}, {Molina}, {Molinaro}, {Moln{\'a}r}, {Montegriffo}, {Mor}, {Morbidelli}, {Morel}, {Morris}, {Mulone}, {Munoz}, {Muraveva}, {Murphy}, {Musella}, {Noval}, {Ord{\'e}novic}, {Orr{\`u}}, {Osinde}, {Pagani}, {Pagano},
  {Palaversa}, {Palicio}, {Panahi}, {Pawlak}, {Pe{\~n}alosa Esteller}, {Penttil{\"a}}, {Piersimoni}, {Pineau}, {Plachy}, {Plum}, {Poggio}, {Poretti}, {Poujoulet}, {Pr{\v{s}}a}, {Pulone}, {Racero}, {Ragaini}, {Rainer}, {Raiteri}, {Rambaux}, {Ramos}, {Ramos-Lerate}, {Re Fiorentin}, {Regibo}, {Reyl{\'e}}, {Ripepi}, {Riva}, {Rixon}, {Robichon}, {Robin}, {Roelens}, {Rohrbasser}, {Romero-G{\'o}mez}, {Rowell}, {Royer}, {Rybicki}, {Sadowski}, {Sagrist{\`a} Sell{\'e}s}, {Sahlmann}, {Salgado}, {Salguero}, {Samaras}, {Sanchez Gimenez}, {Sanna}, {Santove{\~n}a}, {Sarasso}, {Schultheis}, {Sciacca}, {Segol}, {Segovia}, {S{\'e}gransan}, {Semeux}, {Shahaf}, {Siddiqui}, {Siebert}, {Siltala}, {Slezak}, {Smart}, {Solano}, {Solitro}, {Souami}, {Souchay}, {Spagna}, {Spoto}, {Steele}, {Steidelm{\"u}ller}, {Stephenson}, {S{\"u}veges}, {Szabados}, {Szegedi-Elek}, {Taris}, {Tauran}, {Taylor}, {Teixeira}, {Thuillot}, {Tonello}, {Torra}, {Torra}, {Turon}, {Unger}, {Vaillant}, {van Dillen}, {Vanel}, {Vecchiato}, {Viala}, {Vicente},
  {Voutsinas}, {Weiler}, {Wevers}, {Wyrzykowski}, {Yoldas}, {Yvard}, {Zhao}, {Zorec}, {Zucker}, {Zurbach}, \& {Zwitter}}]{GC21}
{Gaia Collaboration}, {Brown}, A.~G.~A., {Vallenari}, A., {et~al.} 2021, \aap, 649, A1

\bibitem[{{Gaia Collaboration} {et~al.}(2016){Gaia Collaboration}, {Prusti}, {de Bruijne}, {Brown}, {Vallenari}, {Babusiaux}, {Bailer-Jones}, {Bastian}, {Biermann}, {Evans}, {Eyer}, {Jansen}, {Jordi}, {Klioner}, {Lammers}, {Lindegren}, {Luri}, {Mignard}, {Milligan}, {Panem}, {Poinsignon}, {Pourbaix}, {Randich}, {Sarri}, {Sartoretti}, {Siddiqui}, {Soubiran}, {Valette}, {van Leeuwen}, {Walton}, {Aerts}, {Arenou}, {Cropper}, {Drimmel}, {H{\o}g}, {Katz}, {Lattanzi}, {O'Mullane}, {Grebel}, {Holland}, {Huc}, {Passot}, {Bramante}, {Cacciari}, {Casta{\~n}eda}, {Chaoul}, {Cheek}, {De Angeli}, {Fabricius}, {Guerra}, {Hern{\'a}ndez}, {Jean-Antoine-Piccolo}, {Masana}, {Messineo}, {Mowlavi}, {Nienartowicz}, {Ord{\'o}{\~n}ez-Blanco}, {Panuzzo}, {Portell}, {Richards}, {Riello}, {Seabroke}, {Tanga}, {Th{\'e}venin}, {Torra}, {Els}, {Gracia-Abril}, {Comoretto}, {Garcia-Reinaldos}, {Lock}, {Mercier}, {Altmann}, {Andrae}, {Astraatmadja}, {Bellas-Velidis}, {Benson}, {Berthier}, {Blomme}, {Busso}, {Carry}, {Cellino}, {Clementini},
  {Cowell}, {Creevey}, {Cuypers}, {Davidson}, {De Ridder}, {de Torres}, {Delchambre}, {Dell'Oro}, {Ducourant}, {Fr{\'e}mat}, {Garc{\'\i}a-Torres}, {Gosset}, {Halbwachs}, {Hambly}, {Harrison}, {Hauser}, {Hestroffer}, {Hodgkin}, {Huckle}, {Hutton}, {Jasniewicz}, {Jordan}, {Kontizas}, {Korn}, {Lanzafame}, {Manteiga}, {Moitinho}, {Muinonen}, {Osinde}, {Pancino}, {Pauwels}, {Petit}, {Recio-Blanco}, {Robin}, {Sarro}, {Siopis}, {Smith}, {Smith}, {Sozzetti}, {Thuillot}, {van Reeven}, {Viala}, {Abbas}, {Abreu Aramburu}, {Accart}, {Aguado}, {Allan}, {Allasia}, {Altavilla}, {{\'A}lvarez}, {Alves}, {Anderson}, {Andrei}, {Anglada Varela}, {Antiche}, {Antoja}, {Ant{\'o}n}, {Arcay}, {Atzei}, {Ayache}, {Bach}, {Baker}, {Balaguer-N{\'u}{\~n}ez}, {Barache}, {Barata}, {Barbier}, {Barblan}, {Baroni}, {Barrado y Navascu{\'e}s}, {Barros}, {Barstow}, {Becciani}, {Bellazzini}, {Bellei}, {Bello Garc{\'\i}a}, {Belokurov}, {Bendjoya}, {Berihuete}, {Bianchi}, {Bienaym{\'e}}, {Billebaud}, {Blagorodnova}, {Blanco-Cuaresma}, {Boch},
  {Bombrun}, {Borrachero}, {Bouquillon}, {Bourda}, {Bouy}, {Bragaglia}, {Breddels}, {Brouillet}, {Br{\"u}semeister}, {Bucciarelli}, {Budnik}, {Burgess}, {Burgon}, {Burlacu}, {Busonero}, {Buzzi}, {Caffau}, {Cambras}, {Campbell}, {Cancelliere}, {Cantat-Gaudin}, {Carlucci}, {Carrasco}, {Castellani}, {Charlot}, {Charnas}, {Charvet}, {Chassat}, {Chiavassa}, {Clotet}, {Cocozza}, {Collins}, {Collins}, {Costigan}, {Crifo}, {Cross}, {Crosta}, {Crowley}, {Dafonte}, {Damerdji}, {Dapergolas}, {David}, {David}, {De Cat}, {de Felice}, {de Laverny}, {De Luise}, {De March}, {de Martino}, {de Souza}, {Debosscher}, {del Pozo}, {Delbo}, {Delgado}, {Delgado}, {di Marco}, {Di Matteo}, {Diakite}, {Distefano}, {Dolding}, {Dos Anjos}, {Drazinos}, {Dur{\'a}n}, {Dzigan}, {Ecale}, {Edvardsson}, {Enke}, {Erdmann}, {Escolar}, {Espina}, {Evans}, {Eynard Bontemps}, {Fabre}, {Fabrizio}, {Faigler}, {Falc{\~a}o}, {Farr{\`a}s Casas}, {Faye}, {Federici}, {Fedorets}, {Fern{\'a}ndez-Hern{\'a}ndez}, {Fernique}, {Fienga}, {Figueras}, {Filippi},
  {Findeisen}, {Fonti}, {Fouesneau}, {Fraile}, {Fraser}, {Fuchs}, {Furnell}, {Gai}, {Galleti}, {Galluccio}, {Garabato}, {Garc{\'\i}a-Sedano}, {Gar{\'e}}, {Garofalo}, {Garralda}, {Gavras}, {Gerssen}, {Geyer}, {Gilmore}, {Girona}, {Giuffrida}, {Gomes}, {Gonz{\'a}lez-Marcos}, {Gonz{\'a}lez-N{\'u}{\~n}ez}, {Gonz{\'a}lez-Vidal}, {Granvik}, {Guerrier}, {Guillout}, {Guiraud}, {G{\'u}rpide}, {Guti{\'e}rrez-S{\'a}nchez}, {Guy}, {Haigron}, {Hatzidimitriou}, {Haywood}, {Heiter}, {Helmi}, {Hobbs}, {Hofmann}, {Holl}, {Holland}, {Hunt}, {Hypki}, {Icardi}, {Irwin}, {Jevardat de Fombelle}, {Jofr{\'e}}, {Jonker}, {Jorissen}, {Julbe}, {Karampelas}, {Kochoska}, {Kohley}, {Kolenberg}, {Kontizas}, {Koposov}, {Kordopatis}, {Koubsky}, {Kowalczyk}, {Krone-Martins}, {Kudryashova}, {Kull}, {Bachchan}, {Lacoste-Seris}, {Lanza}, {Lavigne}, {Le Poncin-Lafitte}, {Lebreton}, {Lebzelter}, {Leccia}, {Leclerc}, {Lecoeur-Taibi}, {Lemaitre}, {Lenhardt}, {Leroux}, {Liao}, {Licata}, {Lindstr{\o}m}, {Lister}, {Livanou}, {Lobel}, {L{\"o}ffler},
  {L{\'o}pez}, {Lopez-Lozano}, {Lorenz}, {Loureiro}, {MacDonald}, {Magalh{\~a}es Fernandes}, {Managau}, {Mann}, {Mantelet}, {Marchal}, {Marchant}, {Marconi}, {Marie}, {Marinoni}, {Marrese}, {Marschalk{\'o}}, {Marshall}, {Mart{\'\i}n-Fleitas}, {Martino}, {Mary}, {Matijevi{\v{c}}}, {Mazeh}, {McMillan}, {Messina}, {Mestre}, {Michalik}, {Millar}, {Miranda}, {Molina}, {Molinaro}, {Molinaro}, {Moln{\'a}r}, {Moniez}, {Montegriffo}, {Monteiro}, {Mor}, {Mora}, {Morbidelli}, {Morel}, {Morgenthaler}, {Morley}, {Morris}, {Mulone}, {Muraveva}, {Musella}, {Narbonne}, {Nelemans}, {Nicastro}, {Noval}, {Ord{\'e}novic}, {Ordieres-Mer{\'e}}, {Osborne}, {Pagani}, {Pagano}, {Pailler}, {Palacin}, {Palaversa}, {Parsons}, {Paulsen}, {Pecoraro}, {Pedrosa}, {Pentik{\"a}inen}, {Pereira}, {Pichon}, {Piersimoni}, {Pineau}, {Plachy}, {Plum}, {Poujoulet}, {Pr{\v{s}}a}, {Pulone}, {Ragaini}, {Rago}, {Rambaux}, {Ramos-Lerate}, {Ranalli}, {Rauw}, {Read}, {Regibo}, {Renk}, {Reyl{\'e}}, {Ribeiro}, {Rimoldini}, {Ripepi}, {Riva}, {Rixon},
  {Roelens}, {Romero-G{\'o}mez}, {Rowell}, {Royer}, {Rudolph}, {Ruiz-Dern}, {Sadowski}, {Sagrist{\`a} Sell{\'e}s}, {Sahlmann}, {Salgado}, {Salguero}, {Sarasso}, {Savietto}, {Schnorhk}, {Schultheis}, {Sciacca}, {Segol}, {Segovia}, {Segransan}, {Serpell}, {Shih}, {Smareglia}, {Smart}, {Smith}, {Solano}, {Solitro}, {Sordo}, {Soria Nieto}, {Souchay}, {Spagna}, {Spoto}, {Stampa}, {Steele}, {Steidelm{\"u}ller}, {Stephenson}, {Stoev}, {Suess}, {S{\"u}veges}, {Surdej}, {Szabados}, {Szegedi-Elek}, {Tapiador}, {Taris}, {Tauran}, {Taylor}, {Teixeira}, {Terrett}, {Tingley}, {Trager}, {Turon}, {Ulla}, {Utrilla}, {Valentini}, {van Elteren}, {Van Hemelryck}, {van Leeuwen}, {Varadi}, {Vecchiato}, {Veljanoski}, {Via}, {Vicente}, {Vogt}, {Voss}, {Votruba}, {Voutsinas}, {Walmsley}, {Weiler}, {Weingrill}, {Werner}, {Wevers}, {Whitehead}, {Wyrzykowski}, {Yoldas}, {{\v{Z}}erjal}, {Zucker}, {Zurbach}, {Zwitter}, {Alecu}, {Allen}, {Allende Prieto}, {Amorim}, {Anglada-Escud{\'e}}, {Arsenijevic}, {Azaz}, {Balm}, {Beck}, {Bernstein},
  {Bigot}, {Bijaoui}, {Blasco}, {Bonfigli}, {Bono}, {Boudreault}, {Bressan}, {Brown}, {Brunet}, {Bunclark}, {Buonanno}, {Butkevich}, {Carret}, {Carrion}, {Chemin}, {Ch{\'e}reau}, {Corcione}, {Darmigny}, {de Boer}, {de Teodoro}, {de Zeeuw}, {Delle Luche}, {Domingues}, {Dubath}, {Fodor}, {Fr{\'e}zouls}, {Fries}, {Fustes}, {Fyfe}, {Gallardo}, {Gallegos}, {Gardiol}, {Gebran}, {Gomboc}, {G{\'o}mez}, {Grux}, {Gueguen}, {Heyrovsky}, {Hoar}, {Iannicola}, {Isasi Parache}, {Janotto}, {Joliet}, {Jonckheere}, {Keil}, {Kim}, {Klagyivik}, {Klar}, {Knude}, {Kochukhov}, {Kolka}, {Kos}, {Kutka}, {Lainey}, {LeBouquin}, {Liu}, {Loreggia}, {Makarov}, {Marseille}, {Martayan}, {Martinez-Rubi}, {Massart}, {Meynadier}, {Mignot}, {Munari}, {Nguyen}, {Nordlander}, {Ocvirk}, {O'Flaherty}, {Olias Sanz}, {Ortiz}, {Osorio}, {Oszkiewicz}, {Ouzounis}, {Palmer}, {Park}, {Pasquato}, {Peltzer}, {Peralta}, {P{\'e}turaud}, {Pieniluoma}, {Pigozzi}, {Poels}, {Prat}, {Prod'homme}, {Raison}, {Rebordao}, {Risquez}, {Rocca-Volmerange}, {Rosen},
  {Ruiz-Fuertes}, {Russo}, {Sembay}, {Serraller Vizcaino}, {Short}, {Siebert}, {Silva}, {Sinachopoulos}, {Slezak}, {Soffel}, {Sosnowska}, {Strai{\v{z}}ys}, {ter Linden}, {Terrell}, {Theil}, {Tiede}, {Troisi}, {Tsalmantza}, {Tur}, {Vaccari}, {Vachier}, {Valles}, {Van Hamme}, {Veltz}, {Virtanen}, {Wallut}, {Wichmann}, {Wilkinson}, {Ziaeepour}, \& {Zschocke}}]{GC16}
{Gaia Collaboration}, {Prusti}, T., {de Bruijne}, J.~H.~J., {et~al.} 2016, \aap, 595, A1

\bibitem[{{Gaia Collaboration} {et~al.}(2023){Gaia Collaboration}, {Vallenari}, {Brown}, {Prusti}, {de Bruijne}, {Arenou}, {Babusiaux}, {Biermann}, {Creevey}, {Ducourant}, {Evans}, {Eyer}, {Guerra}, {Hutton}, {Jordi}, {Klioner}, {Lammers}, {Lindegren}, {Luri}, {Mignard}, {Panem}, {Pourbaix}, {Randich}, {Sartoretti}, {Soubiran}, {Tanga}, {Walton}, {Bailer-Jones}, {Bastian}, {Drimmel}, {Jansen}, {Katz}, {Lattanzi}, {van Leeuwen}, {Bakker}, {Cacciari}, {Casta{\~n}eda}, {De Angeli}, {Fabricius}, {Fouesneau}, {Fr{\'e}mat}, {Galluccio}, {Guerrier}, {Heiter}, {Masana}, {Messineo}, {Mowlavi}, {Nicolas}, {Nienartowicz}, {Pailler}, {Panuzzo}, {Riclet}, {Roux}, {Seabroke}, {Sordo}, {Th{\'e}venin}, {Gracia-Abril}, {Portell}, {Teyssier}, {Altmann}, {Andrae}, {Audard}, {Bellas-Velidis}, {Benson}, {Berthier}, {Blomme}, {Burgess}, {Busonero}, {Busso}, {C{\'a}novas}, {Carry}, {Cellino}, {Cheek}, {Clementini}, {Damerdji}, {Davidson}, {de Teodoro}, {Nu{\~n}ez Campos}, {Delchambre}, {Dell'Oro}, {Esquej},
  {Fern{\'a}ndez-Hern{\'a}ndez}, {Fraile}, {Garabato}, {Garc{\'\i}a-Lario}, {Gosset}, {Haigron}, {Halbwachs}, {Hambly}, {Harrison}, {Hern{\'a}ndez}, {Hestroffer}, {Hodgkin}, {Holl}, {Jan{\ss}en}, {Jevardat de Fombelle}, {Jordan}, {Krone-Martins}, {Lanzafame}, {L{\"o}ffler}, {Marchal}, {Marrese}, {Moitinho}, {Muinonen}, {Osborne}, {Pancino}, {Pauwels}, {Recio-Blanco}, {Reyl{\'e}}, {Riello}, {Rimoldini}, {Roegiers}, {Rybizki}, {Sarro}, {Siopis}, {Smith}, {Sozzetti}, {Utrilla}, {van Leeuwen}, {Abbas}, {{\'A}brah{\'a}m}, {Abreu Aramburu}, {Aerts}, {Aguado}, {Ajaj}, {Aldea-Montero}, {Altavilla}, {{\'A}lvarez}, {Alves}, {Anders}, {Anderson}, {Anglada Varela}, {Antoja}, {Baines}, {Baker}, {Balaguer-N{\'u}{\~n}ez}, {Balbinot}, {Balog}, {Barache}, {Barbato}, {Barros}, {Barstow}, {Bartolom{\'e}}, {Bassilana}, {Bauchet}, {Becciani}, {Bellazzini}, {Berihuete}, {Bernet}, {Bertone}, {Bianchi}, {Binnenfeld}, {Blanco-Cuaresma}, {Blazere}, {Boch}, {Bombrun}, {Bossini}, {Bouquillon}, {Bragaglia}, {Bramante}, {Breedt},
  {Bressan}, {Brouillet}, {Brugaletta}, {Bucciarelli}, {Burlacu}, {Butkevich}, {Buzzi}, {Caffau}, {Cancelliere}, {Cantat-Gaudin}, {Carballo}, {Carlucci}, {Carnerero}, {Carrasco}, {Casamiquela}, {Castellani}, {Castro-Ginard}, {Chaoul}, {Charlot}, {Chemin}, {Chiaramida}, {Chiavassa}, {Chornay}, {Comoretto}, {Contursi}, {Cooper}, {Cornez}, {Cowell}, {Crifo}, {Cropper}, {Crosta}, {Crowley}, {Dafonte}, {Dapergolas}, {David}, {David}, {de Laverny}, {De Luise}, {De March}, {De Ridder}, {de Souza}, {de Torres}, {del Peloso}, {del Pozo}, {Delbo}, {Delgado}, {Delisle}, {Demouchy}, {Dharmawardena}, {Di Matteo}, {Diakite}, {Diener}, {Distefano}, {Dolding}, {Edvardsson}, {Enke}, {Fabre}, {Fabrizio}, {Faigler}, {Fedorets}, {Fernique}, {Fienga}, {Figueras}, {Fournier}, {Fouron}, {Fragkoudi}, {Gai}, {Garcia-Gutierrez}, {Garcia-Reinaldos}, {Garc{\'\i}a-Torres}, {Garofalo}, {Gavel}, {Gavras}, {Gerlach}, {Geyer}, {Giacobbe}, {Gilmore}, {Girona}, {Giuffrida}, {Gomel}, {Gomez}, {Gonz{\'a}lez-N{\'u}{\~n}ez},
  {Gonz{\'a}lez-Santamar{\'\i}a}, {Gonz{\'a}lez-Vidal}, {Granvik}, {Guillout}, {Guiraud}, {Guti{\'e}rrez-S{\'a}nchez}, {Guy}, {Hatzidimitriou}, {Hauser}, {Haywood}, {Helmer}, {Helmi}, {Sarmiento}, {Hidalgo}, {Hilger}, {H{\l}adczuk}, {Hobbs}, {Holland}, {Huckle}, {Jardine}, {Jasniewicz}, {Jean-Antoine Piccolo}, {Jim{\'e}nez-Arranz}, {Jorissen}, {Juaristi Campillo}, {Julbe}, {Karbevska}, {Kervella}, {Khanna}, {Kontizas}, {Kordopatis}, {Korn}, {K{\'o}sp{\'a}l}, {Kostrzewa-Rutkowska}, {Kruszy{\'n}ska}, {Kun}, {Laizeau}, {Lambert}, {Lanza}, {Lasne}, {Le Campion}, {Lebreton}, {Lebzelter}, {Leccia}, {Leclerc}, {Lecoeur-Taibi}, {Liao}, {Licata}, {Lindstr{\o}m}, {Lister}, {Livanou}, {Lobel}, {Lorca}, {Loup}, {Madrero Pardo}, {Magdaleno Romeo}, {Managau}, {Mann}, {Manteiga}, {Marchant}, {Marconi}, {Marcos}, {Marcos Santos}, {Mar{\'\i}n Pina}, {Marinoni}, {Marocco}, {Marshall}, {Martin Polo}, {Mart{\'\i}n-Fleitas}, {Marton}, {Mary}, {Masip}, {Massari}, {Mastrobuono-Battisti}, {Mazeh}, {McMillan}, {Messina}, {Michalik},
  {Millar}, {Mints}, {Molina}, {Molinaro}, {Moln{\'a}r}, {Monari}, {Mongui{\'o}}, {Montegriffo}, {Montero}, {Mor}, {Mora}, {Morbidelli}, {Morel}, {Morris}, {Muraveva}, {Murphy}, {Musella}, {Nagy}, {Noval}, {Oca{\~n}a}, {Ogden}, {Ordenovic}, {Osinde}, {Pagani}, {Pagano}, {Palaversa}, {Palicio}, {Pallas-Quintela}, {Panahi}, {Payne-Wardenaar}, {Pe{\~n}alosa Esteller}, {Penttil{\"a}}, {Pichon}, {Piersimoni}, {Pineau}, {Plachy}, {Plum}, {Poggio}, {Pr{\v{s}}a}, {Pulone}, {Racero}, {Ragaini}, {Rainer}, {Raiteri}, {Rambaux}, {Ramos}, {Ramos-Lerate}, {Re Fiorentin}, {Regibo}, {Richards}, {Rios Diaz}, {Ripepi}, {Riva}, {Rix}, {Rixon}, {Robichon}, {Robin}, {Robin}, {Roelens}, {Rogues}, {Rohrbasser}, {Romero-G{\'o}mez}, {Rowell}, {Royer}, {Ruz Mieres}, {Rybicki}, {Sadowski}, {S{\'a}ez N{\'u}{\~n}ez}, {Sagrist{\`a} Sell{\'e}s}, {Sahlmann}, {Salguero}, {Samaras}, {Sanchez Gimenez}, {Sanna}, {Santove{\~n}a}, {Sarasso}, {Schultheis}, {Sciacca}, {Segol}, {Segovia}, {S{\'e}gransan}, {Semeux}, {Shahaf}, {Siddiqui}, {Siebert},
  {Siltala}, {Silvelo}, {Slezak}, {Slezak}, {Smart}, {Snaith}, {Solano}, {Solitro}, {Souami}, {Souchay}, {Spagna}, {Spina}, {Spoto}, {Steele}, {Steidelm{\"u}ller}, {Stephenson}, {S{\"u}veges}, {Surdej}, {Szabados}, {Szegedi-Elek}, {Taris}, {Taylor}, {Teixeira}, {Tolomei}, {Tonello}, {Torra}, {Torra}, {Torralba Elipe}, {Trabucchi}, {Tsounis}, {Turon}, {Ulla}, {Unger}, {Vaillant}, {van Dillen}, {van Reeven}, {Vanel}, {Vecchiato}, {Viala}, {Vicente}, {Voutsinas}, {Weiler}, {Wevers}, {Wyrzykowski}, {Yoldas}, {Yvard}, {Zhao}, {Zorec}, {Zucker}, \& {Zwitter}}]{GC23}
{Gaia Collaboration}, {Vallenari}, A., {Brown}, A.~G.~A., {et~al.} 2023, \aap, 674, A1

\bibitem[{{Gallart} {et~al.}(2019){Gallart}, {Bernard}, {Brook}, {Ruiz-Lara}, {Cassisi}, {Hill}, \& {Monelli}}]{gallart19}
{Gallart}, C., {Bernard}, E.~J., {Brook}, C.~B., {et~al.} 2019, Nature Astronomy, 3, 932

\bibitem[{{Gallart} {et~al.}(2005){Gallart}, {Zoccali}, \& {Aparicio}}]{gallart05}
{Gallart}, C., {Zoccali}, M., \& {Aparicio}, A. 2005, \araa, 43, 387

\bibitem[{{G{\'o}mez} {et~al.}(2013){G{\'o}mez}, {Helmi}, {Cooper}, {Frenk}, {Navarro}, \& {White}}]{gomez13}
{G{\'o}mez}, F.~A., {Helmi}, A., {Cooper}, A.~P., {et~al.} 2013, \mnras, 436, 3602

\bibitem[{{Gratton} {et~al.}(2003){Gratton}, {Carretta}, {Desidera}, {Lucatello}, {Mazzei}, \& {Barbieri}}]{gratton03}
{Gratton}, R.~G., {Carretta}, E., {Desidera}, S., {et~al.} 2003, \aap, 406, 131

\bibitem[{{GRAVITY Collaboration} {et~al.}(2018){GRAVITY Collaboration}, {Abuter}, {Amorim}, {Anugu}, {Baub{\"o}ck}, {Benisty}, {Berger}, {Blind}, {Bonnet}, {Brandner}, {Buron}, {Collin}, {Chapron}, {Cl{\'e}net}, {Coud{\'e} Du Foresto}, {de Zeeuw}, {Deen}, {Delplancke-Str{\"o}bele}, {Dembet}, {Dexter}, {Duvert}, {Eckart}, {Eisenhauer}, {Finger}, {F{\"o}rster Schreiber}, {F{\'e}dou}, {Garcia}, {Garcia Lopez}, {Gao}, {Gendron}, {Genzel}, {Gillessen}, {Gordo}, {Habibi}, {Haubois}, {Haug}, {Hau{\ss}mann}, {Henning}, {Hippler}, {Horrobin}, {Hubert}, {Hubin}, {Jimenez Rosales}, {Jochum}, {Jocou}, {Kaufer}, {Kellner}, {Kendrew}, {Kervella}, {Kok}, {Kulas}, {Lacour}, {Lapeyr{\`e}re}, {Lazareff}, {Le Bouquin}, {L{\'e}na}, {Lippa}, {Lenzen}, {M{\'e}rand}, {M{\"u}ler}, {Neumann}, {Ott}, {Palanca}, {Paumard}, {Pasquini}, {Perraut}, {Perrin}, {Pfuhl}, {Plewa}, {Rabien}, {Ram{\'\i}rez}, {Ramos}, {Rau}, {Rodr{\'\i}guez-Coira}, {Rohloff}, {Rousset}, {Sanchez-Bermudez}, {Scheithauer}, {Sch{\"o}ller}, {Schuler}, {Spyromilio},
  {Straub}, {Straubmeier}, {Sturm}, {Tacconi}, {Tristram}, {Vincent}, {von Fellenberg}, {Wank}, {Waisberg}, {Widmann}, {Wieprecht}, {Wiest}, {Wiezorrek}, {Woillez}, {Yazici}, {Ziegler}, \& {Zins}}]{gravitycollaboration18}
{GRAVITY Collaboration}, {Abuter}, R., {Amorim}, A., {et~al.} 2018, \aap, 615, L15

\bibitem[{{Grevesse} \& {Sauval}(1998)}]{grevesse1998}
{Grevesse}, N. \& {Sauval}, A.~J. 1998, \ssr, 85, 161

\bibitem[{{Hasselquist} {et~al.}(2021){Hasselquist}, {Hayes}, {Lian}, {Weinberg}, {Zasowski}, {Horta}, {Beaton}, {Feuillet}, {Garro}, {Gallart}, {Smith}, {Holtzman}, {Minniti}, {Lacerna}, {Shetrone}, {J{\"o}nsson}, {Cioni}, {Fillingham}, {Cunha}, {O'Connell}, {Fern{\'a}ndez-Trincado}, {Mu{\~n}oz}, {Schiavon}, {Almeida}, {Anguiano}, {Beers}, {Bizyaev}, {Brownstein}, {Cohen}, {Frinchaboy}, {Garc{\'\i}a-Hern{\'a}ndez}, {Geisler}, {Lane}, {Majewski}, {Nidever}, {Nitschelm}, {Povick}, {Price-Whelan}, {Roman-Lopes}, {Rosado}, {Sobeck}, {Stringfellow}, {Valenzuela}, {Villanova}, \& {Vincenzo}}]{hasselquist21}
{Hasselquist}, S., {Hayes}, C.~R., {Lian}, J., {et~al.} 2021, \apj, 923, 172

\bibitem[{{Hayes} {et~al.}(2018){Hayes}, {Majewski}, {Shetrone}, {Fern{\'a}ndez-Alvar}, {Allende Prieto}, {Schuster}, {Carigi}, {Cunha}, {Smith}, {Sobeck}, {Almeida}, {Beers}, {Carrera}, {Fern{\'a}ndez-Trincado}, {Garc{\'\i}a-Hern{\'a}ndez}, {Geisler}, {Lane}, {Lucatello}, {Matthews}, {Minniti}, {Nitschelm}, {Tang}, {Tissera}, \& {Zamora}}]{hayes2018}
{Hayes}, C.~R., {Majewski}, S.~R., {Shetrone}, M., {et~al.} 2018, \apj, 852, 49

\bibitem[{{Haywood} {et~al.}(2018){Haywood}, {Di Matteo}, {Lehnert}, {Snaith}, {Khoperskov}, \& {G{\'o}mez}}]{haywood2018}
{Haywood}, M., {Di Matteo}, P., {Lehnert}, M.~D., {et~al.} 2018, \apj, 863, 113

\bibitem[{{Helmi}(2020)}]{helmi2020}
{Helmi}, A. 2020, \araa, 58, 205

\bibitem[{{Helmi} {et~al.}(2018){Helmi}, {Babusiaux}, {Koppelman}, {Massari}, {Veljanoski}, \& {Brown}}]{helmi2018}
{Helmi}, A., {Babusiaux}, C., {Koppelman}, H.~H., {et~al.} 2018, \nat, 563, 85

\bibitem[{{Helmi} \& {de Zeeuw}(2000)}]{helmidezeeuw00}
{Helmi}, A. \& {de Zeeuw}, P.~T. 2000, \mnras, 319, 657

\bibitem[{{Helmi} {et~al.}(1999){Helmi}, {White}, {de Zeeuw}, \& {Zhao}}]{helmi99}
{Helmi}, A., {White}, S. D.~M., {de Zeeuw}, P.~T., \& {Zhao}, H. 1999, \nat, 402, 53

\bibitem[{{Horta} {et~al.}(2021){Horta}, {Schiavon}, {Mackereth}, {Pfeffer}, {Mason}, {Kisku}, {Fragkoudi}, {Allende Prieto}, {Cunha}, {Hasselquist}, {Holtzman}, {Majewski}, {Nataf}, {O'Connell}, {Schultheis}, \& {Smith}}]{horta2021}
{Horta}, D., {Schiavon}, R.~P., {Mackereth}, J.~T., {et~al.} 2021, \mnras, 500, 1385

\bibitem[{{Horta} {et~al.}(2023){Horta}, {Schiavon}, {Mackereth}, {Weinberg}, {Hasselquist}, {Feuillet}, {O'Connell}, {Anguiano}, {Allende-Prieto}, {Beaton}, {Bizyaev}, {Cunha}, {Geisler}, {Garc{\'\i}a-Hern{\'a}ndez}, {Holtzman}, {J{\"o}nsson}, {Lane}, {Majewski}, {M{\'e}sz{\'a}ros}, {Minniti}, {Nitschelm}, {Shetrone}, {Smith}, \& {Zasowski}}]{horta23}
{Horta}, D., {Schiavon}, R.~P., {Mackereth}, J.~T., {et~al.} 2023, \mnras, 520, 5671

\bibitem[{{Ibata} {et~al.}(2021){Ibata}, {Malhan}, {Martin}, {Aubert}, {Famaey}, {Bianchini}, {Monari}, {Siebert}, {Thomas}, {Bellazzini}, {Bonifacio}, {Caffau}, \& {Renaud}}]{ibata2021}
{Ibata}, R., {Malhan}, K., {Martin}, N., {et~al.} 2021, \apj, 914, 123

\bibitem[{{Ibata} {et~al.}(1994){Ibata}, {Gilmore}, \& {Irwin}}]{ibata94}
{Ibata}, R.~A., {Gilmore}, G., \& {Irwin}, M.~J. 1994, \nat, 370, 194

\bibitem[{{Johnston} {et~al.}(1996){Johnston}, {Hernquist}, \& {Bolte}}]{johnston1996}
{Johnston}, K.~V., {Hernquist}, L., \& {Bolte}, M. 1996, \apj, 465, 278

\bibitem[{{Kasen} {et~al.}(2017){Kasen}, {Metzger}, {Barnes}, {Quataert}, \& {Ramirez-Ruiz}}]{kasen2017}
{Kasen}, D., {Metzger}, B., {Barnes}, J., {Quataert}, E., \& {Ramirez-Ruiz}, E. 2017, \nat, 551, 80

\bibitem[{{Katz} {et~al.}(2023){Katz}, {Sartoretti}, {Guerrier}, {Panuzzo}, {Seabroke}, {Th{\'e}venin}, {Cropper}, {Benson}, {Blomme}, {Haigron}, {Marchal}, {Smith}, {Baker}, {Chemin}, {Damerdji}, {David}, {Dolding}, {Fr{\'e}mat}, {Gosset}, {Jan{\ss}en}, {Jasniewicz}, {Lobel}, {Plum}, {Samaras}, {Snaith}, {Soubiran}, {Vanel}, {Zwitter}, {Antoja}, {Arenou}, {Babusiaux}, {Brouillet}, {Caffau}, {Di Matteo}, {Fabre}, {Fabricius}, {Fragkoudi}, {Haywood}, {Huckle}, {Hottier}, {Lasne}, {Leclerc}, {Mastrobuono-Battisti}, {Royer}, {Teyssier}, {Zorec}, {Crifo}, {Jean-Antoine Piccolo}, {Turon}, \& {Viala}}]{katz23}
{Katz}, D., {Sartoretti}, P., {Guerrier}, A., {et~al.} 2023, \aap, 674, A5

\bibitem[{{Khoperskov} {et~al.}(2023){Khoperskov}, {Minchev}, {Libeskind}, {Haywood}, {Di Matteo}, {Belokurov}, {Steinmetz}, {Gomez}, {Grand}, {Hoffman}, {Knebe}, {Sorce}, {Spaare}, {Tempel}, \& {Vogelsberger}}]{khoperskov2023}
{Khoperskov}, S., {Minchev}, I., {Libeskind}, N., {et~al.} 2023, \aap, 677, A90

\bibitem[{{Kobayashi} {et~al.}(2020){Kobayashi}, {Karakas}, \& {Lugaro}}]{kobayashi2020}
{Kobayashi}, C., {Karakas}, A.~I., \& {Lugaro}, M. 2020, \apj, 900, 179

\bibitem[{{Kobayashi} \& {Nomoto}(2009)}]{kobayashi2009}
{Kobayashi}, C. \& {Nomoto}, K. 2009, \apj, 707, 1466

\bibitem[{{Koppelman} {et~al.}(2018){Koppelman}, {Helmi}, \& {Veljanoski}}]{koppelman18}
{Koppelman}, H., {Helmi}, A., \& {Veljanoski}, J. 2018, \apjl, 860, L11

\bibitem[{{Koppelman} {et~al.}(2020){Koppelman}, {Bos}, \& {Helmi}}]{koppelman2020}
{Koppelman}, H.~H., {Bos}, R. O.~Y., \& {Helmi}, A. 2020, \aap, 642, L18

\bibitem[{{Koppelman} {et~al.}(2019){Koppelman}, {Helmi}, {Massari}, {Price-Whelan}, \& {Starkenburg}}]{koppelman19}
{Koppelman}, H.~H., {Helmi}, A., {Massari}, D., {Price-Whelan}, A.~M., \& {Starkenburg}, T.~K. 2019, \aap, 631, L9

\bibitem[{{Kruijssen} {et~al.}(2020){Kruijssen}, {Pfeffer}, {Chevance}, {Bonaca}, {Trujillo-Gomez}, {Bastian}, {Reina-Campos}, {Crain}, \& {Hughes}}]{kruijssen2020}
{Kruijssen}, J.~M.~D., {Pfeffer}, J.~L., {Chevance}, M., {et~al.} 2020, \mnras, 498, 2472

\bibitem[{{Kurucz}(2005)}]{kurucz}
{Kurucz}, R.~L. 2005, Memorie della Societa Astronomica Italiana Supplementi, 8, 14

\bibitem[{{Lach} {et~al.}(2020){Lach}, {R{\"o}pke}, {Seitenzahl}, {Cot{\'e}}, {Gronow}, \& {Ruiter}}]{lach2020}
{Lach}, F., {R{\"o}pke}, F.~K., {Seitenzahl}, I.~R., {et~al.} 2020, \aap, 644, A118

\bibitem[{{Leung} \& {Nomoto}(2018)}]{leung2018}
{Leung}, S.-C. \& {Nomoto}, K. 2018, \apj, 861, 143

\bibitem[{{Limberg} {et~al.}(2022){Limberg}, {Souza}, {P{\'e}rez-Villegas}, {Rossi}, {Perottoni}, \& {Santucci}}]{limberg22}
{Limberg}, G., {Souza}, S.~O., {P{\'e}rez-Villegas}, A., {et~al.} 2022, \apj, 935, 109

\bibitem[{{Lind} {et~al.}(2011){Lind}, {Asplund}, {Barklem}, \& {Belyaev}}]{lind2011}
{Lind}, K., {Asplund}, M., {Barklem}, P.~S., \& {Belyaev}, A.~K. 2011, \aap, 528, A103

\bibitem[{{Lind} {et~al.}(2022){Lind}, {Nordlander}, {Wehrhahn}, {Montelius}, {Osorio}, {Barklem}, {Af{\c{s}}ar}, {Sneden}, \& {Kobayashi}}]{lind22}
{Lind}, K., {Nordlander}, T., {Wehrhahn}, A., {et~al.} 2022, \aap, 665, A33

\bibitem[{{Lindegren} {et~al.}(2021){Lindegren}, {Klioner}, {Hern{\'a}ndez}, {Bombrun}, {Ramos-Lerate}, {Steidelm{\"u}ller}, {Bastian}, {Biermann}, {de Torres}, {Gerlach}, {Geyer}, {Hilger}, {Hobbs}, {Lammers}, {McMillan}, {Stephenson}, {Casta{\~n}eda}, {Davidson}, {Fabricius}, {Gracia-Abril}, {Portell}, {Rowell}, {Teyssier}, {Torra}, {Bartolom{\'e}}, {Clotet}, {Garralda}, {Gonz{\'a}lez-Vidal}, {Torra}, {Abbas}, {Altmann}, {Anglada Varela}, {Balaguer-N{\'u}{\~n}ez}, {Balog}, {Barache}, {Becciani}, {Bernet}, {Bertone}, {Bianchi}, {Bouquillon}, {Brown}, {Bucciarelli}, {Busonero}, {Butkevich}, {Buzzi}, {Cancelliere}, {Carlucci}, {Charlot}, {Cioni}, {Crosta}, {Crowley}, {del Peloso}, {del Pozo}, {Drimmel}, {Esquej}, {Fienga}, {Fraile}, {Gai}, {Garcia-Reinaldos}, {Guerra}, {Hambly}, {Hauser}, {Jan{\ss}en}, {Jordan}, {Kostrzewa-Rutkowska}, {Lattanzi}, {Liao}, {Licata}, {Lister}, {L{\"o}ffler}, {Marchant}, {Masip}, {Mignard}, {Mints}, {Molina}, {Mora}, {Morbidelli}, {Murphy}, {Pagani}, {Panuzzo}, {Pe{\~n}alosa
  Esteller}, {Poggio}, {Re Fiorentin}, {Riva}, {Sagrist{\`a} Sell{\'e}s}, {Sanchez Gimenez}, {Sarasso}, {Sciacca}, {Siddiqui}, {Smart}, {Souami}, {Spagna}, {Steele}, {Taris}, {Utrilla}, {van Reeven}, \& {Vecchiato}}]{lindegren21}
{Lindegren}, L., {Klioner}, S.~A., {Hern{\'a}ndez}, J., {et~al.} 2021, \aap, 649, A2

\bibitem[{{L{\"o}vdal} {et~al.}(2022){L{\"o}vdal}, {Ruiz-Lara}, {Koppelman}, {Matsuno}, {Dodd}, \& {Helmi}}]{lovdal22}
{L{\"o}vdal}, S.~S., {Ruiz-Lara}, T., {Koppelman}, H.~H., {et~al.} 2022, \aap, 665, A57

\bibitem[{{Mackereth} {et~al.}(2019){Mackereth}, {Schiavon}, {Pfeffer}, {Hayes}, {Bovy}, {Anguiano}, {Allende Prieto}, {Hasselquist}, {Holtzman}, {Johnson}, {Majewski}, {O'Connell}, {Shetrone}, {Tissera}, \& {Fern{\'a}ndez-Trincado}}]{mackereth19}
{Mackereth}, J.~T., {Schiavon}, R.~P., {Pfeffer}, J., {et~al.} 2019, \mnras, 482, 3426

\bibitem[{{Malhan} {et~al.}(2022){Malhan}, {Ibata}, {Sharma}, {Famaey}, {Bellazzini}, {Carlberg}, {D'Souza}, {Yuan}, {Martin}, \& {Thomas}}]{malhna2022}
{Malhan}, K., {Ibata}, R.~A., {Sharma}, S., {et~al.} 2022, \apj, 926, 107

\bibitem[{{Massari} {et~al.}(2019){Massari}, {Koppelman}, \& {Helmi}}]{massari19}
{Massari}, D., {Koppelman}, H.~H., \& {Helmi}, A. 2019, \aap, 630, L4

\bibitem[{{Matsuno} {et~al.}(2019){Matsuno}, {Aoki}, \& {Suda}}]{matsuno19}
{Matsuno}, T., {Aoki}, W., \& {Suda}, T. 2019, \apjl, 874, L35

\bibitem[{{Matsuno} {et~al.}(2022){Matsuno}, {Koppelman}, {Helmi}, {Aoki}, {Ishigaki}, {Suda}, {Yuan}, \& {Hattori}}]{matsuno22}
{Matsuno}, T., {Koppelman}, H.~H., {Helmi}, A., {et~al.} 2022, \aap, 661, A103

\bibitem[{{Matteucci} \& {Greggio}(1986)}]{matteucci&greggio86}
{Matteucci}, F. \& {Greggio}, L. 1986, \aap, 154, 279

\bibitem[{{McMillan}(2017)}]{mcmillan17}
{McMillan}, P.~J. 2017, \mnras, 465, 76

\bibitem[{{Mikkola} {et~al.}(2023){Mikkola}, {McMillan}, \& {Hobbs}}]{mikkola2023}
{Mikkola}, D., {McMillan}, P.~J., \& {Hobbs}, D. 2023, \mnras, 519, 1989

\bibitem[{{Minelli} {et~al.}(2021){Minelli}, {Mucciarelli}, {Massari}, {Bellazzini}, {Romano}, \& {Ferraro}}]{minelli21}
{Minelli}, A., {Mucciarelli}, A., {Massari}, D., {et~al.} 2021, \apjl, 918, L32

\bibitem[{{Molero} {et~al.}(2023){Molero}, {Magrini}, {Matteucci}, {Romano}, {Palla}, {Cescutti}, {Viscasillas V{\'a}zquez}, \& {Spitoni}}]{molero2023}
{Molero}, M., {Magrini}, L., {Matteucci}, F., {et~al.} 2023, \mnras, 523, 2974

\bibitem[{{Monachesi} {et~al.}(2019){Monachesi}, {G{\'o}mez}, {Grand}, {Simpson}, {Kauffmann}, {Bustamante}, {Marinacci}, {Pakmor}, {Springel}, {Frenk}, {White}, \& {Tissera}}]{monachesi2019}
{Monachesi}, A., {G{\'o}mez}, F.~A., {Grand}, R. J.~J., {et~al.} 2019, \mnras, 485, 2589

\bibitem[{{Montalb{\'a}n} {et~al.}(2021){Montalb{\'a}n}, {Mackereth}, {Miglio}, {Vincenzo}, {Chiappini}, {Buldgen}, {Mosser}, {Noels}, {Scuflaire}, {Vrard}, {Willett}, {Davies}, {Hall}, {Nielsen}, {Khan}, {Rendle}, {van Rossem}, {Ferguson}, \& {Chaplin}}]{montalban2021}
{Montalb{\'a}n}, J., {Mackereth}, J.~T., {Miglio}, A., {et~al.} 2021, Nature Astronomy, 5, 640

\bibitem[{{Monty} {et~al.}(2020){Monty}, {Venn}, {Lane}, {Lokhorst}, \& {Yong}}]{monty2020}
{Monty}, S., {Venn}, K.~A., {Lane}, J. M.~M., {Lokhorst}, D., \& {Yong}, D. 2020, \mnras, 497, 1236

\bibitem[{{Mucciarelli}(2013)}]{4dao}
{Mucciarelli}, A. 2013, arXiv e-prints, arXiv:1311.1403

\bibitem[{{Mucciarelli} {et~al.}(2021{\natexlab{a}}){Mucciarelli}, {Bellazzini}, \& {Massari}}]{mucciarelli21}
{Mucciarelli}, A., {Bellazzini}, M., \& {Massari}, D. 2021{\natexlab{a}}, \aap, 653, A90

\bibitem[{{Mucciarelli} {et~al.}(2021{\natexlab{b}}){Mucciarelli}, {Massari}, {Minelli}, {Romano}, {Bellazzini}, {Ferraro}, {Matteucci}, \& {Origlia}}]{mucciarelli2021NatAs}
{Mucciarelli}, A., {Massari}, D., {Minelli}, A., {et~al.} 2021{\natexlab{b}}, Nature Astronomy, 5, 1247

\bibitem[{{Mucciarelli} {et~al.}(2013){Mucciarelli}, {Pancino}, {Lovisi}, {Ferraro}, \& {Lapenna}}]{gala}
{Mucciarelli}, A., {Pancino}, E., {Lovisi}, L., {Ferraro}, F.~R., \& {Lapenna}, E. 2013, \apj, 766, 78

\bibitem[{{Myeong} {et~al.}(2022){Myeong}, {Belokurov}, {Aguado}, {Evans}, {Caldwell}, \& {Bradley}}]{myeong22}
{Myeong}, G.~C., {Belokurov}, V., {Aguado}, D.~S., {et~al.} 2022, \apj, 938, 21

\bibitem[{{Myeong} {et~al.}(2018){Myeong}, {Evans}, {Belokurov}, {Sanders}, \& {Koposov}}]{myeong18}
{Myeong}, G.~C., {Evans}, N.~W., {Belokurov}, V., {Sanders}, J.~L., \& {Koposov}, S.~E. 2018, \apjl, 856, L26

\bibitem[{{Myeong} {et~al.}(2019){Myeong}, {Vasiliev}, {Iorio}, {Evans}, \& {Belokurov}}]{myeong19}
{Myeong}, G.~C., {Vasiliev}, E., {Iorio}, G., {Evans}, N.~W., \& {Belokurov}, V. 2019, \mnras, 488, 1235

\bibitem[{{Naidu} {et~al.}(2020){Naidu}, {Conroy}, {Bonaca}, {Johnson}, {Ting}, {Caldwell}, {Zaritsky}, \& {Cargile}}]{naidu20}
{Naidu}, R.~P., {Conroy}, C., {Bonaca}, A., {et~al.} 2020, \apj, 901, 48

\bibitem[{{Naidu} {et~al.}(2022){Naidu}, {Conroy}, {Bonaca}, {Zaritsky}, {Ting}, {Caldwell}, {Cargile}, {Speagle}, {Chandra}, {Johnson}, {Woody}, \& {Han}}]{naidu2022}
{Naidu}, R.~P., {Conroy}, C., {Bonaca}, A., {et~al.} 2022, arXiv e-prints, arXiv:2204.09057

\bibitem[{{Newton} {et~al.}(2018){Newton}, {Cautun}, {Jenkins}, {Frenk}, \& {Helly}}]{newton18}
{Newton}, O., {Cautun}, M., {Jenkins}, A., {Frenk}, C.~S., \& {Helly}, J.~C. 2018, \mnras, 479, 2853

\bibitem[{{Nissen} \& {Schuster}(1997)}]{NS97}
{Nissen}, P.~E. \& {Schuster}, W.~J. 1997, \aap, 326, 751

\bibitem[{{Nissen} \& {Schuster}(2010)}]{nissen&schuster2010}
{Nissen}, P.~E. \& {Schuster}, W.~J. 2010, \aap, 511, L10

\bibitem[{{Nomoto} {et~al.}(2013){Nomoto}, {Kobayashi}, \& {Tominaga}}]{nomoto2013}
{Nomoto}, K., {Kobayashi}, C., \& {Tominaga}, N. 2013, \araa, 51, 457

\bibitem[{{Oria} {et~al.}(2022){Oria}, {Tenachi}, {Ibata}, {Famaey}, {Yuan}, {Arentsen}, {Martin}, \& {Viswanathan}}]{oria2022}
{Oria}, P.-A., {Tenachi}, W., {Ibata}, R., {et~al.} 2022, \apjl, 936, L3

\bibitem[{{Pedregosa} {et~al.}(2012){Pedregosa}, {Varoquaux}, {Gramfort}, {Michel}, {Thirion}, {Grisel}, {Blondel}, {M{\"u}ller}, {Nothman}, {Louppe}, {Prettenhofer}, {Weiss}, {Dubourg}, {Vanderplas}, {Passos}, {Cournapeau}, {Brucher}, {Perrot}, \& {Duchesnay}}]{pedregosa12}
{Pedregosa}, F., {Varoquaux}, G., {Gramfort}, A., {et~al.} 2012, arXiv e-prints, arXiv:1201.0490

\bibitem[{{Pietrinferni} {et~al.}(2021){Pietrinferni}, {Hidalgo}, {Cassisi}, {Salaris}, {Savino}, {Mucciarelli}, {Verma}, {Silva Aguirre}, {Aparicio}, \& {Ferguson}}]{pietrinferni2021}
{Pietrinferni}, A., {Hidalgo}, S., {Cassisi}, S., {et~al.} 2021, \apj, 908, 102

\bibitem[{{Pignatari} {et~al.}(2010){Pignatari}, {Gallino}, {Heil}, {Wiescher}, {K{\"a}ppeler}, {Herwig}, \& {Bisterzo}}]{pignatari2010}
{Pignatari}, M., {Gallino}, R., {Heil}, M., {et~al.} 2010, \apj, 710, 1557

\bibitem[{{Pillepich} {et~al.}(2015){Pillepich}, {Madau}, \& {Mayer}}]{pillepich2015}
{Pillepich}, A., {Madau}, P., \& {Mayer}, L. 2015, \apj, 799, 184

\bibitem[{{Reddy} {et~al.}(2006){Reddy}, {Lambert}, \& {Allende Prieto}}]{reddy06}
{Reddy}, B.~E., {Lambert}, D.~L., \& {Allende Prieto}, C. 2006, \mnras, 367, 1329

\bibitem[{{Reddy} {et~al.}(2003){Reddy}, {Tomkin}, {Lambert}, \& {Allende Prieto}}]{reddy2003}
{Reddy}, B.~E., {Tomkin}, J., {Lambert}, D.~L., \& {Allende Prieto}, C. 2003, \mnras, 340, 304

\bibitem[{{Reggiani} {et~al.}(2017){Reggiani}, {Mel{\'e}ndez}, {Kobayashi}, {Karakas}, \& {Placco}}]{reggiani2017}
{Reggiani}, H., {Mel{\'e}ndez}, J., {Kobayashi}, C., {Karakas}, A., \& {Placco}, V. 2017, \aap, 608, A46

\bibitem[{{Reid} \& {Brunthaler}(2004)}]{reidandbrunthaler04}
{Reid}, M.~J. \& {Brunthaler}, A. 2004, \apj, 616, 872

\bibitem[{{Rey} {et~al.}(2023){Rey}, {Agertz}, {Starkenburg}, {Renaud}, {Joshi}, {Pontzen}, {Martin}, {Feuillet}, \& {Read}}]{rey23}
{Rey}, M.~P., {Agertz}, O., {Starkenburg}, T.~K., {et~al.} 2023, \mnras, 521, 995

\bibitem[{{Riello} {et~al.}(2021){Riello}, {De Angeli}, {Evans}, {Montegriffo}, {Carrasco}, {Busso}, {Palaversa}, {Burgess}, {Diener}, {Davidson}, {Rowell}, {Fabricius}, {Jordi}, {Bellazzini}, {Pancino}, {Harrison}, {Cacciari}, {van Leeuwen}, {Hambly}, {Hodgkin}, {Osborne}, {Altavilla}, {Barstow}, {Brown}, {Castellani}, {Cowell}, {De Luise}, {Gilmore}, {Giuffrida}, {Hidalgo}, {Holland}, {Marinoni}, {Pagani}, {Piersimoni}, {Pulone}, {Ragaini}, {Rainer}, {Richards}, {Sanna}, {Walton}, {Weiler}, \& {Yoldas}}]{riello2021}
{Riello}, M., {De Angeli}, F., {Evans}, D.~W., {et~al.} 2021, \aap, 649, A3

\bibitem[{{Roederer} {et~al.}(2014){Roederer}, {Preston}, {Thompson}, {Shectman}, {Sneden}, {Burley}, \& {Kelson}}]{roederer2014}
{Roederer}, I.~U., {Preston}, G.~W., {Thompson}, I.~B., {et~al.} 2014, \aj, 147, 136

\bibitem[{{Romano} {et~al.}(2010){Romano}, {Karakas}, {Tosi}, \& {Matteucci}}]{romano2010}
{Romano}, D., {Karakas}, A.~I., {Tosi}, M., \& {Matteucci}, F. 2010, \aap, 522, A32

\bibitem[{{Ruiz-Lara} {et~al.}(2022){Ruiz-Lara}, {Matsuno}, {L{\"o}vdal}, {Helmi}, {Dodd}, \& {Koppelman}}]{ruiz-lara22}
{Ruiz-Lara}, T., {Matsuno}, T., {L{\"o}vdal}, S.~S., {et~al.} 2022, \aap, 665, A58

\bibitem[{{Sch{\"o}nrich} {et~al.}(2010){Sch{\"o}nrich}, {Binney}, \& {Dehnen}}]{schonrich2010}
{Sch{\"o}nrich}, R., {Binney}, J., \& {Dehnen}, W. 2010, \mnras, 403, 1829

\bibitem[{{Smiljanic} {et~al.}(2016){Smiljanic}, {Romano}, {Bragaglia}, {Donati}, {Magrini}, {Friel}, {Jacobson}, {Randich}, {Ventura}, {Lind}, {Bergemann}, {Nordlander}, {Morel}, {Pancino}, {Tautvai{\v{s}}ien{\.{e}}}, {Adibekyan}, {Tosi}, {Vallenari}, {Gilmore}, {Bensby}, {Fran{\c{c}}ois}, {Koposov}, {Lanzafame}, {Recio-Blanco}, {Bayo}, {Carraro}, {Casey}, {Costado}, {Franciosini}, {Heiter}, {Hill}, {Hourihane}, {Jofr{\'e}}, {Lardo}, {de Laverny}, {Lewis}, {Monaco}, {Morbidelli}, {Sacco}, {Sbordone}, {Sousa}, {Worley}, \& {Zaggia}}]{smiljanic2016}
{Smiljanic}, R., {Romano}, D., {Bragaglia}, A., {et~al.} 2016, \aap, 589, A115

\bibitem[{{Stephens} \& {Boesgaard}(2002)}]{stephens2002}
{Stephens}, A. \& {Boesgaard}, A.~M. 2002, \aj, 123, 1647

\bibitem[{{Stetson} \& {Pancino}(2008)}]{daospec}
{Stetson}, P.~B. \& {Pancino}, E. 2008, \pasp, 120, 1332

\bibitem[{{Strassmeier} {et~al.}(2015){Strassmeier}, {Ilyin}, {J{\"a}rvinen}, {Weber}, {Woche}, {Barnes}, {Bauer}, {Beckert}, {Bittner}, {Bredthauer}, {Carroll}, {Denker}, {Dionies}, {DiVarano}, {D{\"o}scher}, {Fechner}, {Feuerstein}, {Granzer}, {Hahn}, {Harnisch}, {Hofmann}, {Lesser}, {Paschke}, {Pankratow}, {Plank}, {Pl{\"u}schke}, {Popow}, \& {Sablowski}}]{pepsi}
{Strassmeier}, K.~G., {Ilyin}, I., {J{\"a}rvinen}, A., {et~al.} 2015, Astronomische Nachrichten, 336,, 336, 324

\bibitem[{{Tenachi} {et~al.}(2022){Tenachi}, {Oria}, {Ibata}, {Famaey}, {Yuan}, {Arentsen}, {Martin}, \& {Viswanathan}}]{tenachi22}
{Tenachi}, W., {Oria}, P.-A., {Ibata}, R., {et~al.} 2022, \apjl, 935, L22

\bibitem[{{Tinsley}(1979)}]{tinsley79}
{Tinsley}, B.~M. 1979, \apj, 229, 1046

\bibitem[{{Tolstoy} {et~al.}(2009){Tolstoy}, {Hill}, \& {Tosi}}]{tolstoy09}
{Tolstoy}, E., {Hill}, V., \& {Tosi}, M. 2009, \araa, 47, 371

\bibitem[{{Tolstoy} {et~al.}(2023){Tolstoy}, {Sk{\'u}lad{\'o}ttir}, {Battaglia}, {Brown}, {Massari}, {Irwin}, {Starkenburg}, {Salvadori}, {Hill}, {Jablonka}, {Salaris}, {van Essen}, {Olsthoorn}, {Helmi}, \& {Pritchard}}]{tolstoy2023}
{Tolstoy}, E., {Sk{\'u}lad{\'o}ttir}, {\'A}., {Battaglia}, G., {et~al.} 2023, \aap, 675, A49

\bibitem[{{Tonry} \& {Davis}(1979)}]{tonry&davies1979}
{Tonry}, J. \& {Davis}, M. 1979, \aj, 84, 1511

\bibitem[{{Vasiliev}(2019)}]{vasiliev19}
{Vasiliev}, E. 2019, \mnras, 482, 1525

\bibitem[{{Vincenzo} {et~al.}(2019){Vincenzo}, {Spitoni}, {Calura}, {Matteucci}, {Silva Aguirre}, {Miglio}, \& {Cescutti}}]{vincenzo2019}
{Vincenzo}, F., {Spitoni}, E., {Calura}, F., {et~al.} 2019, \mnras, 487, L47

\bibitem[{{Weinberg} {et~al.}(2022){Weinberg}, {Holtzman}, {Johnson}, {Hayes}, {Hasselquist}, {Shetrone}, {Ting}, {Beaton}, {Beers}, {Bird}, {Bizyaev}, {Blanton}, {Cunha}, {Fern{\'a}ndez-Trincado}, {Frinchaboy}, {Garc{\'\i}a-Hern{\'a}ndez}, {Griffith}, {Johnson}, {J{\"o}nsson}, {Lane}, {Leung}, {Mackereth}, {Majewski}, {M{\'e}sz{\'a}ros}, {Nitschelm}, {Pan}, {Schiavon}, {Schneider}, {Schultheis}, {Smith}, {Sobeck}, {Stassun}, {Stringfellow}, {Vincenzo}, {Wilson}, \& {Zasowski}}]{weinberg2022}
{Weinberg}, D.~H., {Holtzman}, J.~A., {Johnson}, J.~A., {et~al.} 2022, \apjs, 260, 32

\bibitem[{{White} \& {Frenk}(1991)}]{white&frenk1991}
{White}, S. D.~M. \& {Frenk}, C.~S. 1991, \apj, 379, 52

\bibitem[{{Woosley} {et~al.}(2002){Woosley}, {Heger}, \& {Weaver}}]{woosley2002}
{Woosley}, S.~E., {Heger}, A., \& {Weaver}, T.~A. 2002, Reviews of Modern Physics, 74, 1015

\bibitem[{{Woosley} \& {Weaver}(1995)}]{woosley&weaver1995}
{Woosley}, S.~E. \& {Weaver}, T.~A. 1995, \apjs, 101, 181

\bibitem[{{Yuan} {et~al.}(2020){Yuan}, {Myeong}, {Beers}, {Evans}, {Lee}, {Banerjee}, {Gudin}, {Hattori}, {Li}, {Matsuno}, {Placco}, {Smith}, {Whitten}, \& {Zhao}}]{yuan2020}
{Yuan}, Z., {Myeong}, G.~C., {Beers}, T.~C., {et~al.} 2020, \apj, 891, 39

\bibitem[{{Zwitter} {et~al.}(2018){Zwitter}, {Kos}, {Chiavassa}, {Buder}, {Traven}, {{\v{C}}otar}, {Lin}, {Asplund}, {Bland-Hawthorn}, {Casey}, {De Silva}, {Duong}, {Freeman}, {Lind}, {Martell}, {D'Orazi}, {Schlesinger}, {Simpson}, {Sharma}, {Zucker}, {Anguiano}, {Casagrande}, {Collet}, {Horner}, {Ireland}, {Kafle}, {Lewis}, {Munari}, {Nataf}, {Ness}, {Nordlander}, {Stello}, {Ting}, {Tinney}, {Watson}, {Wittenmyer}, \& {{\v{Z}}erjal}}]{zwitter18}
{Zwitter}, T., {Kos}, J., {Chiavassa}, A., {et~al.} 2018, \mnras, 481, 645

\end{thebibliography}

\end{document}